\documentclass[aps,prb,twocolumn,showpacs,floatfix]{revtex4-1}

\usepackage{graphicx,color}
\usepackage{amsfonts}
\usepackage[figuresright]{rotating}  
\usepackage{amssymb}
\usepackage{amsmath}

\usepackage{bbold}
\usepackage{wrapfig,lipsum,booktabs}
\usepackage{mathtools}
\usepackage{psfrag}
\usepackage{subfigure}
\usepackage{multirow}
\usepackage{tabularx}
\usepackage{textcomp}
\usepackage{bm}
\usepackage{hyperref}
\usepackage{relsize}
\hypersetup{
 pdfnewwindow=true, colorlinks=true,
 linkcolor=blue, anchorcolor=blue,
 citecolor=blue, filecolor=blue,
 menucolor=blue, urlcolor=blue}

\def\nn{\nonumber}

\def\beq{\begin{eqnarray}}
\def\eeq{\end{eqnarray}}

\def\up{\uparrow}
\def\down{\downarrow}

\renewcommand{\v}[1]{\ensuremath{\mathbf{#1}}} 
 
\let\baraccent=\= 
\renewcommand{\=}[1]{\stackrel{#1}{=}} 

\makeatletter

\begin{document}

\title{Bulk photovoltaic effects in the presence of a static electric field}

\author{Benjamin\ \surname{M. Fregoso}}
\affiliation{Department of Physics, Kent State University, Kent, Ohio 44242, USA}

\begin{abstract}
This paper presents a study of dc photocurrents in biased insulators to the third order in the electric field. We find three photocurrents which are characterized by physical divergences of the third-order free-electron polarization susceptibility. In the absence of momentum relaxation and saturation effects, these dc photocurrents grow as $t^n$  $(n=2,1,0)$ with illumination time. The photocurrents are dubbed \textit{jerk}, third-order injection, and third-order shift current, respectively, and are generalizations of the second-order injection and shift currents of the bulk photovoltaic effect. We also revisit the theory of the bulk photovoltaic effect and include Fermi surface contributions which are important in metals. Finally, we show that injection, shift, and jerk currents admit simple physical interpretations in terms of semiclassical wave packet dynamics in electric fields. Experimental signatures and extensions to higher-order susceptibilities are also discussed.

\end{abstract}

\maketitle

\section{Introduction and main results} 
\label{sec:intro}
Electrons in crystals can exhibit fascinating dynamics in the presence of external electric and magnetic fields. In metals, the anomalous Hall effect~\cite{Karplus1954,Nagaosa2010} or the chiral anomaly in Weyl semimetals~\cite{Armitage2018,Vishwanath2015} are two examples. Insulators, despite lacking a Fermi surface, can also exhibit nontrivial carrier dynamics as in the bulk photovoltaic effects (BPVE).\cite{Auston1972,Glass1974,Koch1976,Belinicher1980,Sturman1992,Belinicher1988,Baltz1981,Laman1999,Laman2005,Sipe2000,Driel2001,Cote2002,Ghalgaoui2018,Bieler2005,Bieler2007,Rioux2011,Rioux2012,Somma2014,Nakamura2016,Holtz2016,Rappe2017,Spanier2016,Tan2016,Rangel2017,Ibanez-Azpiroz2018,Panday,Wang,Fregoso2017,Kushnir2017,Nakamura2017,Ogawa2017,Kushnir2019,Burger2019,Sotome2019,Juan2017,Rees} The BPVE is the generation of a dc photocurrent in homogeneous insulators or semiconductors that lack inversion symmetry. 

In the presence of a static $\v{E}_0$ and an optical field $\v{E}$, the dc current can be expanded in powers of the electric fields. Schematically we can write~\cite{Sturman1992}

\begin{align}
\v{J}_{dc} = \sigma^{(1)}_{dark} \v{E}_0 + \sigma^{(2)}_{dark} \v{E}_0 + \sigma^{(2)}_{bpve} \v{E}^2 + \sigma^{(3)}_{ph} \v{E}^2 \v{E}_0 +\cdots.  
\label{eq:dc_current_taylor}
\end{align}
The first and second terms are the linear and quadratic dc conductivity in the absence of illumination. The third is the BPVE and the fourth is the photoconductivity, i.e., the intensity-dependent dc conductivity. In insulators, the BPVE is usually the dominant contribution. In this article we first review the theory of the BPVE (including Fermi surface contributions), and then we extend it to study the photoconductivity.  
 
The peculiar nature of the BPVE was first noticed by (1) its dependence on the intensity of light, (2) its large open-circuit photovoltages, and (3) its dependence on light polarization~\cite{Auston1972,Glass1974,Koch1976}. (1) indicates that the BPVE is quadratic in the optical field, (2) indicates that the BPVE is an ultrafast effect in which transport occurs before carriers pretermalize at the bottom of the conduction band (top of the valence band), and (3) indicates that the BPVE response tensor is complex and has two components. The real part $\sigma_2$ couples to the real electric fields and the imaginary part $\eta_2$ to the imaginary electric fields, schematically\cite{Sturman1992} 

\begin{align}
\v{J}_{dc,pbve}^{(2)} = \sigma_2 |\v{E}|^2 + \eta_2 \v{E}\times\v{E}^{*}.  
\end{align}
This led to the first successful phenomenological theory of both components of the BPVE, namely, the injection current, also called circular photogalvanic effect (CPGE), represented by $\eta_2$ and shift current represented by $\sigma_2$. 

The lack of inversion symmetry could manifest in two distinct scenarios in the BPVE. In the first scenario, photoexcited carriers relax momentum asymmetrically into $\pm \v{k}$ directions via collisions with other electrons, phonons or impurities. This leads to a polar distribution and a net current~\cite{Belinicher1980,Belinicher1988,Sturman1992}. In the second scenario, the origin of the BPVE is the light-matter interactions not the dynamics of momentum relaxation. In injection current processes, light pumps carriers into velocity-carrying states asymmetrically at $\pm \v{k}$ points in the Brillouin zone (BZ) leading to a polar distribution and a net current~\cite{Sturman1992,Sipe2000}.  Within a simple relaxation time approximation, the steady state injection current is proportional to the first power of the relaxation time constant and vanishes for linearly polarized light.

In shift current processes, inversion symmetry breaking manifests as a separation of the centers of charge of the valence and conduction bands so that charge moves coherently across the unit cell upon carrier photoexcitation from valence to conduction band~\cite{Baltz1981}. The shift current vanishes for circular polarization of light and decays in the time scale of the quantum coherence of the solid.

\begin{table*}
\caption{Summary of bulk photovoltaic effects (BPVEs) obtained from divergences of free electric polarization susceptibilities.  The standard BPVEs are derived from the singularities of $\chi_2$. Higher-order BPVEs can be classified by their dependence on illumination time in the absence of momentum relaxation and saturation effects, e.g., $\eta_2$, $\eta_3$ and $\eta_4$ are all injection current responses and $\sigma_2$, $\sigma_3$ are all shift currents responses. We write susceptibilities as, $\chi_{n}^{abc...}(-\omega_{\Sigma},\omega_{\beta},\omega_{\sigma},...)$ where $b,c...$ are Cartesian indices, $\omega_{\beta},\omega_{\sigma},...$ are frequency components, and $\omega_{\Sigma}=\omega_{\beta}+\omega_{\sigma}+...$ frequency sums~\cite{Boyd2008}. $[X,Y]$~($\{X,Y\}$) indicate commutation (anticommutation) with respect to $b,c$ indices only. 
Other conventions are explained in Sec.~\ref{sec:notation}.}
 \begin{center}
 \scriptsize
 \begin{tabular}{|m{1.6cm}| m{1cm}| m{8.8cm}|m{1.1cm}| m{3.3cm}|m{1.0cm} |} 
   \hline
   BPVE & Symbol & Expression & Time dependence $\sim t^{\alpha}$ & Origin & Ref.  \\ [0.5ex] 
   \hline\hline
	  Injection   & $\eta_2^{abc}$   & 	
		$ \frac{\pi e^3}{2\hbar^2 V}\sum_{nm\v{k}} f_{mn}~ \omega_{nm;a} [r^b_{nm}, r^c_{mn}] \delta(\omega_{nm}-\omega)$                
		                                                & $1$ &\multirow{2}{8em}{$\chi_2(0,\omega,-\omega)~\to~\infty $}  & \onlinecite{Aversa1995} \\ [4ex]
	  Shift       & $\sigma_2^{abc}$ & 
		$\frac{i\pi e^3}{2 \hbar^2 V}\sum_{nm\v{k}} f_{mn}  \{r^c_{nm;a}, r^b_{mn}\} \delta(\omega_{nm}-\omega)$
                                                    & $0$ &                                               & \onlinecite{Aversa1995} \\	[4ex]
   \hline
    Jerk               & $\iota_3^{abcd}$  & 
		$\frac{\pi e^4}{3\hbar^3 V}\sum_{nm\v{k}} f_{mn} \big[ 2\omega_{nm;ad} r^b_{nm}r^c_{mn}+ \omega_{nm;a}  (r^b_{nm}r^c_{mn})_{;d} \big] \delta(\omega_{nm}-\omega)$ 
		                                                & $2$  & \multirow{3}{8em}{$\chi_3(0,\omega,-\omega,0)~\to~\infty$} & \onlinecite{Fregoso2018} \\[4ex]
     Injection & $\eta_3$    & Eq.~\ref{eq:eta3} & $1$   &                                          & present \\[2ex]
	   Shift     & $\sigma_3$  & Eq.~\ref{eq:sigma3} & $0$   &                                                 & present \\[2ex]
%
%
    \hline
    Injection          & $\nu_3$     & Eq.~7         & $1$   & \multirow{2}{8em}{$\chi_3(0,-2\omega,\omega,\omega)~\to~\infty$} 
		                                                                                                                & \onlinecite{Aversa1996} \\[4ex]
    Shift              & $\sigma_3$  & Eq.~8         & $0$   &                                                 & \onlinecite{Aversa1996} \\[4ex]
   \hline
    Snap               & $\varsigma_4$ & Eq.~\ref{eq:Jsnap} & $3$  &  \multirow{4}{20em}{$\chi_4(0,\omega,-\omega,0,0)~\to~\infty$}                                                                                                                 & present \\[4ex]
%
%
    Jerk               & $\iota_4$     &                    & $2$  &                                        &         \\[4ex]
    Injection          & $\eta_4$      &                    & $1$  &                                        &         \\[4ex]
    shift              & $\sigma_4$    &                    & $0$  &                                        &         \\[4ex]

%
%
	\hline
	  Any               &     $a_n$          &   
		$\frac{1}{2\pi i}\oint_{|z|=\rho}dz  \frac{\chi_{n}}{z^{l+1}},~~~z=-i\omega_{\Sigma},~\rho\to 0,~ l=-n,\cdots,-1$, Eq.~\ref{eq:gen_bpve}
								                                       &   $\alpha=n-1,...,0$          & $\chi_n(\omega_{\Sigma},\omega_{\beta},\omega_{\sigma},...)\to\infty$, $\omega_{\Sigma}=\omega_{\beta}+\omega_{\sigma}+\cdots\to 0$    & present \\ [6ex]
   \hline	
 \end{tabular}
 \end{center}
\label{table:chi_n}
\end{table*}

The BPVE has been extensively studied since the 1960s in ferroelectrics mainly in the context of photovoltaic applications. The injection current, the shift current, or both have been reported in many materials,\cite{Auston1972,Glass1974,Koch1976,Laman1999,Laman2005,Cote2002,Ghalgaoui2018,Bieler2005,Bieler2007,Somma2014,Nakamura2016,Holtz2016,Spanier2016,Kushnir2017,Nakamura2017,Ogawa2017,Kushnir2019,Burger2019,Sotome2019,Rees} including GaAs,\cite{Cote2002,Ghalgaoui2018} CdSe,\cite{Laman1999,Laman2005} CdS,\cite{Laman2005} quantum wells,\cite{Bieler2005,Bieler2007} RhSi,\cite{Rees} and Bi$_{12}$GeO$_{20}$.\cite{Burger2019} 
More recently, the BPVE has attracted attention for its promise in novel optoelectric applications;~\cite{Rappe2017,Spanier2016,Tan2016} specifically in two-dimensional (2D) ferroelectrics.\cite{Kushnir2017,Fregoso2017,Rangel2017,Ibanez-Azpiroz2018,Panday,Wang}

Following Sipe and coworkers\cite{Sipe2000}, BPVE response tensors can be derived from the perspective of divergent polarization susceptibilities. In this approach, the BPVE arises from light-matter interactions and not from momentum relaxation processes; the latter are included phenomenologically a \textit{posteriori}. For not too large electric fields, the insulator's response to an external electric field is described perturbatively by susceptibilities $\chi_n$ as
\begin{align}
\v{P} = \v{P}_0 + \chi_1 \v{E} + \chi_2 \v{E}^2 + \chi_3 \v{E}^3 + \cdots,
\end{align}
where $\v{P}_0$ is the electric polarization in the absence of an external electric field,~\cite{King-Smith1993,Resta1994} $\chi_1$ is the linear susceptibility, and $\chi_2, \chi_3,...$ are nonlinear susceptibilities.~\cite{comment_1} 

The electric polarization in insulators is commonly thought to be determined by the off-diagonal elements of the density matrix because these elements describe the displacement of charge from its equilibrium position in the presence of an electric field. Intraband processes, however, have been shown to be important.\cite{Aspnes1972,Aversa1995,Culcer2017} Among other things they cure unphysical divergences in susceptibilities in the dc limit by incorporating the fact that the intraband motion of Bloch electrons cannot accelerate indefinitely in insulators~\cite{Aspnes1972,Aversa1995}. Importantly, when intraband and interband processes are taken into account on an equal footing divergent susceptibilities represent real photocurrents. 

Consider, for example, the dc divergences of $\chi_2$. If we denote the amplitude of the electric field by $E^{b}=\sum_{\beta} E^{b}_{\beta} e^{-i\omega_{\beta} t}$, the polarization to second order

\begin{align}
P^{a(2)} = \sum_{b\beta c\sigma} \chi_{2}^{abc}(-\omega_{\Sigma},\omega_{\beta},\omega_{\sigma}) E^{b}_{\beta} E^{c}_{\sigma}e^{-i\omega_{\Sigma}t},
\end{align}
oscillates with frequency $\omega_{\Sigma}=\omega_{\beta}+\omega_{\sigma}$ in the long-time limit. The intraband part of the susceptibility $\chi_{2i}$ in
\begin{align}
\chi_2=\chi_{2i}+\chi_{2e},
\end{align}
can be expanded in powers of $\omega_{\Sigma}$ as~\cite{Aversa1995,Sipe2000}
\begin{align}
(-i\omega_{\Sigma})^2\chi_{2i} = \eta_2 + (-i\omega_{\Sigma}) \sigma_{2} + \cdots,
\label{eq:chi2_taylor_v1}
\end{align}
or equivalently 

\begin{align}
\chi_{2i} = \frac{\eta_2}{z^2} + \frac{\sigma_{2}}{z} + \cdots,
\label{eq:chi2_laurent}
\end{align}
where $z=-i\omega_{\Sigma}$. Clearly, $\chi_{2i}$ diverges at zero frequency sum. Since $\chi_{2e}$ is regular as $\omega_{\Sigma}\to 0$, $\chi_{2}$ itself diverges at zero frequency sum. As a side note, in \textit{metals}, the Fermi surface adds other divergent dc contributions to the second-order BPVE, see Sec.~\ref{sec:NLHE}.

Assuming a monocromatic optical field and using Maxwell equation  
\begin{align}
\frac{d \v{P}}{d t} &=\v{J},
\label{eq:dP_J_v1}
\end{align} 
Eq.~\ref{eq:chi2_taylor_v1} implies $\eta_2$ and $\sigma_2$ are response functions of the nonlinear currents 
\begin{align}
\frac{d}{dt} J^{a(2)}_{inj} \equiv 2 \sum_{bc} \eta_2^{abc}(0,\omega,-\omega) E^b(\omega) E^c(-\omega),
\label{eq:eta2_timeder_v1}\\
J^{a(2)}_{sh} \equiv 2 \sum_{bc}\sigma_2^{abc}(0,\omega,-\omega) E^b(\omega) E^c(-\omega).
\label{eq:sigma2_timeder_v1} 
\end{align}
$\eta_2$ and $\sigma_2$ are the standard injection and shift current response functions derived from the susceptibility approach.\cite{Sipe2000} Importantly, they vanish for frequencies smaller than the energy gap (they are `resonant'). The dots in Eq.~\ref{eq:chi2_taylor_v1} are associated with the (nonresonant) rectification currents.\cite{Bass1962,Nastos2010} 

In the absence of momentum relaxation and saturation effects the injection and shift currents grow with illumination time as
\begin{align}
|J^{a(2)}_{inj}| &\propto  \eta_2 t,\\
|J^{a(2)}_{sh}| &\propto  \sigma_2.
\end{align}
In this article we study how the injection and shift currents are modified by the presence of a static field, i.e., the fourth term in Eq.~\ref{eq:dc_current_taylor}. We use the method of finding divergences of the free third-order electric polarization susceptibility $\chi_3$. Biased irradiated semiconductors of this kind have been extensively studied numerically using the semiclassical Boltzmann equation.\cite{Jepsen1996} As shown below, this approach misses some important quantum effects which are recovered in the susceptibility approach. In summary, the third order polarization   

\begin{align}
P^{a(3)} = \sum_{b\beta c\sigma d\delta} \chi_{3}^{abcd}(-\omega_{\Sigma},\omega_{\beta},\omega_{\sigma},\omega_{\delta}) E^{b}_{\beta} E^{c}_{\sigma}E^{d}_{\delta} e^{-i\omega_{\Sigma}t},
\end{align}
oscillates with frequency $\omega_{\Sigma}=\omega_{\beta}+\omega_{\sigma} + \omega_{\delta}$ in the long-time limit. We show that the intraband part, $\chi_{3i}$, of $\chi_3 = \chi_{3i} + \chi_{3e}$ admits the Taylor expansion

\begin{align}
(-i\omega_{\Sigma})^3\chi_{3i} = \iota_3 + (-i\omega_{\Sigma})\eta_{3} + (-i\omega_{\Sigma})^2 \sigma_{3} + \cdots,
\label{eq:taylor_chi3}
\end{align}
or alternatively the Laurent series  
\begin{align}
\chi_{3i} = \frac{\iota_3}{z^3} + \frac{\eta_{3}}{z^2} + \frac{\sigma_{3}}{z} +\cdots,
\label{eq:laurent_chi3}
\end{align}
where $z=-i\omega_{\Sigma}$ and $\iota_3,~\eta_3,~\sigma_3$ are (resonant) residues. Clearly, $\chi_{3i}$ diverges in the dc limit ($\omega_{\Sigma}=0$) and similar to $\eta_2$ and $\sigma_2$, $\iota_3$, $\eta_3$ and $\sigma_3$ represent response functions of nonlinear currents

\begin{align}
\frac{d^2}{dt^2} J^{a(3)}_{jerk} \equiv 6 \sum_{bcd} \iota_3^{abcd}(0,\omega,-\omega,0) E^b(\omega) E^c(-\omega) E^d_0 
\label{eq:iota3_intro}\\
\frac{d}{dt} J^{a(3)}_{inj} \equiv 6 \sum_{bcd} \eta_3^{abcd}(0,\omega,-\omega,0) E^b(\omega) E^c(-\omega)E^d_0 
\label{eq:eta3_intro}\\
J^{a(3)}_{sh} \equiv 6 \sum_{bcd}\sigma_3^{abcd}(0,\omega,-\omega,0) E^b(\omega) E^c(-\omega)E^d_0.
\label{eq:sigma3_intro} 
\end{align}
The difference is that a static field (zero frequency) is taken into account in addition to a monochromatic optical field.  In the absence of momentum relaxation and saturation effects the currents vary as $t^2,t,t^0$ with illumination time and we dub them \textit{jerk}, third-order injection current and third-order shift current, respectively. The dots in Eq.~(\ref{eq:laurent_chi3}) represent regular terms associated with rectification currents. 

Since $\chi_{3e}$ is regular in the dc limit, one can write the same expansion as in Eq.~(\ref{eq:laurent_chi3}) for both $\chi_{3i}$ and $\chi_3$. Similarly, the third order conductivity which is defined by

\begin{align}
J^{a(3)} \equiv \sum_{b\beta c\sigma d\delta} \sigma^{abcd(3)}(-\omega_{\Sigma},\omega_{\beta},\omega_{\sigma},\omega_{\delta}) E^{b}_{\beta} E^{c}_{\sigma}E^{d}_{\delta} e^{-i\omega_{\Sigma}t},
\end{align}
admits the expansion

\begin{align}
\sigma^{(3)} = \frac{\iota_3}{z^2} + \frac{\eta_{3}}{z} + \sigma_{3} +\cdots.
\label{eq:laurent_conductivity}
\end{align}
The subsequent evolution of charge distribution in the sample involves not only the above generation processes but also the macroscopic current dynamics in a sample for which momentum and energy relaxation is crucial. In the presence of dissipation, the dc divergences will be cut off by a momentum relaxation time scale, just as the dc divergence of metals in the Drude model is cut off by a momentum relaxation time. In the BPVE, we expect two main relaxation time scales. One is the relaxation time scale of the diagonal elements of the density matrix, $\tau_1$, which $\iota_3$, $\eta_3$, and $\eta_2$ depend on. This could be of the order of 100 fs or longer in clean semiconductors.~\cite{Li2018} The second is the relaxation time scale of the off-diagonal elements of the density matrix, $\tau_2$, which $\sigma_3$ and $\sigma_2$ depend on. Typically, $\tau_2 < \tau_1$, but a recent experiment found $\tau_2$ to be as large as 250 fs.~\cite{Ghalgaoui2018} For weakly disordered semiconductors, the photoconductivity ($\omega_{\Sigma} = 0$) in Eq.~\ref{eq:dc_current_taylor}  becomes

\begin{align}
\sigma^{(3)}_{ph} \sim \tau_1^2 \iota_3 + \tau_1 \eta_{3} + \sigma_{3}.
\label{eq:laurent_conductivity_dissorder}
\end{align}

We can generalize the above results to any power in the electric fields. In general, with each additional power in the electric field, $\chi_{ni}$ has an additional frequency factor in the denominator. This means that the dc singularities of $\chi_{ni}$ are, at most, of the order $n$. We can show that the $n$th order $z=0$ singularities of $\chi_n$ ($n\geq 2$), represent photocurrents which vary as $t^n$ in the absence of momentum relaxation and saturation effects. This occurs  when all but two of the external frequencies are zero. In addition, there is a hierarchy of higher order shift, injection,..., currents which are represented by $z=0$ singularities of order $1,2,3,..n$ of $\chi_{n}$. Formally $\chi_{n}$ (or $\sigma^{(n)}$) can be expanded as
\begin{align}
\chi_{n} = \sum_{l=-n}^{\infty} a_l z^{l},
\end{align} 
where $a_l=0$ for frequencies less than the gap and hence the residues are 
\begin{align}
a_l = \frac{1}{2\pi i}\oint_{|z|=\rho} \frac{\chi_{n} ~d z}{z^{l+1}}.
\label{eq:gen_bpve_1}
\end{align}
The poles of $\chi_{n}$ may be of lower order than $n$ when the optical field is not monocromatic; see, for example, the fourth row in Table~\ref{table:chi_n} where the field's frequencies are $\omega$ and $ 2\omega$. 

Importantly, we give simple physical arguments to explain the microscopic processes involved in $\iota_3,~\eta_3,~\sigma_3$ and $\sigma_2$ and provide explicit expressions in terms of material parameters amenable for first principles computations. To have a sense of the magnitude of these currents, we calculate them in single-layer GeS using a two-dimensional (2D) tight-binding model. 

The article is organized as follows. In Sec.~\ref{sec:notation} we describe the conventions used in this paper. In Sec.~\ref{sec:Ham}, \ref{sec:polarization_op}, and \ref{sec:current_op} we introduce the Hamiltonian, polarization, and current operators. In Sec.~\ref{sec:intra_curr} we revisit the calculation of the intraband current following Sipe and Shkrebtii.\cite{Sipe2000} In Sec.~\ref{sec:chi2} we rederive the expressions for the injection and shift current responses giving simple physical interpretations based on semiclassical wave packet dynamics in electric fields. In Sec.~\ref{sec:NLHE} we include Fermi surface contributions to the second-order BPVE. We then study the physical divergences of $\chi_3$ at zero frequency in  Sec.~\ref{sec:chi3}, \ref{sec:jerk_current}, \ref{sec:modified_injection}, and \ref{sec:mod_shift}. The jerk current has been presented previously and is included here only for completeness~\cite{Fregoso2018}. BPVEs arising from singularities of $\chi_n$ $(n>3)$ are discussed in Sec.~\ref{sec:snap_current}. Experimental signatures of jerk, third-order injection, and third-order shift current in single-layer GeS are summarized in Sec.\ref{sec:exp_sign} A summary of the BPVEs in insulators is presented in Table~\ref{table:chi_n}. Details of the derivations are given in the appendices.

\section{Notation}
\label{sec:notation}
To keep the notation under control we often omit the independent variables such as time, real space position, or crystal momentum, specially in expressions which are diagonal in these variables. 

We use the standard notation for the $n$th electric polarization susceptibility,\cite{Boyd2008} $\chi_{n}^{abc...}(-\omega_{\Sigma},\omega_{\beta},\omega_{\sigma},...)$, where $\omega_{\beta},\omega_{\sigma}$,... label external frequency components, $abc,...$ label Cartesian components, and $\omega_{\Sigma}=\omega_{\beta}+\omega_{\sigma}+...$ the frequency sum. We often write $\chi_n^{abc...}$ or  simply $\chi_n$ for brevity absorbing a free permittivity factor $\epsilon_0$ into the susceptibility.

We adopt a semicolon and subscript, `$_{;a}$', to mean a covariant derivative with respect to crystal momenta with Cartesian component $a=x,y,z$. Unless otherwise specified, we contract spinor indices, e.g., $n\alpha \to n$ in all expressions. A hat on a Hamiltonian, polarization, and current indicates an operator and a lack of a hat means a quantum mechanical average. We do not use hats on the creation or annihilation operators or on the position operator. A bold font indicates a vector or spinor. 

To distinguish the injection current derived from $\eta_3$ from that of $\eta_2$ we often call the former third-order injection current and the latter second-order injection current. Similarly, third-order shift current refers to current derived from $\sigma_3$. We hope the missing details will become clear from the context.

\section{Hamiltonian}
\label{sec:Ham}
We start from a Hamiltonian 
\begin{align}
\hat{H}_0= \int d\v{r} ~\pmb{\psi}^{\dagger}\left(\frac{\hat{p}^2}{2m} + V(\v{r}) + \mu^2_B\v{e}\cdot(\hat{\v{p}}\times \pmb{\sigma})\right)\pmb{\psi},
\label{eq.unper_H}
\end{align}
describing Bloch electrons with spin-orbit (SO) coupling, where $V(\v{r})$ is the periodic potential of the ions, $\hat{\v{p}}=-i\hbar\pmb{\nabla}_{\v{r}}$ is the momentum operator, $\v{e}(\v{r})=-\pmb{\nabla}_{\v{r}} V(\v{r})$ is the SO field from the nucleus,  and $\mu_B = e\hbar/2mc$ is the Bohr magneton. Electron-electron correlations in mean-field theory can be easily included by  renormalizing the parameters of the noninteracting theory in Eq.~\ref{eq.unper_H}. Momentum relaxation is incorporated phenomenologically at the end of the calculation. The electron charge is $e=-|e|$. We define the real space spinor field as

\begin{align}
\pmb{\psi} = 
\begin{pmatrix}
\psi_{\up} \\
\psi_{\down}
\end{pmatrix}.
\end{align}
A classical homogeneous electric field is coupled to the Hamiltonian by minimal substitution, $\hat{\v{p}} \to \hat{\v{p}} - e \v{A}$. After the gauge transformation 

\begin{align}
\tilde{\psi}_{\alpha}=\psi_{\alpha} e^{-ie\v{A}\cdot\v{r}/\hbar},
\label{eq:gauge_transf}
\end{align}
($\alpha$ is the spinor component) the Hamiltonian for the transformed fields becomes 
\begin{align}
\hat{H}(t)= \hat{H}_0 +\hat{H}_{D}(t).
\label{eq:H_tot}
\end{align}
In what follows we omit the tilde above the transformed fields. $\hat{H}_0$ is given by Eq.~(\ref{eq.unper_H}), and the perturbation has the dipole form 
\begin{align}
\hat{H}_D &= -e\int d\v{r} ~\pmb{\psi}^{\dagger}~\v{r}\cdot\v{E}~\pmb{\psi}.
\label{eq:dipole_H}
\end{align}
The electric field is given by  $\v{E}=-\partial \v{A}/\partial t$. The eigenfunctions of $H_0$ can be chosen to be Bloch wavefunctions $\pmb{\psi}^{(\beta)}_n(\v{kr})=\v{u}^{(\beta)}_{n}(\v{kr})e^{-i\v{kr}}$, where $\v{u}^{(\beta)}_{n}(\v{k},\v{r+R})=\v{u}^{(\beta)}_{n}(\v{k},\v{r})$ has the period of a lattice vector $\v{R}$.  $\v{k}$ is the crystal momentum and $\beta=1,2$ is the spinor index. The field operators can then be expanded in Bloch states

\begin{align}
\psi_{\alpha}(\v{r}) = \sum_{n\beta\v{k}} \psi^{(\beta)}_{n\alpha}(\v{k}\v{r}) a_{n\beta}(\v{k}),
\label{eq:expansion_bloch}
\end{align}
where $a^{\dagger}_{n\beta}(\v{k})$ creates a particle in a Bloch state and obeys anticommutation rules $\{a^{\dagger}_{n\alpha}(\v{k}),a_{m\beta}(\v{k}')\}=\delta_{nm}\delta_{\alpha\beta}\delta_{\v{kk'}}$  ($=\delta_{nm}(2\pi)^{3}\delta(\v{k}-\v{k}')/V$ in the thermodynamic limit). In this basis, $H_0$ is diagonal 
\begin{align}
\hat{H}_0 = \sum_{n\beta\v{k}} \hbar \omega_{n\beta} a^{\dagger}_{n\beta} a_{n\beta},
\label{eq:H0_diag}
\end{align}
and $\hbar\omega_{n\beta}(\v{k})$ is the energy of band $n$ and spinor $\beta$. The sum over crystal momenta is confined to the Brillouin Zone (BZ). In the thermodynamic limit in $d$-dimensions the sum becomes $\sum_{\v{k}} \to V \int d^d k/(2\pi)^d$, where $V$ is the volume of the crystal. In what follows we chose the periodic gauge by which Bloch wavefunctions are periodic in reciprocal lattice vectors, $\pmb{\psi}^{(\beta)}_n(\v{k+G},\v{r})=\pmb{\psi}^{(\beta)}_n(\v{k},\v{r})$. 

\section{Polarization operator}
\label{sec:polarization_op}
The many-body polarization operator is well defined in finite systems. It is given by 

\begin{align}
\hat{\v{P}} &= \frac{1}{V}\int d\v{r} ~\pmb{\psi}^{\dagger}~e\v{r}~\pmb{\psi},
\end{align}
where $e \v{r}/V$ is the one-body polarization operator. From Eq.~\ref{eq:dipole_H}, the dipole Hamiltonian becomes simply

\begin{align}
\hat{H}_D &= - V\hat{\v{P}}\cdot\v{E}.
\end{align}
In periodic systems, $H_D$  is given in terms of Bloch operators as

\begin{align}
\hat{\v{P}} = \frac{e}{V} \sum_{nm\v{k}\v{k}'} \langle n\v{k}|\v{r}|m\v{k}' \rangle  a^{\dagger}_{n}(\v{k}) a_{m}(\v{k}').
\label{eq:pol_gen}
\end{align}
Because the position operator is unbounded and the Bloch wavefunctions extend to infinity, the matrix elements (restoring spinor indices)
\begin{align}
\langle n \v{k}|\v{r}|m \v{k}' \rangle  \to& \langle n\alpha\v{k}|\v{r}|m\beta\v{k}' \rangle \nn\\
&=  \int d\v{r}~ \pmb{\psi}^{(\alpha)\dagger}_{n}(\v{k}\v{r})\v{r}~ \pmb{\psi}^{(\beta)}_{m}(\v{k}'\v{r}),
\end{align}
are singular. Fortunately, this singularity does not propagate to observables such as the spontaneous polarization~\cite{Sipe2000} if we separate the singularity by the well-known identity~\cite{Karplus1954,Blount1962}

\begin{align} 
\langle n\v{k}|\v{r}|m\v{k}' \rangle = \delta_{nm}[\delta(\v{k}-\v{k}')\v{\xi}_{nn} + i \pmb{\nabla}_{\v{k}}\delta(\v{k}-\v{k}')]+ \nn \\
(1-\delta_{nm})\delta(\v{k}-\v{k}') \v{\xi}_{nm}.
\label{eq:blount_position_op}
\end{align}
Here $\v{\xi}_{nm}$ are the Berry connections 
\begin{align}
\xi_{n m} \to \xi_{n\alpha m\beta} = \int d\v{r}~\v{u}^{(\alpha)\dagger}_{n}~i\pmb{\nabla}_{\v{k}}~ \v{u}^{(\beta)}_{m}.
\label{eq:Berry_conn}
\end{align} 
The polarization operator can then be separated into interband component proportional to ($1- \delta_{nm}$), and intraband component proportional to $\delta_{nm}$. To tighten the notation let us define the dipole matrix elements as

\begin{align}
\v{r}_{nm} &\equiv \pmb{\xi}_{nm} ~~~~ n\neq m \nn\\
&\equiv 0 ~~~~~~~~\textrm{otherwise}.
\label{eq:r_inter}
\end{align}
The polarization is then~\cite{Sipe2000}
\begin{align}
\hat{\v{P}} = \hat{\v{P}}_e + \hat{\v{P}}_i,
\label{eq:Ptot}
\end{align}
where 
\begin{align}
\hat{\v{P}}_e &= \frac{e}{V}\sum_{nm\v{k}} \v{r}_{nm} a^{\dagger}_{n} a_{m},
\label{eq:pe_and_pi1}\\
\hat{P}^{b}_i &= \frac{ie}{V}\sum_{n\v{k}}  a^{\dagger}_{n} a_{n;b},
\label{eq:pe_and_pi2}
\end{align}
and $b=x,y,z$. The intraband polarization depends on the covariant derivative of $a_n$ 
\begin{align}
a_{n;b} \equiv \big(\frac{\partial}{\partial k^b}  - i\xi^{b}_{nn}\big) a_n,
\label{eq:cov_der1}
\end{align}
which transforms as a scalar, $a_{n;b} \to a_{n;b} e^{i\phi_{n}^{\beta}}$, under local gauge transformations $\psi^{(\beta)}_{n} \to \psi^{(\beta)}_{n}e^{i\phi_{n}^{(\beta)}}$.  This should be contrasted with the transformation of $\partial a_n/\partial k^b$ which acquires a gauge-dependent contribution and hence it cannot represent a physical observable. 

From Eq.~\ref{eq:Ptot}, the susceptibility also naturally separates into intraband and interband contributions as

\begin{align}
\chi= \chi_{i} + \chi_{e}.
\end{align}

\section{Current operator}
\label{sec:current_op}
The current density is given by 

\begin{align}
\hat{\v{J}} &= \frac{e}{V} \int d\v{r}~ \pmb{\psi}^{\dagger} \hat{\v{v}}\pmb{\psi},
\label{eq:current_total}
\end{align} 
where $\hat{\v{v}} =[\v{r},\hat{H}_0]/i\hbar = \hat{\v{p}}/m+ \mu^2_{B}\pmb{\sigma}\times \v{e}$ is the electron's velocity. In the presence of light, the momentum changes to $\hat{\v{p}}\to \hat{\v{p}}-e\v{A}$, but after the gauge transformation~(\ref{eq:gauge_transf}), the current has the same expression. In terms of Bloch operators it becomes
\begin{align}
\hat{\v{J}} &= \frac{e}{V}\sum_{nm\v{k}} \v{v}_{nm} a^{\dagger}_n a_m,
\label{eq:current}
\end{align}
where $\v{v}_{nm}\equiv \langle n\v{k}|\hat{\v{v}} |m\v{k}\rangle$. The current satisfies charge conservation and Maxwell's equation

\begin{align}
\nabla\cdot \hat{\v{j}} + \frac{\partial \hat{\rho}}{\partial t} &=0 \\
\frac{d \hat{\v{P}}}{d t} &=\hat{\v{J}},
\label{eq:dP_J}
\end{align} 
where $\hat{\rho} =e\pmb{\psi}^{\dagger}\pmb{\psi}$ is the local charge density, $\hat{\v{j}}=(e/2) \pmb{\psi}^{\dagger}\hat{\v{v}}\pmb{\psi} +(e/2) (\hat{\v{v}}\pmb{\psi})^{\dagger}\pmb{\psi}$ is the local charge current, and $\hat{\v{P}}$ is the polarization given by Eq.~\ref{eq:Ptot}. Local particle conservation follows from the equation of motion (EOM) of $\hat{\rho}$ in the standard way.  Maxwell's equation is established as follows. From Eqs.~\ref{eq:H_tot} and \ref{eq:Ptot} and $i\hbar d\hat{\v{P}}/dt =[\hat{\v{P}},\hat{H}]$, we obtain

\begin{align}
i\frac{d \hat{P}^{a}}{dt}= \frac{e}{V}\sum_{nm\v{k}}\big( i\omega_{n;a}\delta_{nm}+ \omega_{mn} r^{a}_{nm} \big)  a^{\dagger}_{n} a_{m}
\label{eq:pe_H0} 
\end{align} 
where $\omega_{nm} \equiv \omega_n - \omega_m$. We define the covariant derivative of the matrix element $O_{nm}\equiv \langle n\v{k}|O|m\v{k} \rangle$ between Bloch states $n,m$ at a single crystal momentum by
\begin{align}
O_{nm;b}\equiv \bigg[\frac{\partial}{\partial k^b} - i(\xi^b_{nn}-\xi^b_{mm})\bigg]O_{nm},
\label{eq:cov_der_2}
\end{align}
which can be shown to transform as a tensor under gauge transformations. Since the energy bands are the diagonal matrix elements of the Hamiltonian, their covariant derivative reduces to the standard derivative $\omega_{n;a} = \partial \omega_n/\partial k^a =v^a_n= p_{n}^a/m + \mu^2_{B}(\pmb{\sigma}\times \v{e})^{a}_{nn}$. On the right hand side of Eq.~\ref{eq:pe_H0}, we recognize the diagonal and off-diagonal matrix elements of the velocity. The off-diagonal matrix elements are obtained by taking Bloch matrix elements on both sides of $\hat{\v{v}}=[\v{r},\hat{H}]/i\hbar$. Comparing with Eq.~\ref{eq:current}, the Maxwell's equation is established in the basis of Bloch operators.
 
The intraband polarization operator defines the intraband current operator which, as shown below, connects the semiclassical wave packet dynamics and the BPVEs.

\subsection{Intraband current}
\label{sec:intra_curr}
We define the intraband current operator as the time derivative of the intraband polarization operator $\hat{\v{J}}_i=d \hat{\v{P}}_i/dt$. Similarly, the interband current is  $\hat{\v{J}}_e=d \hat{\v{P}}_e/dt$. The total current is the sum of the two
\begin{align}
\hat{\v{J}} = \hat{\v{J}}_i + \hat{\v{J}}_e. 
\end{align}
Let us first calculate $\hat{\v{J}}_e$ from 
\begin{align}
i\hbar \frac{d \hat{P}^a_e}{dt}= [\hat{P}^a_e,\hat{H_0}] - V\sum_b[\hat{P}^{a}_e, \hat{P}^{b}_i + \hat{P}^{b}_e ]E^b.
\label{eq:Je}
\end{align}  
The first term has been computed in Eq.~(\ref{eq:pe_H0}). The second term is 

\begin{align}
[\hat{P}^a_e,\hat{P}^b_i + \hat{P}^b_e] &=\nn\\
-\frac{ie^2}{V^2} \sum_{nm\v{k}} \big(r^a_{nm;b}& + i\sum_p [r^{a}_{np} r^{b}_{pm}-r^{b}_{np}r^{a}_{pm}]\big)a^{\dagger}_{n}a_{m}.
\label{eq:pa_pi_pe}
\end{align}  
To make progress we now invoke a sum rule first discussed by Sipe and coworkers.\cite{Aversa1995} It derives from taking matrix elements of

\begin{align}
[r^a,r^b]=0,
\label{eq:sum_rule_2}
\end{align}  
and carefully separating the interband and intraband parts of the position operator shown in Eq.~\ref{eq:blount_position_op}. It is easy to show that such procedure works for spinor matrix elements too. Two cases are of interest follow. Taking diagonal matrix elements $(n=m)$ of Eq.~\ref{eq:sum_rule_2} gives

\begin{align}
\Omega^{ba}_n \equiv \frac{\partial \xi^{a}_{nn}}{\partial k^b}-\frac{\partial \xi^{b}_{nn}}{\partial k^a}  
=-i \sum_l [r^a_{nl}r^b_{ln} -r^{b}_{nl}r^{a}_{ln}],
\label{eq:sum_rule_diag}
\end{align}
and off-diagonal elements ($m\neq n$) gives
\begin{align}
r_{nm;b}^a - r_{nm;a}^{b} = -i \sum_l [r^a_{nl}r^b_{lm} -r^{b}_{nl}r^{a}_{lm}].
\label{eq:sum_rule_nondiag}
\end{align}
It is customary, in analogy with electrodynamics, to define a gauge field tensor $\Omega^{ab}_n$ derived from the Berry vector potential of band $n$. The Berry curvature $\pmb{\Omega}_{n}=\pmb{\nabla}\times \pmb{\xi}_{nn}$ is related to the gauge field by $\Omega^{ab}_n=\sum_{e}\epsilon_{abe}\Omega^{e}_n$. We now separate the diagonal from the nondiagonal matrix elements in Eq.~\ref{eq:pa_pi_pe} and use Eqs.~(\ref{eq:sum_rule_diag},\ref{eq:sum_rule_nondiag}) to obtain
    
\begin{align}
-V\sum_b [\hat{P}^a_e,\hat{P}^b_i + \hat{P}^b_e]E^b &=\nn\\
\frac{ie^2}{V} \sum_{n\v{k}} (\v{E}\times\pmb{\Omega}_n)^{a} a^{\dagger}_{n}a_{n}& + \frac{ie^2}{V} \sum_{nm\v{k}b} E^b r^b_{nm;a}~ a^{\dagger}_{n}a_{m}.
\label{eq:pa_pi_pe_final}
\end{align}  
Subtracting $\hat{\v{J}}_e$ (Eq.~\ref{eq:Je}) from $\hat{\v{J}}$ (Eq.~\ref{eq:pe_H0}) we obtain $\hat{\v{J}}_i$

\begin{align}
\hat{J}^{a}_i = \frac{e}{V}\sum_{nm\v{k}}\bigg[ \omega_{n;a}\delta_{nm}- &\frac{e}{\hbar} (\v{E}\times\pmb{\Omega}_n)^{a}\delta_{nm} \nn \\ 
&-\frac{e}{\hbar}  \v{E}\cdot \v{r}_{nm;a}\bigg]~ a^{\dagger}_{n}a_{m}.
\label{eq:J_intra}
\end{align}
This is an important result. The first term is the standard group velocity (renormalized by the SOC) of an electron wave packet in band $n$, $\omega_{n;a}=v^a_{n}$. As shown below, this term gives rise to the injection current contribution to the BPVE. The second term depends on the Berry curvature $\pmb{\Omega}_n$ and is often called `anomalous' velocity. It gives rise to many topological effects in condensed matter physics. For example, it gives rise to the (intrinsic) anomalous Hall conductivity in metallic ferromagnets,\cite{Karplus1954,Haldane2004} and, as shown in Sec.~\ref{sec:NLHE}, to the (intrinsic) nonlinear Hall effect in nonmagnetic metals.\cite{Sodemann2015,Moore2010} In insulators, this term contributes to third order in the electric field but not to second order.

The third term resembles a small dipole created by the external electric field. Just as the standard momentum derivative of Bloch energies leads to the usual group velocity, the (covariant) derivative of the dipole energy $U_{nm}=e \v{E} \cdot \v{r}_{nm}$, can be thought of as a group velocity 
\begin{align}
v^{a}_{dip,nm}= -\frac{e}{\hbar} \v{E}\cdot \v{r}_{nm;a}
\label{eq:dip_vel}
\end{align}
associated with a \textit{pair} of wave packets in distinct bands. 

The first two integrands in Eq.(\ref{eq:J_intra}) are gauge invariant and are usually interpreted as velocity contributions of electron wave packets.~\cite{Xiao2010} The dipole velocity, on the other hand, is not gauge invariant and hence is not a physical velocity. However, the product of the dipole velocity and the density matrix \textit{is} gauge invariant and in this context the dipole velocity can be given the interpretation of the velocity of pairs of wave packets. As shown below, the dipole velocity gives rise to the shift current contribution to the BPVE. Indeed, the intraband current unifies the well-known semiclassical dynamics of wave packets in electric fields with the BPVEs.

What is the physical interpretation of the interband current? The fact that the interband polarization is regular in the dc limit ($\omega_{\Sigma}\to 0$) implies the interband current vanishes in this limit. This suggests that the interband current captures electron oscillations about their equilibrium positions but not their uniform acceleration. 

Up to this point, the above formalism is valid for metals and insulators. Except for Sec.~\ref{sec:NLHE}, we will focus on the short time response of insulators, discarding Fermi surface contributions and momentum relaxation. By `short time' we mean shorter than momentum relaxation characteristic time ($\sim$100 fs) but longer than the period of light ($\sim$2 fs).

\section{Perturbation theory}
\label{sec:perturbation_th}
Let us define the single-particle density matrix
\begin{align}
\rho_{mn}\equiv\langle a^{\dagger}_{n} a_{m}\rangle,
\end{align}  
where the $a_n$ operators are in the Heisenberg representation. The quantum average is over the ground state defined with all the valence bands filled and all conduction bands empty. Being noninteracting, the system is completely characterized by the single-particle density matrix. The amplitude of the electric field is 

\begin{align}
E^{b} = \sum_{\beta} E^b_{\beta} e^{-i(\omega_{\beta}+i\epsilon) t},
\label{eq:E_field}
\end{align}
where $\beta=1,2,...$ labels the frequency components of the field. The dipole Hamiltonian is treated as a perturbation with the electric field being turned on slowly in the infinite past so that all the transients effects have vanished. As usual, this is accomplished by taking the limit $\epsilon\to 0$ at the end of the calculation. To find the density matrix we first compute its EOM~\cite{Sipe2000} 

\begin{align}
\frac{\partial \rho_{mn}}{\partial t} + i \omega_{mn} \rho_{mn} = \frac{e}{i\hbar}\sum_{lb} E^{b}(\rho_{ml}r^{b}_{ln}-r^{b}_{ml}\rho_{ln}) \nn \\
-\frac{e}{\hbar} \sum_b  E^b \rho_{mn;b}.
\label{eq:eom_rho}
\end{align}
The first term on the right comes from interband processes as can be recognized by the presence of $\v{r}_{nm}$. The second term comes from intraband processes which involve the covariant derivative of the density matrix
\begin{align}
\rho_{mn;b}\equiv \bigg[\frac{\partial}{\partial k^b} - i(\xi^b_{mm}-\xi^b_{nn})\bigg]\rho_{mn}, 
\end{align}
Only when the intraband and interband motion is considered on an equal footing, the EOM reduces to the Boltzmann equation (in the one-band limit) with no collision integral.
\subsection{Zeroth order}
If $\v{E}=0$ the solution of Eq.(\ref{eq:eom_rho}) is simply $\rho^{(0)}_{mn} = \delta_{nm}f_{n}$, where $f_n\equiv f(\epsilon_{n}(\v{k}))= 0,1$ is the Fermi occupation of band $n$ at zero temperature. 

\subsection{First order}
Substituting the zeroth-order solution into the right-hand side of Eq.(\ref{eq:eom_rho}) and solving for $\rho_{mn}^{(1)}$ we obtain
\begin{align}
\rho^{(1)}_{mn} &=\sum_{b\beta} \bar{\rho}^{(1)b\beta}_{mn} E^{b}_{\beta} e^{-i\omega_{\beta}t} \\
 &= \frac{e}{\hbar}\sum_{b\beta} \frac{r_{mn}^b f_{nm}}{\omega_{mn}-\omega_{\beta}} E^{b}_{\beta} e^{-i\omega_{\beta}t}.
\label{eq:1st_pt}
\end{align}
where we defined $f_{nm}\equiv f_n-f_m$. Note that to first order only  interband processes are allowed in insulators.
 
\subsection{Second order}
To second order we have

\begin{align}
\rho^{(2)}_{mn} = \sum_{b\beta} \sum_{c\sigma} \bar{\rho}^{(2)b\beta c\sigma}_{mn} E^{b}_{\beta} E^{c}_{\sigma} e^{-i\omega_{\Sigma} t},
\label{eq:2nd_pt_v1}
\end{align}
where
\begin{align}
\bar{\rho}^{(2)b\beta c\sigma}_{mn} &= \frac{i e}{\hbar(\omega_{mn}-\omega_{\Sigma})}\bigg[  \bar{\rho}^{(1)b\beta}_{mn;c} \nn\\
&~~~~~~~+ i\sum_{l} \big(\bar{\rho}^{(1)b\beta}_{ml} r^{c}_{nl} - r_{ml}^{c} \bar{\rho}^{(1)b\beta}_{ln} \big)\bigg],
\label{eq:2nd_pt}
\end{align}
and $\omega_{\Sigma}=\omega_{\beta}+\omega_{\sigma}$. The covariant derivative of a quotient in $\bar{\rho}^{(1)b\beta}_{mn;c}$ is simply

\begin{align}
\left(\frac{r_{mn}^{a} f_{nm}}{\omega_{mn}-\omega_{\alpha}}\right)_{;b} = \frac{r_{mn;b}^{a} f_{nm}}{\omega_{mn}-\omega_{\alpha}} - 
\frac{r_{mn}^{a}f_{nm}\omega_{mn;b}}{(\omega_{mn}-\omega_{\alpha})^2}  
\end{align}

\subsection{nth-order}
In the long-time limit, we expect harmonic solutions of the form
\begin{align}
\rho^{(n)}_{mn} = \sum_{a_1\alpha_1,...} \bar{\rho}^{(n)a_1\alpha_1,..}_{mn} E^{a_1}_{\alpha_1} \cdots E^{a_{n}}_{\alpha_{n}} e^{-i\omega^{(n)}_{\Sigma}t},
\end{align}
where $\omega^{(n)}_{\Sigma} = \omega_{\alpha_1}+\cdots+\omega_{\alpha_{n}} $. Substituting into Eq.(\ref{eq:eom_rho}) and iterating we obtain an equation for $\bar{\rho}^{(n+1)}_{mn}$ in terms of $\bar{\rho}^{(n)}_{mn}$. Omitting the supercripts $a_1\alpha_1,...$ for clarity we obtain
\begin{align}
\bar{\rho}^{(n+1)}_{mn} = \frac{ie}{\hbar (\omega_{mn}- \omega^{(n+1)}_{\Sigma}) }\bigg[ i\sum_l &(\bar{\rho}^{(n)}_{ml} r^{a_{n+1}}_{ln} -r^{a_{n+1}}_{ml}\bar{\rho}^{(n)}_{ln})\nn \\
&+ \bar{\rho}^{(n)}_{mn;a_{n+1}} \bigg].
\label{eq:nth_pt}
\end{align}  
Note that at every order in perturbation theory there are interband (first term) and intraband (second term) contributions. In general, the $n$th-order $\rho^{(n)}$ ($n\geq 1$) has $2^{n-1}$ intraband and $2^{n-1}$ interband contributions.

\section{Physical divergences of $\chi_2$}
\label{sec:chi2}
The susceptibility and conductivity response tensors to second order are defined by 
\begin{align}
P^{a(2)} &= \sum_{b\beta c\sigma} \chi_2^{abc}(-\omega_{\Sigma},\omega_{\beta},\omega_{\sigma}) E^{b}_{\beta} E^{c}_{\beta} e^{-i \omega_{\Sigma} t}, \\
J^{a(2)} &= \sum_{b\beta c\sigma} \sigma^{abc(2)}(-\omega_{\Sigma},\omega_{\beta},\omega_{\sigma}) E^{b}_{\beta} E^{c}_{\beta} e^{-i \omega_{\Sigma} t},
\end{align}
where $\omega_{\Sigma}=\omega_{\beta}+\omega_{\sigma}$. They are related by $d P^{a(2)}/dt = J^{a(2)}$. $\chi_2$ can be split into interband and intraband components, $\chi_2 = \chi_{2e}+\chi_{2i}$, using Eqs.(\ref{eq:pe_and_pi1}),(\ref{eq:J_intra}),(\ref{eq:1st_pt}), and (\ref{eq:2nd_pt_v1}). The result is~\cite{Aversa1995}
\begin{align}
\frac{\chi^{abc}_{2e}}{C_2} &=i \sum_{nm\v{k}} \frac{r_{nm}^a f_{nm}}{\omega_{mn}-\omega_{\Sigma}} \left(\frac{r^b_{mn}}{\omega_{mn}-\omega_{\beta}}\right)_{;c}\nn\\
-& \sum_{n l m\v{k}} \frac{r_{nm}^a }{\omega_{mn}-\omega_{\Sigma}} \left(  \frac{r^b_{ml} r^c_{ln} f_{lm}} {\omega_{ml}-\omega_{\beta}} -
 \frac{r^c_{ml} r^b_{ln} f_{nl}} {\omega_{ln}-\omega_{\beta}} \right),\\
\frac{\chi^{abc}_{2i}}{C_2} &=\frac{i}{\omega_{\Sigma}^2} \sum_{nm\v{k}} \frac{\omega_{nm;a} r_{nm}^b r_{mn}^c f_{mn}}{\omega_{nm}-\omega_{\beta}}\nn\\
& ~~~~~~~~~~~~~~+\frac{1}{i\omega_{\Sigma}} \sum_{nm\v{k}} \frac{r_{nm;a}^c r_{mn}^b f_{nm}}{\omega_{mn}-\omega_{\beta}},
\label{eq:chi2i}
\end{align}
where we defined $C_2=e^3/\hbar^2 V$. These expressions need to be symmetrized with respect to exchange of indices $b\beta \leftrightarrow c\sigma$. We note that $\chi_{2i}$ is easier to calculate from $\v{J}^{(2)}_i$ rather than directly from $\v{P}^{(2)}_i$.

The Taylor expansion of $\chi_{2i}$ in Eq.~\ref{eq:chi2_taylor_v1}~\cite{Aversa1995,Sipe2000} means that $\chi_{2i}$ diverges as $\omega_{\Sigma}\to 0$ and that the injection $\eta_2$ and shift $\sigma_2$ response tensors can be obtained from this expansion, see Appendix~\ref{sec:derivation_eta2_sigma2}. Here we derive these tensors from a slightly different perspective that exposes the analytic properties of $\chi_{2i}$. Let us assume $\chi_{2i}$ admits a Laurent series 

\begin{align}
\chi_{2i} = \frac{\eta_2}{z^2} + \frac{\sigma_2}{z} + \cdots
\end{align}
where $z=-i\omega_{\Sigma}$. Then $\eta_2$ is given by 
\begin{align}
\eta_2 = \frac{1}{2\pi i} \oint_{|z|=\rho} dz~ z\chi_{2i},
\label{eq:eta2_pole}
\end{align}
where $\rho$ is the radius of convergence. All the frequencies are parametrized in terms of $\omega_{\Sigma}=iz$. One such parametrization is

\begin{align}
\omega_{\beta}&= \omega + n_{\beta}\omega_{\Sigma}\\
\omega_{\sigma}&= -\omega + n_{\sigma}\omega_{\Sigma},
\end{align} 
where $n_{\beta}+n_{\sigma}=1$. The manifold where $\omega_{\Sigma}=0$ is a line of singular points $(\omega_{\beta},\omega_{\sigma})=(\omega,-\omega)$, parametrized by a single frequency $\omega>0$. Symmetrizing $\chi_{2i}$ with respect to exchange of indices $b\beta \leftrightarrow c\sigma$ and using Eq.~\ref{eq:eta2_pole} we obtain $\eta_2^{abc}(0,\omega,-\omega)$ as
\begin{align}
\eta^{abc}_2&= \frac{\pi e^3}{\hbar^2 V}\sum_{nm\v{k}} f_{mn}~ \omega_{nm;a} r^b_{nm} r^c_{mn} \delta(\omega_{nm}-\omega),
\label{eq:eta2} 
\end{align}
or equivalently
\begin{align}
\eta^{abc}_2= \frac{\pi e^3}{2\hbar^2 V}\sum_{nm\v{k}} f_{mn} \omega_{nm;a} (r^b_{nm} r^c_{mn}- r^c_{nm} r^b_{mn}) \nn\\
\times \delta(\omega_{nm}-\omega),  
\end{align}
which is independent of the parameters $n_{\beta},n_{\sigma}$. In calculating $\eta_2$ we take the limit $\rho \to 0$ before the limit $\epsilon \to 0$. This corresponds to the physical situation where $\omega_{\Sigma}= 0 $ in the infinite past. Similarly, $\sigma_2^{abc}(0,\omega,-\omega)$ is given by
\begin{align}
\sigma_2 = \frac{1}{2\pi i} \oint_{|z|=\rho} dz~ \chi_{2i}.
\end{align}
An explicit integration gives
\begin{align}
\sigma^{abc}_2&= \frac{i\pi e^3}{2 \hbar^2 V}\sum_{nm\v{k}} f_{mn}  (r^c_{nm;a} r^b_{mn} \nn\\
&~~~~~~~~~~~~~~~~~~~~~~~~~~~~- r^c_{nm} r^b_{mn;a} )  \delta(\omega_{nm}-\omega).
\label{eq:sigma2}
\end{align}
In calculating $\sigma_2$ we took $n_{\beta}=n_{\sigma}=1/2$ to eliminate a resonant imaginary  term which depends on $n_{\beta}-n_{\sigma}$. This term does not arise in the standard method~\cite{Aversa1995,Sipe2000} because there the prescription is to Taylor expand only the real parts. Taking $n_{\beta}=n_{\sigma}$ means we are approaching the line of singularities at right angle.

Eqs.(\ref{eq:eta2}) and (\ref{eq:sigma2}) are the well-known injection and shift current tensors. $\eta_2$ is pure imaginary and antisymmetric in the $b,c$ indices and hence vanishes for linear polarization. $\sigma_2$, on the other hand, is real, symmetric in $b,c$ indices and hence vanishes for circular polarization. The corresponding injection and shift currents are given by
\begin{align}
J^{a(2)}_{sh} &\equiv \sum_{b\beta c\sigma}\sigma_2^{abc}(-\omega_{\Sigma},\omega_{\beta},\omega_{\sigma}) E^b_{\beta} E^c_{\sigma}e^{-i\omega_{\Sigma}t}, \\
\frac{d}{dt} J^{a(2)}_{inj} &\equiv \sum_{b\beta c\sigma}\eta_2^{abc}(-\omega_{\Sigma},\omega_{\beta},\omega_{\sigma}) E^b_{\beta} E^c_{\sigma}e^{-i\omega_{\Sigma}t},
\end{align}
subject to $\omega_{\Sigma}=0$. Assuming a monocromatic source $E^b= E^{b}(\omega) e^{-i\omega t} +c.c.$ and performing the frequency sums keeping only dc terms ($\omega_{\Sigma}=0$), we obtain
\begin{align}
J^{a(2)}_{sh} = 2\sum_{bc}\sigma_2^{abc}(0,\omega,-\omega) E^b(\omega) E^c(-\omega) 
\label{eq:sigma2_timeder} \\
\frac{d}{dt} J^{a(2)}_{inj} = 2\sum_{bc}\eta_2^{abc}(0,\omega,-\omega) E^b(\omega) E^c(-\omega),
\label{eq:eta2_timeder}
\end{align}
where the factor of 2 is from the intrinsic permutation symmetry of susceptibilities.~\cite{Boyd2008} Being quadratic in the fields the injection and shift current vanish for centrosymmetric systems. The above expressions indicate the injection and shift currents vary as 
\begin{align}
|\v{J}^{(2)}_{inj}(t)| \sim \eta_2 t
\label{eq:eta2_lin_in_time}\\
|\v{J}^{(2)}_{sh}(t)|\sim \sigma_2 
\label{eq:sigma2_contant_time}
\end{align}
with illumination time in the absence of momentum relaxation and saturation effects. 

\subsection{Physical interpretation of injection and shift current}
In this section we show that the injection and shift currents can be understood from simple semiclassical wave packet dynamics in electric fields.

\subsubsection{Injection current}

The microscopic origin of the injection current from light-matter interactions is well known. It arises from the asymmetry in the carrier injection rate at time-reversed momenta in the BZ~\cite{Sturman1992,Driel2001}. To see this, let us consider an electron wave packet with velocity $v^a_n$. From the first term in Eq.~\ref{eq:J_intra} the current is 
\begin{align}
J^{a} = \frac{e}{V}\sum_{n\v{k}} f_n v^a_n,
\label{eq:j_standard}
\end{align}
where $f_n \equiv \rho^{(0)}_{nn}$. The effect of an optical field is to \textit{inject} carriers into the current-carrying states in the conduction bands. Taking a time derivative of the occupations we obtain 
\begin{align}
\frac{d}{dt} J_{inj}^{a} = \frac{e}{V}\sum_{n\v{k}}\frac{d f_n}{dt} v^a_n.
\label{eq:dj_dt_pheno}
\end{align}
For low intensity, the Fermi's Golden Rule gives the one-photon absorption rate~\cite{Driel2001}  
\begin{align}
\frac{d f_v}{dt} &= -\frac{2\pi e^2}{\hbar^2}\sum_{c} | \v{E}(\omega)\cdot \v{r}_{cv} |^{2}\delta(\omega_{cv}-\omega), \nn \\
\frac{d f_c}{dt} &= \frac{2\pi e^2}{\hbar^2}\sum_{v} | \v{E}(\omega)\cdot  \v{r}_{cv} |^{2}\delta(\omega_{cv}-\omega),
\label{eq:FG_rule}
\end{align}
where $c,v$ labels a conduction or a valence band, respectively. For complex fields, e.g, circularly polarized or elliptically polarized, the carrier injection rate at time-reversed points $\pm \v{k}$ in the BZ is not the same

\begin{align}
\frac{d}{dt} f_{c}(-\v{k}) \neq \frac{d}{dt} f_{c}(\v{k}),
\end{align}
leading to a polar distribution of Bloch velocity states. This is the microscopic origin of the injection current and, as we show below, of many higher-order injection currents. Substituting into Eq.(\ref{eq:dj_dt_pheno}) we obtain 
\begin{align}
\frac{d}{dt} J^{a(2)}_{inj} = \frac{2\pi e^3}{\hbar^2 V} \sum_{b'c'} \sum_{cv\v{k}} \omega_{cv;a} r^{b'}_{vc} r^{c'}_{cv} \delta(\omega_{cv}-\omega)\nn\\
\times E^{b'}(\omega)E^{c'}(-\omega),
\label{eq:inj_current}
\end{align}
or 

\begin{align}
\frac{d}{dt} J^{a(2)}_{inj} = \frac{2\pi e^3}{\hbar^2 V} \sum_{bc} \sum_{nm\v{k}} f_{mn}\omega_{nm;a} r^{b}_{nm} r^{c}_{mn} \delta(\omega_{nm}-\omega)\nn\\ 
\times E^{b}(\omega)E^{c}(-\omega),
\end{align}
which is the standard injection current shown in Eq.(\ref{eq:eta2_timeder}). 

\subsubsection{Shift current}
\label{sec:shift_current}
\begin{figure}
\includegraphics[width=.46\textwidth]{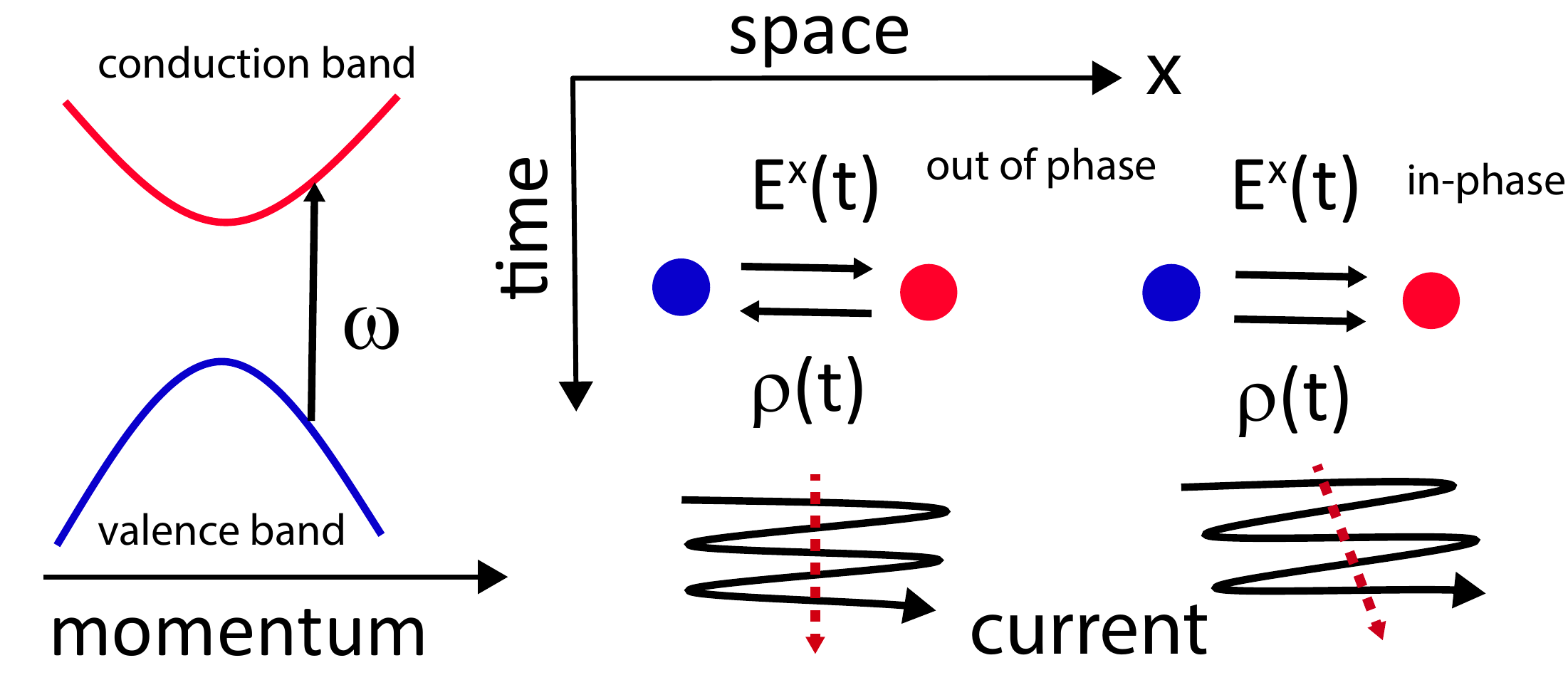}
\caption{Intuitive picture of microscopic generation of shift current. The wiggle lines represent interband coherence oscillations between the valence and conduction band centers of charge (circles) which are spatially separated. The quantum interference between population oscillations $\rho_{nm}(t)$ and dipole velocity oscillations $\v{E}(t)\cdot \v{r}_{mn;a}$ gives rise to a shift current.}
\label{fig:shift_intuitive}  
\end{figure}

Injection current is proportional to the momentum relaxation time and hence explicitly breaks time-reversal symmetry. In the scenario where a shift current originates from light-matter interactions,~\cite{Baltz1981} the shift current does not require the presence of momentum relaxation to break time reversal symmetry. How is time-reversal symmetry broken in shift current processes? It is broken at the time of photon absorption which is an irreversible process. 

Materials that exhibit shift current have valence and conduction band centers spatially separated within the unit cell and hence charge is \textit{shifted} upon photon absorption. This process depends only on the off-diagonal elements of the density matrix and hence it requires quantum coherence as has been extensively documented. Here we propose that shift current arises from the quantum interference of two distinct microscopic processes involving wave packet oscillations in the presence of an electric field. To see this consider the dipole current  in Eq.(\ref{eq:J_intra}) to second order
\begin{align}
J^{a(2)}_{dip}= -\frac{e^2}{\hbar V}\sum_{nm\v{k}} \v{E}(t)\cdot \v{r}_{nm;a}~\rho^{(1)}_{mn}(t).
\label{eq:shift_pheno}
\end{align}
The current is the sum of dipole velocities of each pair of wave packets in bands $n,m$ weighted by the probability $\rho^{(1)}_{mn}$ of being occupied. From Eq.~\ref{eq:1st_pt} we have 
\begin{align}
J^{a(2)}_{dip}= -\frac{e^3}{\hbar^2 V}\sum_{b\beta c\sigma} \sum_{nm\v{k}} \frac{r^{b}_{nm;a} r^{c}_{mn} f_{nm}}{\omega_{mn}-\omega_{\sigma}} E^b_{\beta} E^c_{\sigma} e^{-i\omega_{\Sigma}t},
\label{eq:Jdip_rho1}
\end{align}
where $\omega_{\Sigma}=\omega_{\beta}+\omega_{\sigma}$. Symmetrizing with respect to exchange of indices $b\beta \leftrightarrow c\sigma$, assuming a monocromatic field $E^b=E^b(\omega)e^{-i\omega t}+c.c.$, and keeping only the dc resonant terms we obtain
\begin{align}
J^{a(2)}_{sh}= \frac{i\pi e^3}{\hbar^2 V}\sum_{bc} \sum_{nm\v{k}} f_{mn}( r^{b}_{nm;a} r^{c}_{mn} +r^{c}_{nm;a} r^{b}_{mn}) \nn \\
\times\delta(\omega_{nm}-\omega) E^b(\omega) E^c(-\omega),
\end{align}
which is the standard expression for the shift current in Eq.~\ref{eq:sigma2_timeder}.  This calculation suggests that the constructive \textit{quantum} mechanical interference of interband coherence oscillations and dipole velocity oscillations is the microscopic origin of the shift current, see Fig.~\ref{fig:shift_intuitive}. We note that electron oscillations between centers of charge, alone, do not lead to a dc current. However, the directionality of the electron oscillations combined with an isotropic relaxation (due to, e.g., randomized collisions) could, in principle, also lead to a dc current. In this scenario momentum relaxation would play a significant role in the origin of the current.  

Before showing how the injection and shift currents are modified by the presence of a static electric field, we discuss Fermi surface contributions to the BPVE to second order.

\section{The BPVE in metals}
\label{sec:NLHE}

The Fermi surface  of metals gives rise to two additional contributions to the second-order BPVE. The first is the nonlinear Hall effect (NLHE) discussed by Sodemann and Fu~\cite{Sodemann2015} and the second is a \textit{metallic} jerk current discussed recently by Matsyshyn and Sodemann~\cite{Matsyshyn}. Here we show that these photocurrents can be obtained from Eq.~\ref{eq:J_intra} and simple physical assumptions.

To begin note that $\rho^{(1)}_{nm}$ has a Fermi surface contribution. Substituting $\rho^{(0)}_{nm}= f_n \delta_{nm}$ into the right-hand side of Eq.~\ref{eq:eom_rho} two terms are obtained. The first is an interband contribution given by Eq.~\ref{eq:1st_pt}. The second is the intraband contribution 
    
\begin{align}
\rho^{(1)}_{nm,i}=-\delta_{nm}\frac{ie}{\hbar} \sum_{b \beta} \frac{1}{\omega_{\beta}}\frac{\partial f_n}{\partial k^b} E^{b}_{\beta} e^{-i \omega_{\beta} t},
\end{align}
which depends explicitly on the presence of a Fermi surface via $\partial f_n/\partial k^b$. The dc divergence is cut off by the momentum relaxation time scale $\tau_1$ as

\begin{align}
\frac{1}{-i\omega_{\beta}} \to \frac{1}{\frac{1}{\tau_1} - i \omega_{\beta}}.
\end{align}
From the second term in Eq.~\ref{eq:J_intra} we have

\begin{align}
J^{a(2)}_{nlhe} = -\frac{e^2}{\hbar V}\sum_{b\beta} \sum_{n\v{k}e}  \epsilon_{abe}E^{b}_{\beta}e^{-i\omega_{\beta}}\pmb{\Omega}_{n}^{e}\rho^{(1)}_{nn,i},
\end{align} 
after symmetrizing with respect to exchanges of field indices, and performing the frequency sums the dc current is

\begin{align}
J^{a(2)}_{dc,nlhe} = \sum_{bc} \sigma_{nlhe}^{abc(2)}(0,\omega,-\omega) E^{b}(\omega) E^{c}(-\omega) +c.c.
\end{align} 
where

\begin{align}
\sigma_{nlhe}^{abc(2)}(0,\omega,-\omega) \equiv \frac{e^3}{\hbar^2 V}\frac{\tau_1}{1 + i\omega\tau_1} \sum_{n\v{k}e} \epsilon_{abe} \pmb{\Omega}_{n}^{e} \frac{\partial f_n}{\partial k^c},
\end{align} 
is the known response tensor for the dc nonlinear Hall effect.~\cite{Sodemann2015} From a semiclassical point of view the dc NLHE arises from the quantum interference of (intraband) oscillations of excitations across the Fermi surface and oscillations of the anomalous velocity of wave packets. A key difference with injection current is that NLHE response tensor is antisymmetric in the first two indices, rather than the last two.

Similarly, inspection of Eq.~\ref{eq:J_intra} shows that another dc current to second-order is possible by taking two time derivatives of the velocity in the first term and the equilibrium density matrix

\begin{align}
\frac{d^2 J^{a}}{dt^2} =\frac{e}{V}\sum_{n\v{k}} f_n \frac{d^2 v^{a}}{dt^2}.
\end{align} 
The two derivatives of the velocity can be computed in powers of the electric field. The result is similar to Eq.~\ref{eq:d2vdt2} but with the optical fields replacing the static fields. Symmetrizing and performing the frequency sums we obtain

\begin{align}
J^{a(2)}_{2ndjerk} = \sum_{bc} \iota_2^{abc}(0,\omega,-\omega) E^{b}(\omega) E^{c}(-\omega)
\end{align} 
where

\begin{align}
\iota_{2}^{abc}(0,\omega,-\omega) \equiv \frac{2e^3}{\hbar^2 V} \sum_{n\v{k}} f_n \omega_{n;abc}.
\end{align} 
From the semiclassical perspective, the second-order jerk current arises from the \textit{constant} acceleration of a wave packet. The acceleration can be constant because of the (classical) interference of the oscillating electric field with itself. The response tensor is symmetric under exchange of $b,c$ indices and hence vanish for circular polarization. The second-order jerk current varies as $t^2$ in the absence of momentum dissipation and saturation effects.

The responses at $2\omega$ can be calculated similarly. The results are

\begin{align}
J^{a(2)}_{2\omega,nlhe} &= \sum_{bc} \sigma_{nlhe}^{abc(2)}(-2\omega,\omega,\omega) E^{b}(-\omega) E^{c}(-\omega) +c.c.,
\end{align} 
where $\sigma_{nlhe}^{abc(2)}(-2\omega,\omega,\omega) = \sigma_{nlhe}^{abc(2)}(0,\omega,-\omega)$ and 

\begin{align}
J^{a(2)}_{2ndjerk} = \sum_{bc} \iota_2^{abc}(-2\omega,\omega,\omega) E^{b}(\omega) E^{c}(-\omega) + c.c.
\end{align} 
where $\iota_2^{abc}(-2\omega,\omega,\omega) =(1/2) \iota_2^{abc}(0,\omega,-\omega)$. Note that both, the NLHE and the second-order jerk current can produce current transverse to the polarization of the optical field. This should be taken into account in interpreting experiments in metals. A summary of the metallic BPVEs is given in Table~\ref{table:bpve_metal}.

\begin{table}
\caption{Second-order BPVEs in metals. intra=intraband, inter=interband, FS= Fermi surface. $\omega_{\Sigma}=\omega_{\beta}+ \omega_{\sigma}$, frequency sum of two optical fields, $^{*}$in the absence of momentum relaxation.    }
\begin{tabular}{|m{1.3cm}|m{1.1cm}|m{1.0cm}|m{1.4cm}|m{1.2cm}|m{1.2cm}|} 
\hline 
dc         & Symbol   & Time           & intra. vs& Sing.$^{*}$ & Mom.    \\ 
 current   &          & dep.$^{*}$     & inter. vs&             & relax.  \\
           &          &                & FS        &            &         \\
\hline\hline 
Injection  & $\eta_2$ & $t$            & intra & $\omega_{\Sigma}=0$& $\tau_1$  \\[1ex]
\hline
Shift      &$\sigma_2$& const.         & inter & $\omega_{\Sigma}= 0$& $\tau_2$  \\[1ex]
\hline
 NLHE      &$\sigma_{nlhe}^{(2)}$& const.&intra,FS & $\omega_{\beta}= 0$      & $\tau_1$ \\[1ex]
\hline
Jerk       &$\iota_2$& $t^2$             & intra,FS &                 &  $\tau_1^2$  \\[1ex]
\hline
\end{tabular}
\label{table:bpve_metal}
\end{table}

\section{Physical divergences of $\chi_3$}
\label{sec:chi3}
The susceptibility and conductivity response tensors to third order are defined by 
\begin{align}
P^{a(3)} &= \sum_{b\beta c\sigma d\delta} \chi_3^{abcd}(-\omega_{\Sigma},\omega_{\beta},\omega_{\sigma},\omega_{\delta}) E^{b}_{\beta} E^{c}_{\sigma} E^{d}_{\delta} e^{-i \omega_{\Sigma} t}, \\
J^{a(3)} &= \sum_{b\beta c\sigma d\delta} \sigma^{abcd(3)}(-\omega_{\Sigma},\omega_{\beta},\omega_{\sigma},\omega_{\delta}) E^{b}_{\beta} E^{c}_{\sigma} E^{d}_{\delta} e^{-i \omega_{\Sigma} t},
\end{align}
where $\omega_{\Sigma}=\omega_{\beta}+\omega_{\sigma} + \omega_{\delta}$. They are related by $d \v{P}^{(3)}/dt = \v{J}^{(3)}$. $\chi_3$ can be split into interband and intraband components $\chi_3 = \chi_{3e}+\chi_{3i}$. Expanding the intraband component in powers of $\omega_{\Sigma}$ gives
 
\begin{align}
(-i\omega_{\Sigma})^{3}\chi_{3i} = \iota_3 + (-i\omega_{\Sigma})\eta_3 + (-i\omega_{\Sigma})^{2}\sigma_3 +\cdots.
\label{eq:taylor_chi3_v2} 
\end{align}
See Appendix~\ref{app:chi_3}. Eq.(\ref{eq:taylor_chi3_v2}) is equivalent to 
\begin{align}
\chi_{3i} = \frac{\iota_3}{z^3} + \frac{\eta_{3}}{z^2}+ \frac{\sigma_{3}}{z}+\cdots 
\label{eq:Laurent_chi3}
\end{align}
where $z\equiv -i\omega_{\Sigma}$. Since $\chi_{3e}$ is regular, Eq.~\ref{eq:Laurent_chi3} implies that the conductivity in the limit of no momentum relaxation is 

\begin{align}
\sigma^{(3)} = \frac{\iota_3}{z^2} + \frac{\eta_{3}}{z}+ \sigma_{3}+ z(reg),
\label{eq:conductivity_3rd_gen}
\end{align}
where $reg$ represents the remaining regular terms (as $z\to 0$). The residues $\iota_3$, $\eta_3$ and $\sigma_3$ define various current contributions as follows. The limit 

\begin{align}
\textrm{lim}_{z\to 0} z^2 \sigma^{(3)} = \iota_3,
\end{align}
or equivalently 

\begin{align}
\textrm{lim}_{\omega_{\Sigma}\to 0} \frac{d^2 }{dt^2} J^{a(3)}  &\equiv \frac{d^2}{dt^2} J^{a(3)}_{jerk}\nn\\ 
=&\sum_{b\beta c\sigma d\delta} \iota_3^{abcd}(0,\omega_{\beta},\omega_{\sigma},\omega_{\delta}) E^{b}_{\beta} E^{c}_{\beta} E^{d}_{\delta},
\label{eq:jerk_current_from_limit}
\end{align}
(subject to $\omega_{\Sigma}=0$) defines the jerk current. Similarly the limits 

\begin{align}
\textrm{lim}_{z\to 0} z \left[\sigma^{(3)} - \frac{\iota_3}{z^2}\right] &= \eta_3, \\
\textrm{lim}_{z\to 0}  \left[\sigma^{(3)} - \frac{\iota_3}{z^2} - \frac{\eta_3}{z} \right] &= \sigma_3, 
\end{align}
define higher order injection and shift currents (respectively) in the presence of a static electric field: 

\begin{align}
\frac{d}{dt} J^{a(3)}_{inj} \equiv\sum_{b\beta c\sigma d\delta} \eta_3^{abcd}(0,\omega_{\beta},\omega_{\sigma},\omega_{\delta}) E^{b}_{\beta} E^{c}_{\beta} E^{d}_{\delta}, \\
 J^{a(3)}_{sh} \equiv\sum_{b\beta c\sigma d\delta} \sigma_3^{abcd}(0,\omega_{\beta},\omega_{\sigma},\omega_{\delta}) E^{b}_{\beta} E^{c}_{\beta} E^{d}_{\delta},
\end{align}
subject to $\omega_{\Sigma}=0$. We now analyze each of these currents in detail.

\section{Jerk current}
\label{sec:jerk_current}
\subsection{Hydrodynamic model}
In an isotropic system the current is
\begin{align}
J^a_{clas}=e n v^a,
\end{align}
where $n$ is the carrier density. Taking two derivatives we obtain
\begin{align}
\frac{d^2}{dt^2} J^a_{clas} = e\frac{d^2 n}{dt^2} v^a + 2e \frac{dn }{dt}\frac{dv^a}{dt} + en\frac{d^2 v^a}{dt^2}.
\end{align}
If the rate of carrier injection $dn/dt=g$ and acceleration $e E_0^a/m^{*}$ are constant in time then 
\begin{align}
\frac{d^2}{dt^2} J^a_{clas} = \frac{2e^2 g E_0^a}{m^{*}}=constant,
\label{eq:jerk_classical} 
\end{align}
leads to a current varying quadratically with illumination time. This effect has been extensively studied in the context of the THz generation in bias semiconductor antennas using semiclassical kinetic equations, see for example Ref.~\onlinecite{Jepsen1996}. However, the static field modifies the carrier injection rate giving rise to novel contributions. We now discuss this effect. 

\subsection{Susceptibility divergence}
We find $\iota_3$ from the limit $\textrm{lim}_{\omega_{\Sigma}\to 0} (-i\omega_{\Sigma})^3\chi_{3i} =\iota_3$. The details of the derivation are outlined in Appendix~\ref{app:iota3}. $\iota^{abcd}_3(0,\omega,-\omega,0)$ is given by~\cite{Fregoso2018}
\begin{align}
\iota_3^{abcd} = \frac{2\pi e^4}{6\hbar^3 V}&\sum_{nm\v{k}} f_{mn} \big[ 2\omega_{nm;ad} r^b_{nm}r^c_{mn}\nn\\
&+ \omega_{nm;a}  (r^b_{nm}r^c_{mn})_{;d} \big] \delta(\omega_{nm}-\omega),
\label{eq:jerk}
\end{align}
where $\omega_{nm;ad}=\partial^2 \omega_{nm}/\partial k^d \partial k^a= \partial^2 \omega_{n}/\partial k^d \partial k^a-\partial^2 \omega_{m}/\partial k^d \partial k^a $. 

Assuming time-reversal symmetry in the ground state we can choose $\v{r}_{nm}(-\v{k})=\v{r}_{mn}(\v{k})$ to show that $\iota_3$ is real, symmetric in the $b,c$ indices, and satisfies $[\iota_{3}^{abcd}(0,\omega,-\omega,0)]^{*}= \iota_{3}^{acbd}(0,\omega,-\omega,0)=\iota_{3}^{abcd}(0,-\omega,\omega,0)$. From Eq.~\ref{eq:jerk_current_from_limit}, we see that $\iota_3$ controls the current 
\begin{align}
\frac{d^2}{dt^2} J^{a(3)}_{jerk}=\sum_{b\beta c\gamma d\delta} \iota_{3}^{abcd}(-\omega_{\Sigma},\omega_{\beta},\omega_{\gamma},\omega_{\delta}) E^b_{\beta} E^c_{\gamma} E^d_{\delta} e^{-i\omega_{\Sigma}t},
\label{eq:Jjerk_gen}
\end{align}
subject to $\omega_{\Sigma}=0$. Performing the sum over frequencies we obtain 
\begin{align}
\frac{d^2}{dt^2} J^{a(3)}_{jerk}=6\sum_{bcd} \iota_{3}^{abcd}(0,\omega,-\omega,0) E^b(\omega) E^c(-\omega) E^d_0,
\label{eq:Jacc}
\end{align}
where $E^d_0$ is a static external field. The factor of $6=3!$ is the number of pair-wise exchanges of field indices $(b\beta),(c\sigma),(d\delta)$.~\cite{Boyd2008} The jerk current vanishes for frequencies smaller than the energy band gap. Eq.(\ref{eq:Jacc}) indicates that the jerk current grows quadratically with illumination time
\begin{align}
|\v{J}^{(3)}_{jerk}(t)|\sim \iota_3 t^2,
\end{align}
in the absence of momentum relaxation and saturation effects. In analogy with second derivative of velocity which is called 'jerk' we dub it \textit{jerk} current. This should be compared and contrasted with injection current which grows linearly with illumination time (Eq.~\ref{eq:eta2_lin_in_time}) and shift current which is constant (Eq.~\ref{eq:sigma2_contant_time}). 

\subsection{Materials}
In general, the 81 components of $\iota_3$ are finite in both centrosymmetric and noncentrosymmetry crystal structures. In practice, the symmetries of the 32 crystal classes greatly reduce the number of independent components. For example, GaAs has $\bar{4}3m$ point group, with 21 nonzero components and 4 independent components.~\cite{Boyd2008} However, $\iota_3$ is symmetric under exchange of $bc$ which reduces the number of independent component to 3. In 2D materials the number of components of $\iota_3$ is also small. For example, single-layer GeS has $mm2$ point group which contains a mirror-plane symmetry and a 2-fold axis. In this case $\iota_3$ has only six independent components.
  
In general, linear, circular or unpolarized light will produce jerk current along the direction of the static field. Current transverse to the static field may not be generated with unpolarized or circular polarization. 

\subsection{Physical interpretation of jerk current}
The terms in Eq.(\ref{eq:jerk}) are hard to interpret physically. We now rederive the same result in a physically more transparent way using a phenomenological model.~\cite{Fregoso2018} Consider an electron wave packet in band $n$ subject to a static electric field $E_0^d$. The electron's wavevector obeys
\begin{align}
\hbar\frac{ d \v{k}}{dt}=-e \frac{\partial \v{A}_0}{\partial t},
\end{align}
where the vector potential $\v{A}_{0}$ gives the static electric field $E^d_0=-\partial A_{0}^d/\partial t$. The Bloch velocity of the electron $\v{v}_{n}(\v{k} - e\v{A}_{0}/\hbar)$ can be expanded in powers of $\v{A}_0$. Its first and second time derivatives are given by  
\begin{align}
\frac{d v^a_n}{d t} &=\frac{e}{\hbar} \sum_{d} \omega_{n;ad}  E_0^d, 
\label{eq:dvdt}
\\
\frac{d^2 v^a_n}{d t^2} &=\frac{e^2}{\hbar^2} \sum_{de} \omega_{n;ade}  E_0^d E_0^e. 
\label{eq:d2vdt2}
\end{align}
Now, taking two time derivatives of Eq.(\ref{eq:j_standard})

\begin{align}
\frac{d^2 J^{a}}{dt^2} = \frac{e}{V}\sum_{n\v{k}} \left( \frac{d^2 f_n}{dt^2} v_{n}^{a} +  2\frac{d f_n}{dt} \frac{dv_{n}^{a}}{dt}
+ f_n \frac{d^2 v_{n}^{a}}{dt^2} \right),
\label{eq:d2j}    
\end{align}
and using Eq.~\ref{eq:FG_rule} and Eq.~\ref{eq:dvdt} we have (to linear order in $E_0^d$)
\begin{align}
&\frac{d^2}{dt^2} J^{a(3)}_{jerk} =\nn\\
& \frac{2\pi e^4}{\hbar^3 V}\sum_{b c'd}\sum_{cv\v{k}}2 \omega_{cv;ad} r_{vc}^b r_{cv}^{c'} \delta(\omega_{cv}-\omega) E^{b}(\omega) E^{c'}(-\omega) E_0^d \nn\\
&+\frac{2\pi e^4}{\hbar^3 V}\sum_{bc'd} \sum_{cv\v{k}}\omega_{cv;a} \frac{\partial (r_{vc}^b r_{cv}^{c'})}{\partial k^d} \delta(\omega_{cv}-\omega) \nn\\
&~~~~~~~~~~~~~~~~~~~~~~~~~~~~~~~~~~~~\times E^{b}(\omega) E^{c'}(-\omega) E_0^d.
\end{align}
Since $\omega>0$ we can extend the sums over to all bands and recover Eq.(\ref{eq:Jacc}). An important point of this calculation is to show that the physical origin of the first term in Eq.(\ref{eq:jerk}) comes from the acceleration of carriers in the static electric field. The second contribution comes from a nonconstant carrier injection rate $d^2 f_n/dt^2\neq 0$ which is missing in the standard semiclassical approach.\cite{Jepsen1996}

\subsection{Jerk Hall current}

\begin{figure}
\includegraphics[width=.45\textwidth]{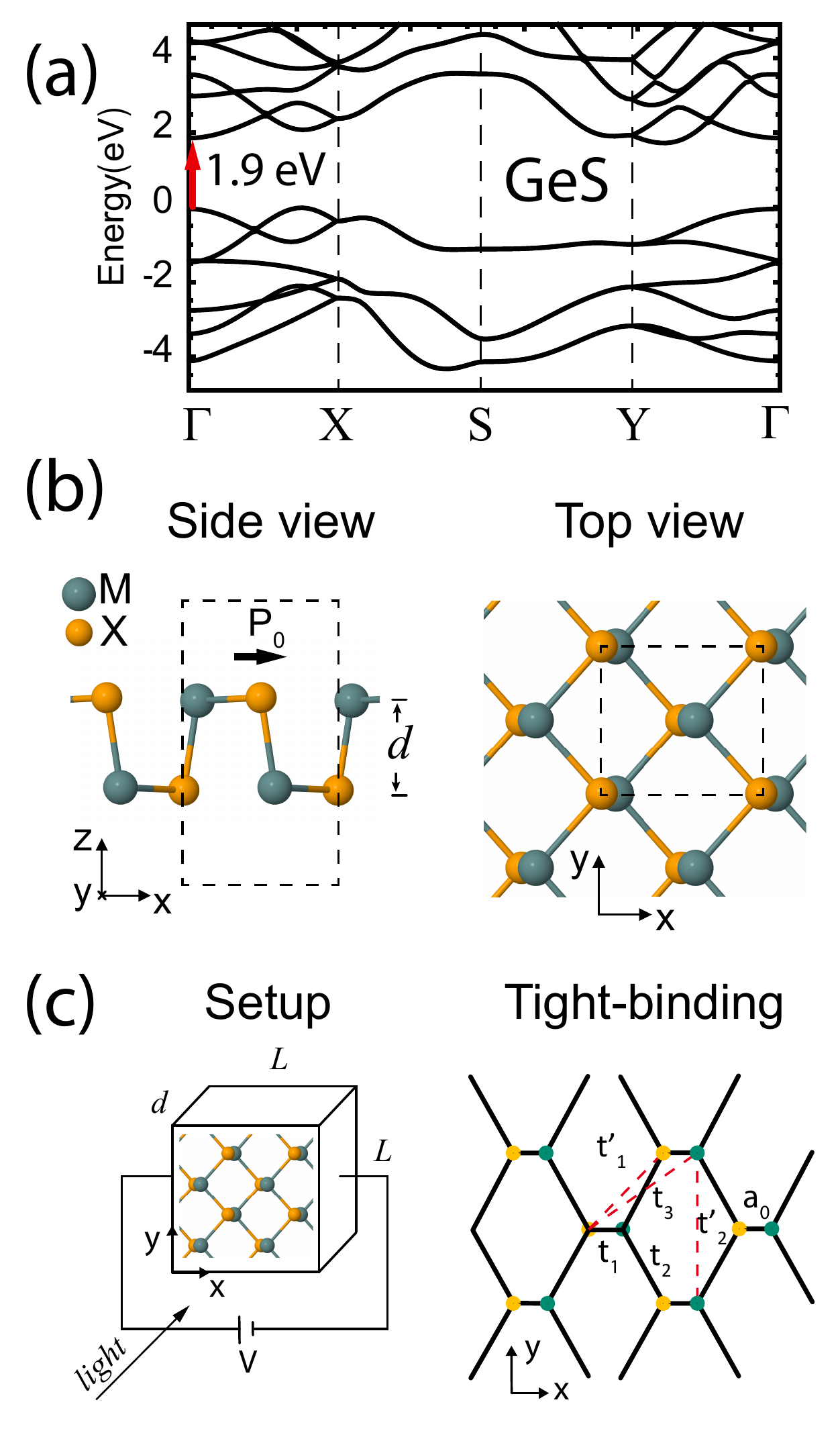}
\caption{(a) Band structure of single-layer GeS~\cite{Gomes2015,Rangel2017} indicating transitions near the band edge (red arrow). (b) crystal structure of single-layer GeS, (c) sample setup and two-dimensional, two-band tight binding model of single-layer GeS which reproduces the nonlinear optical response of near the band edge. The hopping parameters considered are indicated. See main text for more details.}
\label{fig:tb_model}  
\end{figure}
In an isotropic medium, charge carriers move parallel to the electric field. The jerk current, on the other hand, can flow transverse to the static electric field in a rotationally symmetric medium.  To see this, let us assume a sample biased in the $x$-direction and compute the current in the $y$-direction. An optical field $\v{E}=\hat{\v{x}} E^x(\omega)e^{-i\omega t} + \hat{\v{y}} E^y(\omega)e^{-i\omega t} +c.c.$, with $E^a(\omega)=|E^a(\omega)| e^{-i\phi_a}$, is incident perpendicular to the sample surface which defines the $xy$-plane. The current in the $y-$direction is 

\begin{align}
\frac{d^2}{dt^2} J^{y(3)}_{jerk} = \varsigma_{3jH}^{yx} E_{0}^{x} ,
\label{eq:jerk_hall}
\end{align}
where the effective jerk Hall (jH) conductivity is 
\begin{align}
\varsigma_{3jH}^{yx}&\equiv 6\iota_{3}^{yxxx} |E^{x}(\omega)|^2  +  6\iota_{3}^{yyyx} |E^{y}(\omega)|^2  \nn\\ 
&~~~~+12\iota_{3}^{yyxx}|E^x(\omega)| |E^y(\omega)| \cos(\phi_x - \phi_y).
\label{eq:hall_jerk_cond}
\end{align}
In a simple relaxation time approximation the jerk conductivity (see Eq.~\ref{eq:conductivity_3rd_gen}) is cut off by a relaxation time $\tau_1$ as

\begin{align}
\frac{\iota_3}{(-i\omega_{\Sigma})^2} \to \frac{\iota_3}{(\frac{1}{\tau_1}-i\omega_{\Sigma})^2} 
\end{align}
Hence, the jerk current is 
\begin{align}
J^{y(3)}_{jerk} \sim \frac{\tau^{2}_{1}}{(1-i\omega_{\Sigma}\tau_1)^2} \varsigma_{3jH}^{yx} E_{0}^{x} ,
\label{eq:jerk_with_diss}
\end{align}
where $\tau_{1}$ is the relaxation time of the diagonal elements of the density matrix. In the dc limit the current is proportional to the square of the momentum relaxation. For frequencies larger than the Drude peak but smaller than interband transitions the current is independent of the scattering time and it is a measure of the geometry of the Bloch wavefunctions.

The dependence on light's polarization as $\cos(\phi_x-\phi_y)$ and the square of the momentum relaxation are unique characteristics of the jerk current which can be used to distinguish it from $\eta_3$ and $\sigma_3$.

The symmetries of the crystal can also constrain the contributions to the jerk current, e.g., if the crystal has mirror symmetry $y \to -y$ the first and second terms in Eq.~\ref{eq:hall_jerk_cond} vanish. In addition, for circular polarization $\phi_x-\phi_y=\pm \pi/2$ the last term vanishes. An estimate of the jerk current in realistic materials is given next.

\subsection{Example: Jerk current in single-layer GeS}
\label{sec:jerk_GeS}
\begin{figure}
\includegraphics[width=.47\textwidth]{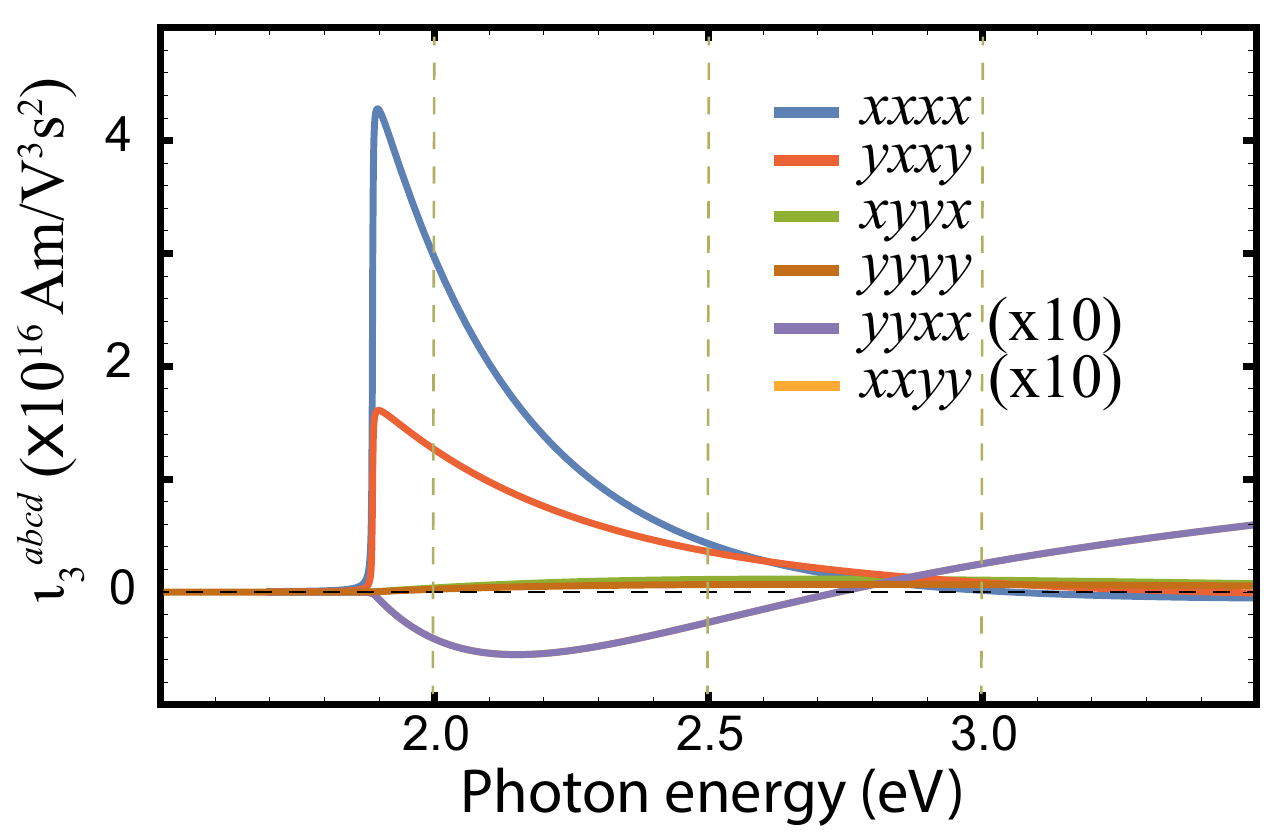}
\caption{Jerk current response tensor of single-layer GeS near the band edge. The two-band model used is shown in Fig.~\ref{fig:tb_model}c. The tensor vanishes for photon energies lower than the energy band gap ($\sim 1.9$ eV~\cite{Cook2017,Rangel2017}). The strongest component is along the polar axis $xxxx$. The components $yyxx,xxyy$, describe a Hall-like response and are an order of magnitude smaller. For added clarity, these components are multiplied by 10.}
\label{fig:jerk_2D_model}  
\end{figure}
To get a sense of the magnitude of the jerk current in real materials we now calculate it for single-layer GeS. Single-layer GeS is of  great interest for its predicted in-plane spontaneous ferroelectric polarization, suitable energy band gap in the visible spectrum ($\sim$ 1.9 eV) and large nonlinear optical response.\cite{Kushnir2017,Rangel2017,Panday2017,Ibanez-Azpiroz2018,Wang2017,Panday,Wang}

We consider a 2D, two-band tight-binding model of single-layer GeS shown in Fig.~\ref{fig:tb_model}c. The details of the model are presented in Appendix~\ref{sec:tb_GeS}. The model has been shown to reproduce the \textit{ab-initio} shift and injection current of  single-layer GeS near the band edge,~\cite{Cook2017,Rangel2017,Panday} specifically in the energy range 1.9-2.14 eV.  Since the model is 2D, we divide the model's 2D current by the thickness of the GeS layer ($d\sim 2.56$\AA) to obtain an effective bulk value.
  
Because of the mirror symmetry $y \to -y$ of the crystal, only six tensor component are independent. As seen in Fig.~\ref{fig:jerk_2D_model}, the strongest is along the polar (chosen along $x$-axis) of magnitude $\sim 10^{16}$ Am/V$^{3}$s$^{2}$. The current transverse to the static electric field, described by the component $\iota_{3}^{yyxx}$ (see Eq.~\ref{eq:jerk_hall}), is an order of magnitude smaller.

The sample is rectangular of dimensions $L\times L$ and thickness $d=2.56$ \AA ~and is biased by an external battery of voltage $V$, as seen Fig.~\ref{fig:tb_model}c. Let us assume the optical field is incident perpendicularly to the plane of single-layer GeS as
\begin{align}
\v{E}(t) &= \hat{\v{x}} E^{x}(\omega) e^{-i\omega t} + \hat{\v{y}} E^{y}(\omega)  e^{-i\omega t} + c.c.\\
\v{E}_0 &= \hat{\v{x}} E^{x}_0.
\end{align}
where $E^{x}(\omega)=E^{0}(\omega)\cos\theta e^{-i\phi_x}$, $E^{y}(\omega)=E^{0}(\omega)\sin\theta e^{-i\phi_y}$, $\theta$ is the angle with the $x$-axis.  The longitudinal and transverse currents  are 

\begin{figure}
\includegraphics[width=.45\textwidth]{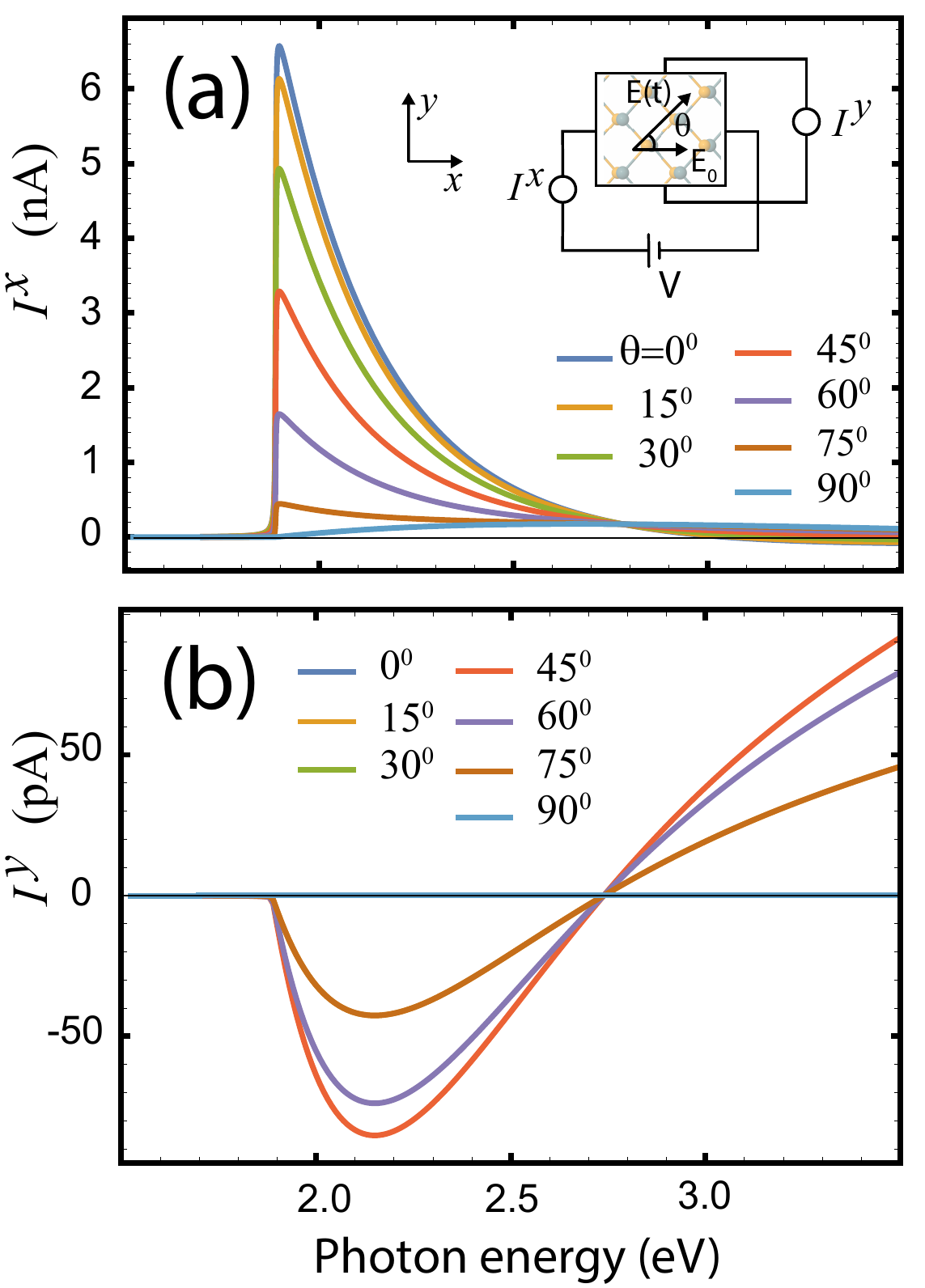}
\caption{Jerk current in single-layer GeS with linear polarization at various angles  $\theta$ with respect to the polar axis. (a) current parallel to the polar axis and (b) current perpendicular to the polar axis. The transverse current is largest at $\theta=45^{0}$ whereas the parallel current is largest when the light's polarization is along the polar axis. The inset shows the top view of the sample.}
\label{fig:iota3_current_tb_ges}  
\end{figure}

\begin{align}
I_{jerk}^{x(3)} &= 6  A\tau_1^2 (\iota_3^{xxxx} |E^{x}(\omega)|^2 + \iota_3^{xyyx} |E^{y}(\omega)|^2 )E_0^{x},\\
I_{jerk}^{y(3)} &= 12 A \tau_1^2 \iota_3^{yyxx} |E^{x}(\omega)| |E^{y}(\omega)|\cos(\phi_x-\phi_y) E_0^{x},
\label{eq:trans_jerk_ges}
\end{align}
where $A=Ld$ is the transverse area of the sample. Note that the current along the polar ($x$)-axis is independent of the polarization of light. Hence, the polar component of the current will not vanish even for unpolarized light. The transverse component of the current, on the other hand, vanishes for circularly polarized (and unpolarized) light and is maximum for linearly polarized light. 

We choose the optical field to be linearly polarized ($\phi_x=\phi_y$) at an angle $\theta$ with the polar axis as shown in the inset to Fig.~\ref{fig:iota3_current_tb_ges}a. The figure shows the jerk current induced as a function of $\theta$. We assumed semiconductor parameters typically found in the laboratory: $L=100 \mu$m, $V=1 $V, $E_0^x=V/L=10^{4}$ V/m, amplitude of the optical field $E^{0}=10^{5}$ V/m, and $\tau_1 = 100$ fs~\cite{Li2018}. 

First note that the magnitude of the current is of the order of pA-nA which is within experimental reach. $I^x$ is maximum when the polarization of light coincides with the polar axis and decreases monotonically as the polarization turns away towards the $y$-axis. $I^y$, on the other hand, is nonmonotonic; it is zero when the light polarization and the polar axis coincide, then rises to a maximum at $45^{0}$, and then decreases to zero for light polarized perpendicular to the polar axis. 

In ultrafast pulsed experiments, the THz radiation emitted by the currents can be analyzed to study the nonlinear optical response of the system without need of mechanical contact. In this scenario the system does no have time to decay and the response is determined mainly by the laser pulse characteristics not by the momentum dissipation mechanism. The above results indicate that the crystal structure, the geometry of the setup and light polarization can be used to uniquely characterize the jerk current tensor components. Injection and shift currents has been reported in THz spectroscopy in various materials~\cite{Cote2002,Laman2005,Bieler2005,Sun2010,Bas2015,Bas2016}. 

\section{third-order injection current}
\label{sec:modified_injection}
An explicit calculation of $\eta_3$  is given in Appendix~\ref{app:details_eta3}. The result is 

\begin{widetext}
\begin{align}
\eta_3^{abcd} (0,\omega,-\omega,0) &=-\frac{\pi e^4}{6\hbar^3 V}\sum_{nm\v{k}} f_{mn} \bigg( \Omega^{ad}_{nm} [r^b_{nm},r^c_{mn}] + i [r^b_{nm},r^c_{mn;a}]_{;d}- 2i\omega_{nm;a} \big[\left(\frac{r_{mn}^d}{\omega_{nm}}\right)_{;b}, r_{nm}^c\big] \bigg)\delta(\omega_{nm}-\omega) \nn \\ 
&~~~-\frac{\pi e^4}{3\hbar^3 V}\sum_{nml\v{k}} f_{mn} \omega_{nm;a} \frac{r_{ln}^d}{\omega_{nl}} [ r_{nm}^b, r_{ml}^c] D_{-}(\omega_{nm},\omega).
\label{eq:eta3} 
\end{align}
\end{widetext}
We defined $\Omega^{ad}_{nm} \equiv \Omega^{ad}_{n}- \Omega^{ad}_{m}$ as the difference of Berry vector potentials. The Berry potential is related to the Berry curvature by $\Omega^{ad}_n = \sum_e \epsilon_{ade} \Omega^{e}_n$. 

The covariant derivative of $r_{mn}^{d}/\omega_{nm}$ is with respect to the gauge dependent $r_{mn}^{d}$ (see for example Eq.~\ref{eq:cov_der_quot}). The product $r_{nm}^{b}r_{mn;a}^{c}$ is gauge invariant and hence its covariant derivative reduces to the standard derivative ( see for example Eq.~\ref{eq:cov_der_prod}). To simplify notation we also defined 
\begin{align}
[O(b),P(c)] &\equiv O(b)P(c)- O(c)P(b),\\ 
D_{\pm}(\omega_{nm},\omega) &\equiv \delta(\omega_{nm}-\omega) \pm \delta(\omega_{nm}+\omega),
\label{eq:D_pm}
\end{align}
where $O,P$ are arbitrary matrix elements which depend on the cartesian indices $b,c$. For example

\begin{align}
[r_{nm}^b,r_{mn}^{c}] \equiv r_{nm}^b r_{mn}^{c} - r_{nm}^c r_{mn}^{d}.
\end{align}
One can see that $\eta_3$ in Eq.~\ref{eq:eta3} is manifestly antisymmetric under exchange of $b,c$. In addition, it is easy to show that $\eta_3^{abcd}(0,\omega,-\omega,0)$ is pure imaginary and satisfies $[\eta_{3}^{abcd}(0,\omega,-\omega,0)]^{*}=- \eta_{3}^{abcd}(0,\omega,-\omega,0)=\eta_{3}^{abcd}(0,-\omega,\omega,0)$. The antisymmetry in the $b,c$ indices implies that $\eta_3$ vanishes for linearly polarized light. $\eta_3$ represents the current  
\begin{align}
\frac{d}{dt} J^{a(3)}_{3i}=6\sum_{bcd} \eta_{3}^{abcd}(0,\omega,-\omega,0) E^b(\omega) E^c(-\omega) E^d_0,
\label{eq:mi_current}
\end{align}
which varies as
\begin{align}
|\v{J}^{(3)}_{3i}|\sim \eta_3 t,
\end{align}
in the absence of momentum relaxation and saturation effects.

\subsection{Materials}
In general, the 81 components of $\eta_3$ are finite in both centrosymmetric and noncentrosymmetry crystal structures. In practice, the symmetries of the 32 crystal classes greatly reduce the number of independent components. For example, GaAs has $\bar{4}3m$ point group, with 21 nonzero components and 4 independent components.~\cite{Boyd2008} However, $\eta_3$ is antisymmetric under exchange of $b,c$ which reduces the number of independent components to 1. In 2D materials the number of components of $\eta_3$ is also small. For example, single-layer GeS has $mm2$ point group which contains a mirror-plane symmetry and a 2-fold axis. In this case $\eta_3$ has only 2 independent components.
  
In general, circular or unpolarized light will produce third-order injection current along the direction of the static field. Current transverse to the static field may not be generated with unpolarized or linear polarization. 

\subsection{Physical interpretation of third-order injection current}
The presence of a static field gives rise to new physical processes which we now describe in detail.

\subsubsection{First term}
The first term in Eq.(\ref{eq:eta3}) arises from the asymmetric injection of carriers into anomalous velocity states. To see this, let us consider an electron wave packet in band $n$ subject to a static field $E_0^d$. The static field induces an anomalous contribution to the electron's velocity which generates a current given by (Eq.~\ref{eq:J_intra})
\begin{align}
\v{J}_{3i,1}= -\frac{e^2}{\hbar V}\sum_{n\v{k}} f_n \v{E}_0\times \pmb{\Omega}_{n},
\end{align}
where we used $f_n=\rho^{(0)}_{nn}$. Taking a time derivative of the occupations we obtain
\begin{align}
\frac{d}{dt}\v{J}_{3i,1}= -\frac{e^2}{\hbar V}\sum_{n\v{k}} \frac{df_n}{dt} \v{E}_0\times \pmb{\Omega}_{n}.
\label{eq:dj_anom}
\end{align}
This expression means that when the optical field is turned on electrons will be excited from the valence into anomalous conduction states. To lowest order, i.e., second order in the optical field and first in the static field, Fermi's Golden rule gives the one-photon injection rate shown in Eq.(\ref{eq:FG_rule}). Using Eqs.(\ref{eq:FG_rule}) we obtain
\begin{align}
\frac{d}{dt}J^{a(3)}_{3i,1}= -\frac{2\pi e^4}{\hbar^3}\sum_{b' c' d} \sum_{vc\v{k}} \Omega_{cv}^{ad} r_{cv}^{b'} r_{vc}^{c'} &\delta(\omega_{cv}-\omega) \times\nn\\
&E^{b'}(\omega) E^{c'}(-\omega) E_0^{d}. 
\label{eq:Janom}
\end{align}
Using the fact that $\omega>0$ we can extend the sum to all bands and recover the first term in Eq.~\ref{eq:eta3}. 

\begin{figure}
\includegraphics[width=.47\textwidth]{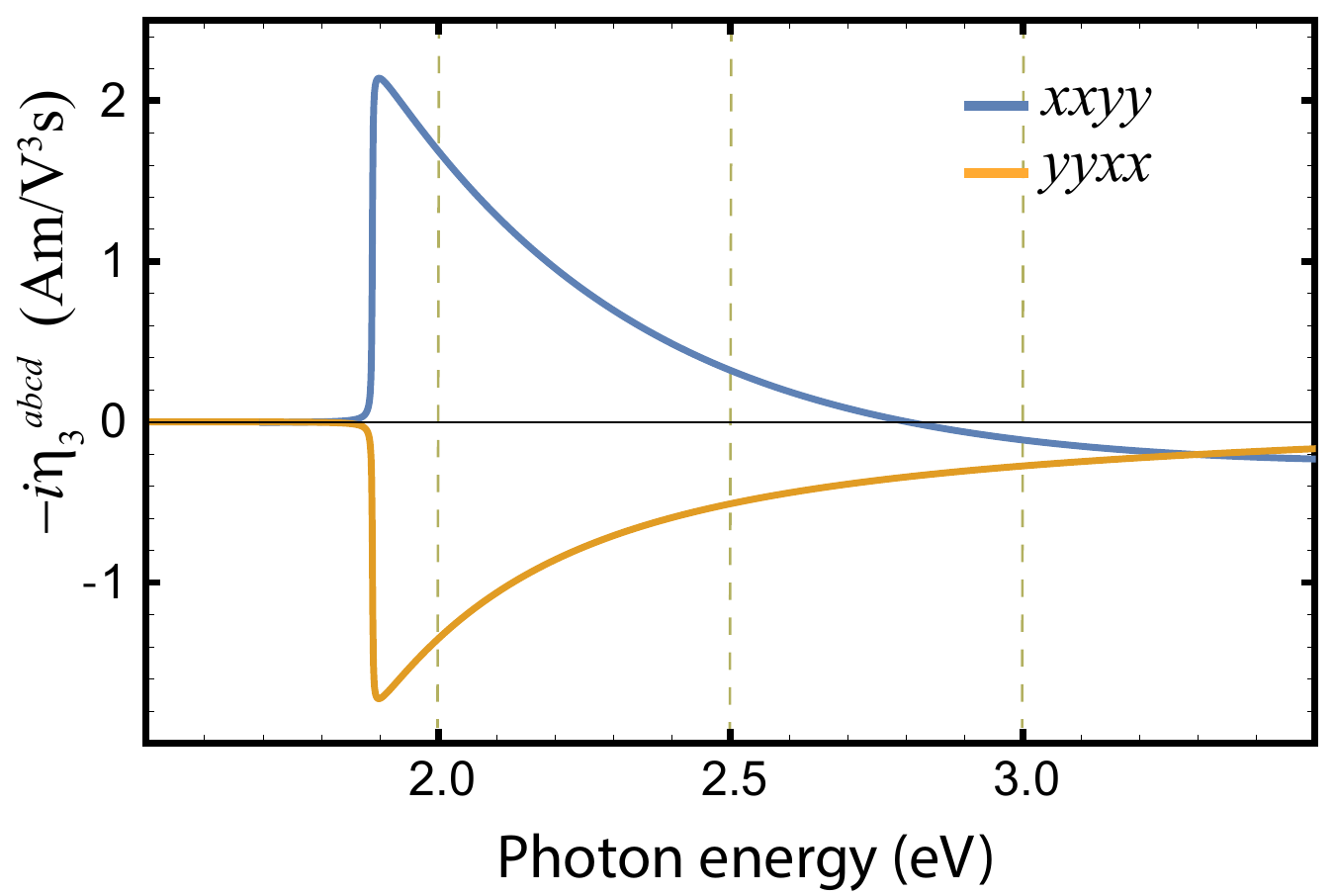}
\caption{Injection current response tensor $\eta_3^{abcd}$ in single-layer GeS near the band edges. $\eta_3$ gives rise only to current transverse to the static field and vanishes for linearly polarized light. The tight-binding model parameters are described in Sec.~\ref{sec:jerk_GeS}.}
\label{fig:eta3_2D_model}  
\end{figure}

\subsubsection{Second term}
In the presence of a static field a wave packet drifts in the BZ giving rise to a current. Similarly, a pair of wave packets could drift coherently in the presence of a static field giving rise to a dipole current. To see this, consider the dipole velocity contribution to the current in Eq.~\ref{eq:Jdip_rho1}. Writing explicitly the small imaginary part of the external frequencies and taking the resonant part we obtain
\begin{align}
J^{a(2)}_{3i,2} =& -\frac{i\pi e^3}{\hbar^2 V}\sum_{bc} \sum_{nm\v{k}} f_{nm}\big[ r^{b}_{nm;a} r^{c}_{mn}\delta(\omega_{mn}+\omega) \nn\\
&+ r^{c}_{nm;a} r^{b}_{mn}\delta(\omega_{mn}-\omega) \big] E^b(\omega) E^c(-\omega).
\end{align}
Taking a time derivative of the dipole matrix elements, exchanging $n,m$ indices, and making $\v{k} \to -\v{k}$, we obtain 
\begin{align}
\frac{d}{dt}J^{a(3)}_{3i,2} =& -\frac{i\pi e^4}{\hbar^3 V}\sum_{bcd} \sum_{nm\v{k}} f_{nm} \frac{\partial}{\partial k^d}\big( r^{b}_{nm;a} r^{c}_{mn} \nn\\
&- r^{c}_{nm;a} r^{b}_{mn} \big)\delta(\omega_{nm}-\omega) E^b(\omega) E^c(-\omega) E_{0}^d,
\end{align}
which can be recognized as the second term in Eq.~\ref{eq:eta3}.

\begin{figure}
\includegraphics[width=.43\textwidth]{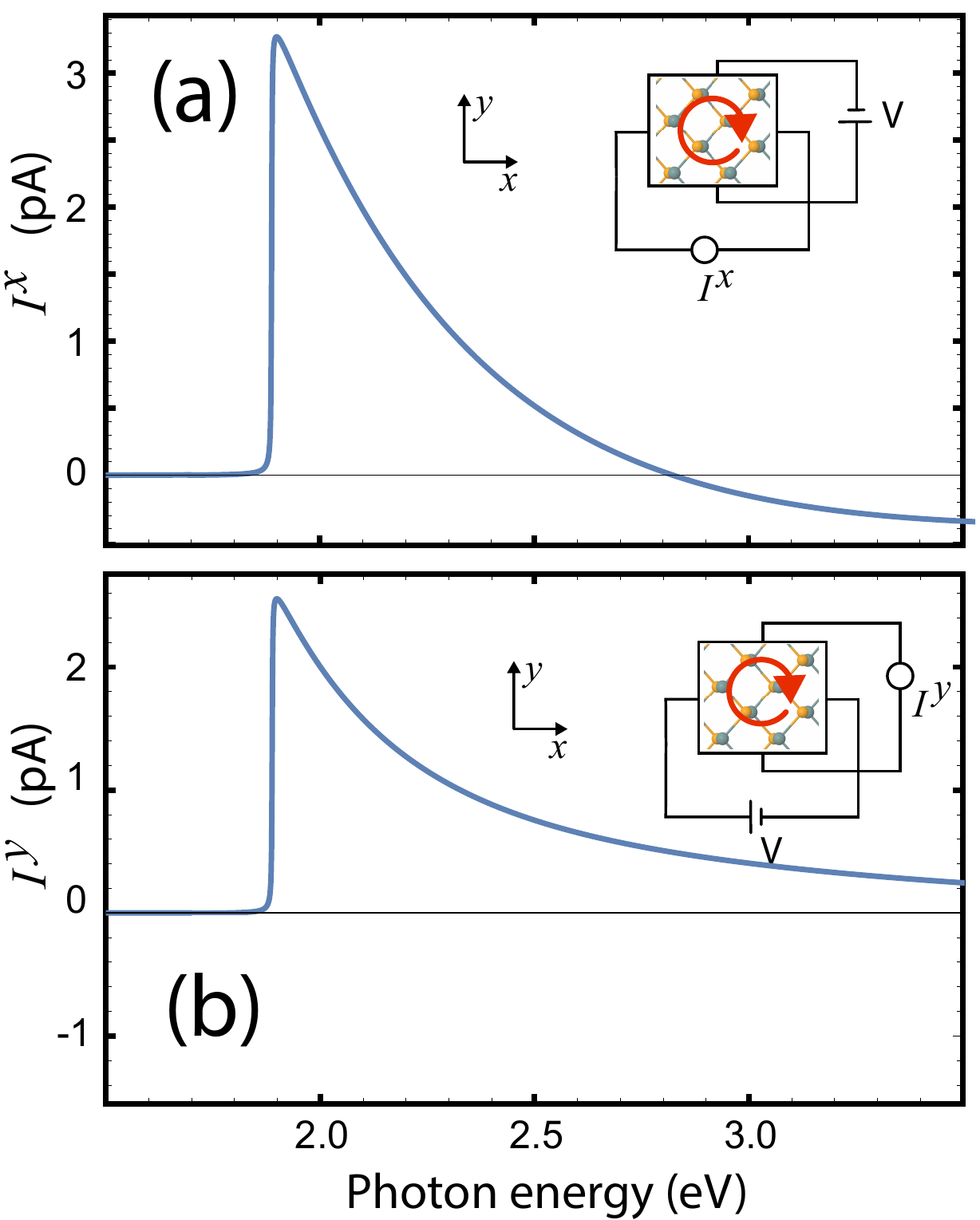}
\caption{$\eta_3$-injection current in single-layer GeS near the band edge. (a) shows the current parallel to the polar axis, $I^{x}$ and (b) the current transverse to the polar axis $I^{y}$. Light is circularly polarized and incident perpendicular to the plane of the GeS in both sample setups.}
\label{fig:3icurrent_ges}  
\end{figure}

\subsubsection{Third term}
The third term takes into account the change of the electron distribution due to the static field. To see this, let us consider the current of an electron wave packet in band $n$ to third order in the electric field. From Eq.~\ref{eq:J_intra}
\begin{align}
J^{a(3)}_{3i,3}=\frac{e}{V}\sum_{n\v{k}} v^{a}_{n} \rho^{(3)}_{nn}.
\end{align} 
Taking a time derivative of the density matrix gives
\begin{align}
\frac{d}{dt}J^{a(3)}_{3i,3}=\frac{e}{V}\sum_{n\v{k}} v^{a}_{n} \frac{\partial \rho^{(3)}_{nn}}{\partial t}.
\end{align} 
From Eq.~\ref{eq:eom_rho} the time derivative of the density matrix is 

\begin{align}
\frac{\partial \rho^{(3)}_{nn}}{\partial t} =  \frac{e}{i\hbar}\sum_{bm} E^{b}(\rho^{(2)}_{nm}r^{b}_{mn}-r^{b}_{nm}\rho^{(2)}_{mn}) 
-\frac{e}{\hbar} \sum_b  E^b \rho^{(2)}_{nn;b}.
\label{eq:eom_rho2}
\end{align}
Now consider the intraband part of the second order density matrix obtained from Eq.~\ref{eq:2nd_pt} 

\begin{align}
\rho^{(2)}_{nm,i} = \frac{i e}{\hbar} \sum_{b\beta c\sigma} \frac{\bar{\rho}^{(1)b\beta}_{nm;c}}{\omega_{mn}-\omega_{\Sigma}}E^{b}_{\beta} E^{c}_{\sigma} e^{-i\omega_{\Sigma} t},
\end{align}
where $\omega_{\Sigma}=\omega_{\beta} + \omega_{\sigma}$. The first order density matrix in the presence of a static field is (see Eq.~\ref{eq:1st_pt})

\begin{align}
\bar{\rho}^{(1)d0}_{nm} = \frac{e}{\hbar} f_{mn} \frac{r_{nm}^d}{\omega_{nm}}.
\label{eq:rho1_static}
\end{align}
Substituting the above equations into Eq.~\ref{eq:eom_rho2} and taking the resonant part we recover the third term in Eq.~\ref{eq:eta3}.  The factor of two in Eq.~\ref{eq:eta3} is due to two possible choices for the static electric field.

\subsubsection{Fourth term} 
This contribution arises from electrons excited from the valence to conduction bands via an intermediate state $l$. These new states are generated by the presence of static field. 

\subsection{Third-order injection Hall current}
\label{sec:inj_Hall}
Let us assume a static field is in the $x$-direction and compute the current in the $y$-direction. An optical field of the form $\v{E}=\hat{\v{x}} E^x(\omega)e^{-i\omega t} + \hat{\v{y}} E^y(\omega)e^{-i\omega t} +c.c.$ is incident perpendicular to the sample surface which we take to define the $xy$-plane. From Eq.~\ref{eq:mi_current} the current transverse to the static field is
\begin{align}
\frac{d}{dt} J^{y(3)}_{3iH} =\varsigma_{3iH}^{yx} E_{0}^{x}
\end{align}
where $E^a(\omega)=|E^a(\omega)| e^{-i\phi_a}$ and the Hall coefficient is 
\begin{align}
\varsigma_{3iH}^{yx} \equiv 12 i \eta_{3}^{yyxx}|E^x(\omega)| |E^y(\omega)| \sin(\phi_x - \phi_y).
\end{align}
Similar to $\eta_2$, $\eta_3$ vanishes for linear polarization  $\phi_x=\phi_y$ and is maximum for circularly polarized light. In a simple relaxation time approximation, the dc singularity in the conductivity (see Eq.~\ref{eq:conductivity_3rd_gen}) is cut off by a phenomenological relaxation time $\tau_1$ as

\begin{align}
\frac{\eta_3}{-i\omega_{\Sigma}} \to \frac{\eta_3}{\frac{1}{\tau_1}-i\omega_{\Sigma}}.
\end{align}
The current is  

\begin{align}
J^{y(3)}_{3iH} \sim \frac{\tau_1}{1-i\omega_{\Sigma}\tau_1} \varsigma_{3iH}^{yx} E_{0}^{x},
\end{align}
If $\omega_{\Sigma}=0$, the current is proportional to $\tau_1$ the relaxation of the diagonal elements of the density matrix. For frequencies larger than the Drude peak $\omega_{\Sigma}\tau_1 \gg 1$ but smaller than interband transitions the current is independent of the scattering time and hence is a measure of the geometry of the Bloch wavefunctions. 

\subsection{Example: Third order injection current in single-layer GeS}
To get a sense of the $\eta_3$-injection current in real materials we now calculate it for single-layer GeS. We use the same 2-band, 2D tight-binding model of single-layer GeS and same sample geometry as in Sec.~\ref{sec:jerk_GeS}. 

Out of the 16 tensor components the antisymmetry in the $b,c$ indices and the mirror symmetry $y\to -y$ of the crystal leaves only two independent components, $yyxx$ and $xxyy$, shown in Fig.~\ref{fig:eta3_2D_model}. These components allow current to flow only perpendicular to the static electric. The second term in $\eta_3$ is the dominant term followed by the third, and the first terms which are one and two orders of magnitude smaller respectively.

We chose the optical field to be circularly polarized and the static field is either along the polar axis of the sample or perpendicular to it
 
\begin{align}
\v{E}(t) &= \hat{\v{x}} E^{0}(\omega) e^{-i\omega t} + \hat{\v{y}} E^{0}(\omega) e^{-i\omega t} + c.c.,\\
\v{E}_0 &= \hat{\v{x}} E^{x}_0,~\textrm{or} ~\hat{\v{y}} E^{y}_0,
\end{align}
where $\phi_x-\phi_y=\pi/2$.  The transverse currents are given by 

\begin{align}
I_{3i}^{x(3)} &= 12 A\tau_1  i \eta_{3}^{xxyy} |E^{x}(\omega)| |E^{y}(\omega)| E_{0}^{x} \sin(\phi_y-\phi_x) \\
I_{3i}^{y(3)} &= 12 A\tau_1  i \eta_{3}^{yyxx} |E^{x}(\omega)| |E^{y}(\omega)| E_{0}^{x} \sin(\phi_x-\phi_y) 
\label{eq:t_curr_eta3_ges}
\end{align}
where $A=Ld$ is the transverse area of the sample. Note that the current vanishes for linearly polarized light but is maximum for circular polarization. The calculated induced current is shown in Fig.~\ref{fig:3icurrent_ges}a and \ref{fig:3icurrent_ges}b. Note that the photocurrent is of the order of pA and of the same sign.

\section{Third-order shift current}
\label{sec:mod_shift}
Explicit calculation of $\sigma_3$ gives 

\begin{widetext}
\begin{align}
\sigma_{3}^{abcd}(0,\omega,-\omega,0)=\frac{\pi e^4}{6\hbar^3 V} \sum_{nm\v{k}} &f_{mn} \big[ \{ \bigg(\frac{r_{mn}^d}{\omega_{nm}}\bigg)_{;c}, r_{nm;a}^b\} -\{ \bigg(\frac{r_{mn;a}^d}{\omega_{nm}}\bigg)_{;c}, r_{nm}^b\} \big]\delta(\omega_{nm}-\omega) \nn\\
&-\frac{i\pi e^4}{6\hbar^3 V}\sum_{nml\v{k}} \frac{f_{ln}}{\omega_{mn}} \left[ \{ r_{nl}^c, (r_{mn}^d r_{lm}^b)_{;a}\} - r^d_{mn}\{ r_{nl;a}^c, r_{lm}^{b}\} \right]D_{+}(\omega_{nl},\omega).
\label{eq:sigma3}
\end{align}
\end{widetext}
For details see Appendix~\ref{app:sigma3}. In Eq.~\ref{eq:sigma3} we defined the anticummutator with respect to the $b,c$ indices as 

\begin{align}
\{O(b),P(c)\} \equiv O(b)P(c)+ O(c)P(b)
\end{align}
where $O,P$ are arbitrary matrix elements. For example

\begin{align}
\{ \left(\frac{r_{mn}^d}{\omega_{nm}}\right)_{;c},r_{nm;a}^{b} \} \equiv \left(\frac{r_{mn}^d}{\omega_{nm}}\right)_{;c}r_{nm;a}^{b} + \left(\frac{r_{mn}^d}{\omega_{nm}}\right)_{;b}r_{nm;a}^{c}.
\end{align}
$D_{+}$ is defined in Eq.~\ref{eq:D_pm}. Clearly, $\sigma_3$ is symmetric under exchange of $b,c$, pure real, and satisfies $\sigma_{3}^{abcd}(0,\omega,-\omega,0)=\sigma_{3}^{acbd}(0,-\omega,\omega,0)$. The tensor defines the nonlinear current 

\begin{align}
J_{sh}^{a(3)} = 6\sum_{bcd} \sigma_{3}^{abcd}(0,\omega,-\omega,0) E^b(\omega) E^c(-\omega) E^d_0,
\label{eq:Jx}
\end{align}
which, in the absence of momentum relaxation and saturation effects, is constant with illumination time (if quantum coherence is maintained).

\subsection{Materials}
In general, the 81 components of $\sigma_3$ are finite in both centrosymmetric and noncentrosymmetry crystal structures. In practice, the symmetries of the 32 crystal classes greatly reduce the number of independent components. For example, GaAs has $\bar{4}3m$ point group, with 21 nonzero components and 4 independent components~\cite{Boyd2008}. However, $\sigma_3$ is symmetric under exchange of $b,c$ which reduces the number of independent components to 3. In 2D materials the number of components of $\sigma_3$ is also small. For example, single-layer GeS has $mm2$ point group which contains a mirror-plane symmetry and a 2-fold axis. In this case $\sigma_3$ has only 6 independent components.
  
In general, linear, circular or unpolarized light will produce third-order shift current along the direction of the static field. Current transverse to the static field may not be generated with unpolarized or circular polarization.

\subsection{Physical interpretation of the third-order shift current}
\subsubsection{First term}
The first term in $\sigma_3$ arises from the quantum interference of the dipole velocity and interband band coherences. To see this, note that an oscillating external field creates a dipole with wave packets in two distinct bands. Because the field oscillates the dipole velocity also oscillates, see Eq.~\ref{eq:dip_vel}. If the occupations of the bands, described by the density matrix, oscillate 180$^{0}$ out of phase with respect to the velocity oscillations, a dc current can be established. The process is mediated by the intraband part of the (second order) density matrix 

\begin{align}
J^{a(3)}_{3sh,1}=-\frac{e^2}{\hbar V} \sum_{nm\v{k}} \v{E}(t)\cdot \v{r}_{nm;a}\rho_{mn,i}^{(2)}(t),
\end{align}
where $\rho_{mn,i}^{(2)}$ is the first term in Eq.~\ref{eq:2nd_pt} which clearly represents the intraband part of $\rho^{(2)}_{mn}$. Setting one of the fields in $\rho^{(2)}$ to be static (say $E^{d}_\delta \to E^{d}_0$) we have

\begin{align}
J^{a(3)}_{3sh,1}=-\frac{i e^4}{\hbar^3 V} \sum_{b\beta c\sigma d} \sum_{nm\v{k}} \frac{r^b_{nm;a}}{\omega_{mn}-\omega_{\sigma}} \left( \frac{r_{mn}^d f_{nm}}{\omega_{mn}} \right)_{;c}\nn\\
\times E^{b}_{\beta} E^{c}_{\sigma} E^{d}_{0} e^{-i\omega_{\Sigma}t} 
\end{align}
where $\omega_{\Sigma}=\omega_{\beta} + \omega_{\sigma}$. Symmetrizing with respect electric field indices, substituting $\omega_{\beta}=\pm \omega$ and $\omega_{\sigma}=\mp \omega$, and keeping only resonant terms we recover the first term in Eq.~\ref{eq:sigma3}.

\subsubsection{Second term}
 The second term in $\sigma_3$ arises from the quantum interference of interband coherences. To see this, note that a static external field creates a dipole with wave packets in two distinct bands. Because the field is static, the dipole velocity is constant. The static dipole velocity together with a nonoscillating interband occupation, can generate a dc current. This process is also mediated by the (static) intraband part of the (second order) density matrix 

\begin{align}
J^{a(3)}_{3sh,2}=-\frac{e^2}{\hbar V}  \sum_{d} \sum_{nm\v{k}} E^{d}_{0} r^{d}_{nm;a}\rho_{mn,i}^{(2)}.
\end{align}
Following the same procedure as above and after an integration by parts it is easy to show that we recover the second term in Eq.~\ref{eq:sigma3}.

\subsubsection{Third and fourth term}
The third and fourth terms in $\sigma_3$ are not easily derived from a simple model. These processes involve virtual transitions to intermediate bands created by the static external field and involve the interband part of the second order density matrix. 

\begin{figure}
\includegraphics[width=.47\textwidth]{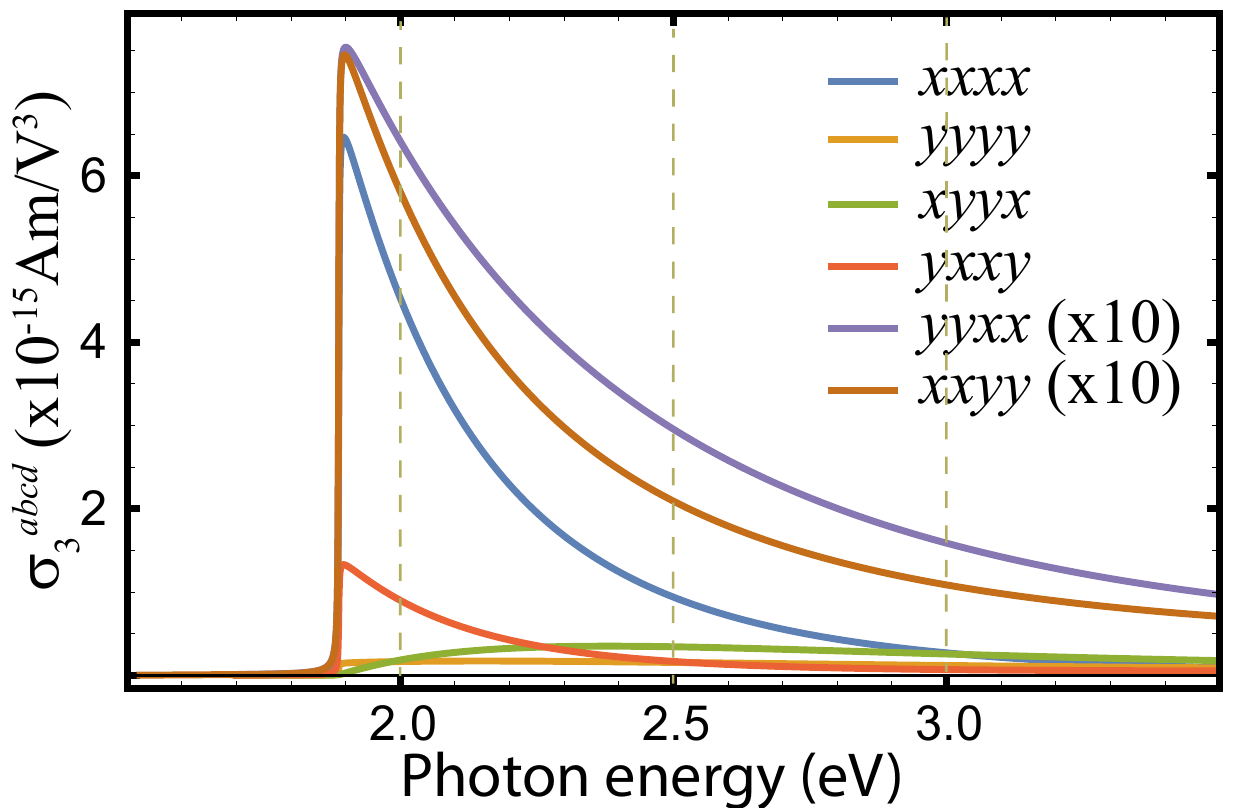}
\caption{Shift current response tensor $\sigma_3^{abcd}$ of single-layer GeS near the band edges. The model parameters are the same as in Sec.~\ref{sec:jerk_GeS}. The largest response is along the polarization axis $x$. The transverse response governed by $xxyy$ and $yyxx$ is an order of magnitude smaller.}
\label{fig:sigma3_tb_model_ges}  
\end{figure}

\subsection{Third-order shift Hall current}
\label{sec:shift_hall}
Let us assume a static field is in the $x$-direction and compute the shift current in the $y$-direction. An optical field of the form $\v{E}=\hat{\v{x}} E^x(\omega)e^{-i\omega t} + \hat{\v{y}} E^y(\omega)e^{-i\omega t} +c.c.$ is incident perpendicular to the sample surface which we take as the $xy$-plane. The current transverse to the static field is
\begin{align}
J^{y(3)}_{3shH} =\varsigma_{3shH}^{yx} E_{0}^{x}
\end{align}
where $E^a(\omega)=|E^a(\omega)| e^{-i\phi_a}$ and the effective Hall conductivity is 
\begin{align}
\varsigma_{3shH}^{yx} \equiv 12 \sigma_{3}^{yyxx}|E^x(\omega)| |E^y(\omega)| \cos(\phi_x - \phi_y).
\end{align}
Similar to $\sigma_2$, $\sigma_3$ vanishes for circular polarization and is maximum for linear polarization $\phi_x=\phi_y$ at 45$^{0}$ with respect to the $x$-axis. Contrary to injection current, the shift current does not have a Drude-like dc divergence but rather gives a finite contribution in this limit (while quantum coherence is maintained).

\subsection{Example: third-order shift current in single-layer GeS}

\begin{figure}
\includegraphics[width=.43\textwidth]{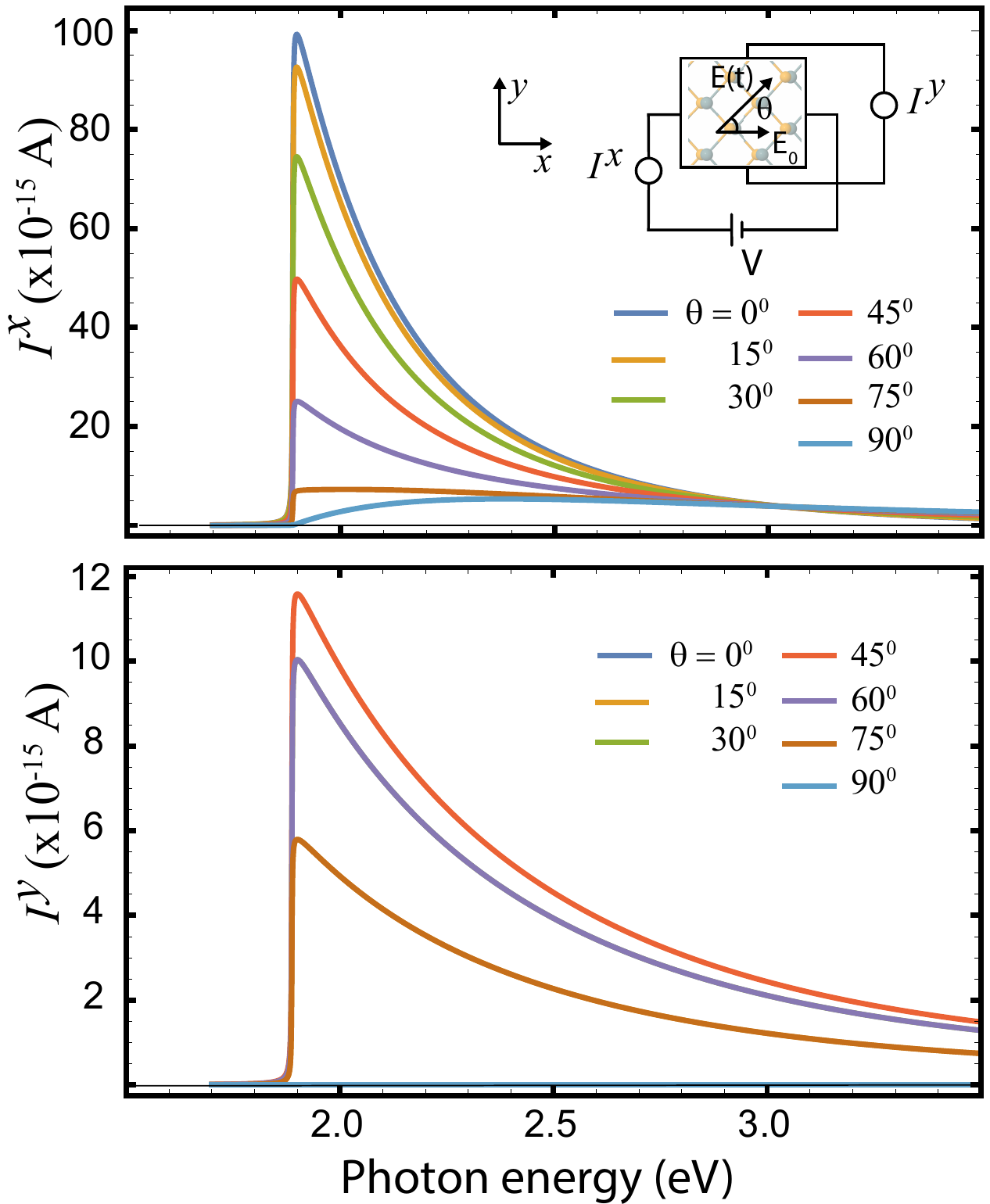}
\caption{$\sigma_3$-shift current in single-layer GeS near the band edge for light linearly polarized at various angles $\theta$ with respect to the polar axis. (a) current parallel to the polar axis $I^{x}$ is largest when the light's polarization lies along the polar axis, and (b) current transverse to the polar axis $I^{y}$ is largest at $\theta= 45^{0}$.}
\label{fig:sigma3_current_ges}  
\end{figure}

To get a sense of the third-order shift current in real materials we now calculate it for single-layer GeS. We use the same setup and tight-binding model of single-layer GeS as in Sec.~\ref{sec:jerk_GeS}. 

Because of the mirror symmetry $y \to -y$ of the model, only six tensor components are independent. As seen in Fig.~\ref{fig:sigma3_tb_model_ges}, the strongest is along the polar axis of magnitude $~\sim 5\times 10^{-15}$ Am/V$^3$. The component transverse to the static electric field $\sigma_{3}^{yyxx}$ (see Sec.~\ref{sec:shift_hall}) is an order of magnitude smaller.

The sample is rectangular of dimensions $L\times L$ and thickness $d=2.56$ \AA~ and is biased by an external battery of voltage $V$ as seen in Fig.~\ref{fig:tb_model}c. For concreteness let us assume the optical field is incident perpendicularly to the plane of single-layer GeS as 
\begin{align}
\v{E}(t) &= \hat{\v{x}} E^{x}(\omega) e^{-i\omega t} + \hat{\v{y}} E^{y}(\omega)  e^{-i\omega t} + c.c.,\\
\v{E}_0 &= \hat{\v{x}} E^{x}_0.
\end{align}
The longitudinal and transverse currents  are

\begin{align}
I_{sh}^{x(3)} &= 6 A (\sigma_3^{xxxx} |E^{x}(\omega)|^2 + \sigma_3^{xyyx} |E^{y}(\omega)|^2 )E_0^{x}\\
I_{sh}^{y(3)} &= 6 A \sigma_3^{yyxx} |E^{x}(\omega)| |E^{y}(\omega)|\cos(\phi_x-\phi_y) E_0^{x}
\label{eq:t_sh_c_ges}
\end{align}
where $E^{x}(\omega)=E^{0}(\omega)\cos\theta e^{-i\phi_x}$, $E^{y}(\omega)=E^{0}(\omega)\sin\theta e^{-i\phi_y}$, $\theta$ is the angle with the $x$-axis, and $A=Ld$ is the transverse area of the sample. Note that the current along the polar $x$-axis is independent of the polarization of light and hence, it will not vanish even for unpolarized light. The transverse component of the current, on the other hand, vanishes for circularly polarized (and unpolarized) light and is maximum for linearly polarized light. 

We choose the optical field to be linearly polarized ($\phi_x=\phi_y$) at an angle $\theta$ with the polar axis as shown in the inset to Fig.~\ref{fig:sigma3_current_ges}a. The figure shows the current along $x$ and $y$-axis induced as a function of $\theta$. We assumed the same  semiconductor parameters as before, e.g., $L=100 \mu$m, $V=1 $V, $E_0^x=V/L=10^{4}$ V/m, amplitude of the optical field $E^{0}=10^{5}$ V/m, and $\tau_1 = 100$ fs. 

First note that the magnitude of the currents is of the order of pA-fA. $I^x$ is maximum when the polarization of light coincides with the polar axis and decreases monotonically as the polarization turns away towards the $y$-axis. $I^y$, on the other hand, is nonmonotonic: it is zero when the polarization and the polar axis coincide, then rises to a maximum at $45^{0}$ and then decreases to zero again for light polarized perpendicular to the polar axis.

\section{Generalizations}
\subsection{Snap current}
\label{sec:snap_current}
By power counting it is easy to see that the leading divergence of $\chi_4$ is of order $\omega^{-4}_{\Sigma}$, and that it occurs when all but two of the external frequencies are zero. Proceeding as before we calculate the corresponding response tensor $\varsigma_{4}^{abcde}(0,\omega,-\omega,0,0)$. Taking three derivatives of Eq.~\ref{eq:j_standard} and using Eqs.~\ref{eq:FG_rule}, \ref{eq:dvdt}, and \ref{eq:d2vdt2} we obtain
\begin{align}
 \varsigma_{4}^{abcde}&= \frac{2\pi e^5}{4! \hbar^4 V}\sum_{nm\v{k}} f_{mn}\big[ 3\omega_{nm;ade} r_{nm}^b r_{mn}^c \nn\\
&~~+ 3\omega_{nm;ad} ( r_{nm}^b r_{mn}^c)_{;e} \nn\\
&~~~+ \omega_{nm;a} (r_{nm}^b r_{mn}^c)_{;de} \big]\delta(\omega_{nm}-\omega).
\label{eq:Jsnap}
\end{align}
The tensor is symmetric in the $b,c$ indices and represents a third derivative of the nonlinear current 
\begin{align}
\frac{d^3 J^{a(4)}_{sp}}{dt^3}=4!\sum_{bcde} \varsigma_{4}^{abcde}(0,\omega,-\omega,0,0) E^b(\omega) E^c(-\omega) E^d_0 E^e_0,
\label{eq:Jsnap_time}
\end{align}
where $E^d_0,E^e_0$ represent static fields. By analogy with a particle's third derivative of its velocity we dub it \textit{snap} current. The current grows as $\sim t^3$ with illumination time in the absence of momentum relaxation and saturation effects. Hence, it is proportional the third power of the relaxation time $\tau_1$

\begin{align}
J^{a(4)}_{sp}\sim \tau_1^3 4!\sum_{bcde} \varsigma_{4}^{abcde} E^b(\omega) E^c(-\omega) E^d_0 E^e_0.
\label{eq:Jsnap_steady}
\end{align}
Note that we can think of the snap current as a second order photoconductivity.

\begin{table*}
\caption{Summary of nonlinear Hall-like responses of single-layer GeS near the band edge. A static electric field is present along $x$ which is taken to define the polar axis of GeS. In addition, an optical electric field is incident perpendicular to the plane of the sample which defines the $xy$-plane. The Hall current is in the $y$-axis. The sample geometry is shown in Fig.~\ref{fig:tb_model}c and the details are in Sec.~\ref{sec:jerk_GeS}. I $\equiv$ inversion symmetry, no I $\equiv$ no inversion symmetry. $I^x$ is the current along the $x$-axis. $^{*}$For comparison, $\eta_2$- and $\sigma_2$-current is given for the same parameters. w.r.t. stands for `with respect to'.}
\begin{tabular}{|m{1.2cm}|m{1.8cm}|m{1.8cm}|m{1.0cm}|m{2.2cm}|m{2.5cm}|m{2.0cm}|m{1.0cm}|m{1.8cm}|m{0.8cm}|} 
\hline 
Current  & Momentum     & Dependence & I vs.& Hall current & Hall current & Hall current &Sign of  & Hall current   & Ref. \\ 
 ~~~~$I^y$ & relaxation & on $E_0^x$ & no I & dependence on& vanishes for & maximum for  &$I^x,I^y$& magnitude$^{*}$& Eq.  \\
           &            &            &      & polarization& polarization  & polarization &         &                &   \\[1ex]
\hline\hline 
$\iota_3$-jerk &$\tau_1^2$& linear&I, no I&$\cos(\phi_x-\phi_y)$& circular,&linear at $45^{0}$& +,-&$10^{-8}$ A&\ref{eq:trans_jerk_ges} \\[2ex]
               &          &            &       &                     & linear $\v{E}(t)\parallel x,y$& w.r.t. $x$-axis&    &&  \\
\hline
$\eta_3$-injection&$\tau_1$& linear    &I, no I    &$\sin(\phi_x-\phi_y)$& linear    & circular          & +,+    & $10^{-12}$ A& \ref{eq:t_curr_eta3_ges} \\[2ex]
\hline
$\sigma_3$-shift  &$\tau_2$& linear    &I, no I    &$\cos(\phi_x-\phi_y)$& circular, &linear at $45^{0}$ & +,+    & $10^{-14}$ A&\ref{eq:t_sh_c_ges} \\[2ex]
                  &        &           &           &                     &linear $\v{E}(t)\parallel x,y$ & w.r.t. $x$-axis&   & &  \\
\hline
$\eta_2$-injection&$\tau_1$& No        &no I       &                   &         &            &           & $10^{-6}$ A$^{*}$& \\[2ex]
\hline
$\sigma_2$-shift  &$\tau_2$& No        &no I       &                   &         &            &           & $10^{-8}$ A$^{*}$& \\[2ex]
\hline
\end{tabular}
\label{table:Hall_ges}
\end{table*}

\subsection{Higher-order singularities}
One can show that the leading physical divergence of $\chi_{ni}$ represents, in general, the $n-1$-th time derivative of a current and that these occur when all but two of the external frequencies are set to zero. They are obtained from the leading term in the Taylor expansions  
\begin{align}
(-i\omega_{\Sigma})^3\chi_{3i} &= \iota_3 + (-i\omega_{\Sigma}) \eta_{3} + (-i\omega_{\Sigma})^2\sigma_{3}+ ... \\
(-i\omega_{\Sigma})^4\chi_{4i} &= \varsigma_4 + (-i\omega_{\Sigma}) \iota_{4} + (-i\omega_{\Sigma})^2\eta_{4}+ ... \\
(-i\omega_{\Sigma})^5\chi_{5i} &= \kappa_5 + (-i\omega_{\Sigma}) \varsigma_{5} +(-i\omega_{\Sigma})^2\iota_5 +\cdots\\
(-i\omega_{\Sigma})^6\chi_{6i} &= \varpi_6 + (-i\omega_{\Sigma}) \kappa_{6} +(-i\omega_{\Sigma})^2\varsigma_6 +\cdots.
\label{eq:snap_pop} \\
\vdots\nn
\end{align}
These higher-order analogs of the injection current are named by analogy with the time derivatives of a particle's velocity, e.g., \textit{jerk}, \textit{snap}, \textit{crackle}, \textit{pop},...,etc. and denote them by, $\iota$, $\varsigma$, $\kappa$,  $\varpi$,.. respectively. Their physical origin is similar to the injection current namely the rate of carrier injection at current carrying states at time-reserved points in the BZ is asymmetric creating a polar distribution. 
   
An alternative formulation is the Laurent series for $\chi_{ni}$ (or $\chi_{n}$ since $\chi_{ne}$ is regular or $\sigma^{(n)}$) as
\begin{align}
\chi_{ni} = \sum_{l=-n}^{\infty} a_l z^{l}
\end{align} 
where $z=-i\omega_{\Sigma}$ and $a_l=0$ for frequencies less than the gap. The residues $a_{-1}=\eta$, $a_{-2}=\sigma$, $a_{-3}=\iota$, etc., are formally given by 
\begin{align}
a_l = \frac{1}{2\pi i}\oint_{|z|=\rho} \frac{\chi_{ni} ~d z}{z^{l+1}},
\label{eq:gen_bpve}
\end{align}
$\rho$ is the radius of convergence of the $1/z$ series. In these calculations the limit $\rho \to 0$ is taken before the limit $\epsilon \to 0$.

In general, if more than two frequencies are distinct~\cite{Aversa1995,Atanasov1996,Driel2001} (but $\omega_{\Sigma}=0$), the series starts from $l>-n$. 

\section{Experimental signatures}
\label{sec:exp_sign}
In real materials, the measured current will be limited by momentum relaxation mechanisms due to collisions with other electrons, phonons, or impurities. For weak disordered insulators we expect the dc divergence of the conductivity in Eq.~\ref{eq:conductivity_3rd_gen} will be cut off by a relaxation time constant as 
\begin{align}
\sigma^{(3)} = \frac{\iota_3}{(\frac{1}{\tau_1} -i\omega_{\Sigma})^2} + \frac{\eta_{3}}{\frac{1}{\tau_1} -i\omega_{\Sigma}}+ \sigma_{3}+\cdots
\label{eq:conductivity_dissipation}
\end{align}
assuming quantum coherence (time scale $\tau_2$) is maintained. Calculation of $\tau_2$ requires a microscopic model of momentum relaxation which will be presented elsewhere.

We have estimated the current of each contribution assuming it can be detected separately. This is a challenge in itself as is well documented in the literature.~\cite{Burger2019} Here we propose to use ultrafast THz spectroscopy together with the symmetry of the crystal, the geometry of the setup, and the polarization of light to isolate these components. In ultrafast experiments, momentum relaxation plays a minor role (at least at short time scales) and the magnitude of the current is determined by the parameters of the lasers. For example the shift current magnitude follows the envelope of the pulse~\cite{Laman1999,Laman2005,Cote2002,Ghalgaoui2018}. Recently, the second-order injection, shift or both currents have been reported via THz radiation~\cite{Laman1999,Laman2005,Cote2002,Bieler2005,Sun2010,Bas2015,Bas2016,Ghalgaoui2018}. In Table~\ref{table:Hall_ges} we present a summary of the jerk, injection and shift Hall-like responses of single-layer of GeS near the band edge. As we can see, either the dependence on polarization, the linearity of the static field, the order of magnitude of the induced current, or the momentum relaxation time scale can be used to distinguish them apart.

\section{Conclusions}
\label{sec:conclusions}
The second-order injection and shift currents are archetypical examples of nontrivial carrier dynamics in insulators and semiconductors.  In this paper we revisited the derivation of the second-order BPVE adding Fermi surface contributions to the theory and proposed a microscopic interpretation of various BPVEs based on the coherent motion of pairs of wave packets in the presence of electric fields.

We also studied the photoconductivity, i.e., a photocurrent second-order in an optical and first order in a static field, from the perspective of the third order electric polarization susceptibility. Three new bulk photovoltaic effects are found. We dub them jerk, third-order injection and third-order shift currents, respectively. The jerk current and third-order injection currents can be thought of as a higher order versions of the standard second-order injection current and have essentially the same microscopic origin, namely, the asymmetric rate of population of current-carrying states at time-reversed points in the BZ. The presence of the electric field, however, gives rise the new contributions due to the anomalous and dipole velocity which are absent in the second-order injection current. 
 
The third-order shift current can be thought as a higher order version of the second-order shift current. It involves the coherent motion of pairs of wave packets across the BZ. We showed that all photocurrents can be understood using semiclassical wave packet dynamics and showed that generalizations to higher order BPVEs are possible. Explicit expressions for the photocurrents amenable for first-principles computations are given. 

\section{acknowledgments}
We thank J.E. Sipe, R.A. Muniz, Y. Lin and C. Aversa for useful discussions. We acknowledge support from DOE-NERSC Contract No. DE-AC02-05CH11231.

\appendix

\section{List of identities}
\label{sec:identities}
Some definitions used in this paper are:

\begin{align}
\v{v}_{nn}(\v{k})& =\langle u_{n}|\v{v}|u_{n} \rangle \equiv \v{v}_n(\v{k}), \label{eq:vel_matrix_elem} \\
f_{n}&\equiv f(\epsilon_{n}(\v{k})), \label{eq:fermi_dis} \\
f_{nm}&\equiv f_{n} - f_{m}, \label{eq:fermi_diff} \\
\pmb{\xi}_{nm}(\v{k}) &\equiv i\left\langle  u_{n}|\pmb{\nabla}_{\v{k}}|u_{m} \right\rangle, \label{eq:berry_def_app} \\
\v{r}_{nm}(\v{k})&\equiv \pmb{\xi}_{nm}(\v{k}),~~(m\neq n)\label{eq:position_matrix} \\
\v{r}_{nn}(\v{k})&\equiv 0 \label{eq:diagonal_position_matrix} \\
\omega_{nm}&\equiv \omega_n - \omega_m. \label{eq:omega_difference}
\end{align}
They describe velocity matrix elements (\ref{eq:vel_matrix_elem}), Fermi distribution (\ref{eq:fermi_dis}), Fermi function differences (\ref{eq:fermi_diff}), Berry connection (\ref{eq:berry_def_app}) , off-diagonal (\ref{eq:position_matrix}) and diagonal dipole matrix elements (\ref{eq:diagonal_position_matrix}), respectively, and frequency band differences (\ref{eq:omega_difference}). $u_{n}$ is the periodic part of the Bloch wave function (spinor index contracted). The covariant derivative of the dipole matrix elements is defined as

\begin{align}
\v{r}_{nm;a}&\equiv \left[\frac{\partial}{\partial k^{a}} - i(\xi^{a}_{nn}-\xi^{a}_{mm})\right]\v{r}_{nm},
\end{align}
or generally of any Bloch matrix element $O_{nm}$ as

\begin{align}
O_{nm;a}\equiv \left[\frac{\partial}{\partial k^a} - i(\xi^a_{nn}-\xi^a_{mm})\right]O_{nm}.
\end{align}
We also defined the commutator and anticommutator with respect to the Cartesian indices $b,c$ as

\begin{align}
[ O(b), K(c)]   \equiv O(b) K(c) - O(c) K(b) \\
\{ O(b), K(c)\} \equiv O(b) K(c) + O(c) K(b) 
\end{align}
where $O,K$ are any Bloch matrix elements.  Some identities used in this paper are:
\begin{align}
\omega_n(-\v{k})&=\omega_n(\v{k})\\
\omega_{n;a}(-\v{k})&=-\omega_{n;a}(\v{k})=-\frac{\partial }{\partial k^a} \omega_{n}(\v{k})\\
\v{v}_{nm}(-\v{k})&=-\v{v}_{mn}(\v{k})=- [\v{v}_{nm}(\v{k})]^{*} \\ 
\v{r}_{nm}(-\v{k})&=\v{r}_{mn}(\v{k})=(\v{r}_{mn}(-\v{k}))^{*}\\
\v{r}_{nm;a}(-\v{k})&=-\v{r}_{mn;a}(\v{k})=-(\v{r}_{nm;a}(\v{k}))^{*}\\
\omega_{nm;a}(\v{k})&= v^a_n(\v{k})-v^a_m(\v{k})=-\omega_{nm;a}(-\v{k})\nn\\
&=\omega_{mn;a}(-\v{k})\\
\pmb{\Omega}_{n}(-\v{k})&=- \pmb{\Omega}_{n}(\v{k})=-(\pmb{\Omega}_{n}(\v{k}))^{*}.
\label{eq:sym_properties}
\end{align}
They arise from the hermicity of operators and the assumptions of time-reversal invariance of the ground state. $\hbar\omega_n$ and $\pmb{\Omega}_{n}$ denote the band energy and Berry curvature of band $n$. 

\section{Derivation of $\eta_2$ and $\sigma_2$ from Taylor expansion of $\chi_2$}
\label{sec:derivation_eta2_sigma2}
To compute $\eta_2^{abc}(0,\omega,-\omega)$ and $\sigma_2^{abc}(0,\omega,-\omega)$ from Eq.~\ref{eq:chi2_taylor_v1}, start from Eq.~\ref{eq:chi2i} and symmetrize  $(-i\omega_{\Sigma})^2\chi_{2i}$ with respect to pair-wise exchanges of electric fields indices $b,\beta$ and $c,\sigma$.\cite{Boyd2008} Then write explicitly the small imaginary part of frequencies, $\omega_{\beta} \to \omega_{\beta} + i\epsilon$,  $\omega_{\sigma} \to \omega_{\sigma} + i\epsilon$ and let $1/(x-i\epsilon)= 1/x + i\pi\delta(x)$. Next, set $\omega_\beta = \omega + n_{\beta}\omega_{\Sigma}, \omega_\sigma = -\omega + n_{\sigma}\omega_{\Sigma}$, $1=n_\beta+n_{\sigma}$, and Taylor expand real parts up to first order in $\omega_{\Sigma}$. It is easy to show that the nonresonant terms cancel and we obtain Eq.~\ref{eq:eta2} and \ref{eq:sigma2} as claimed. In this calculation we used
\begin{align}
(r_{nm}^c r_{mn}^b)_{;a} &= r_{nm;a}^c r_{mn}^b + r_{nm}^c r_{mn;a}^b \nn \\
&= \frac{\partial}{\partial k^a} (r_{nm}^c r_{mn}^b)
\label{eq:cov_der_prod}
\end{align}
and some identities listed in Appendix~\ref{sec:identities}. Note that the expression $r^c_{nm}r^b_{mn}$ is gauge invariant and hence  the covariant derivative reduces to the standard crystal momentum derivative.

\section{Expansion of $\chi_{3i}$}
\label{app:chi_3}

\begin{figure*}
\includegraphics[width=.66\textwidth]{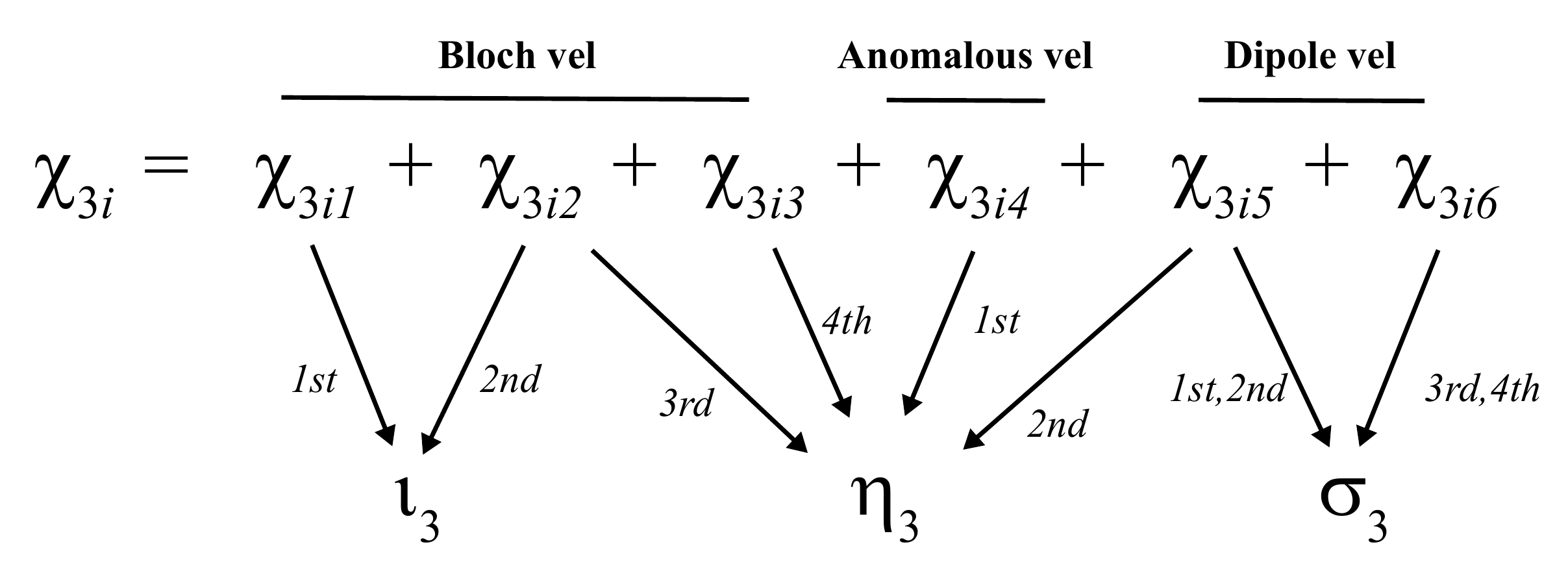}
\caption{Origin of the first, second,..., contributions to the expressions for $\iota_3$-jerk (Eq.~\ref{eq:jerk}), $\eta_3$-injection (Eq.~\ref{eq:eta3}) and $\sigma_3$-shift (Eq.~\ref{eq:sigma3}) response tensors. Each of the 6 terms in the $\chi_{3i}$ originates from either the Bloch velocity (first three), anomalous velocity (fourth) and the dipole velocity (5th and 6th). Due to the structure of the poles in $\chi_{3i}$, the Bloch velocity and dipole velocity contribute to multiple response functions.}
\label{fig:chi3i_conceptual}  
\end{figure*}

 Using Eqs.~\ref{eq:pe_and_pi1}, \ref{eq:J_intra}, \ref{eq:2nd_pt}, and \ref{eq:nth_pt} the third order susceptibility  $\chi_3^{abcd}(-\omega_{\Sigma},\omega_{\beta},\omega_{\sigma},\omega_{\Delta})$ can be written as $\chi_3 = \chi_{3e}+\chi_{3i}$ where  
\begin{widetext}
\begin{align}
\frac{\chi_{3e}}{C_3}= -\sum_{nm\v{k}} &\frac{r_{nm}^a}{\omega_{mn}-\omega_{\Sigma}}\left[\frac{1}{(\omega_{mn}-\omega_2)}\left( \frac{r^b_{mn} f_{nm}}{\omega_{mn}-\omega_{\beta}}\right)_{;c} \right]_{;d}\hspace{-10pt}
-i\sum_{nmp\v{k}} \frac{r_{nm}^a}{\omega_{mn}-\omega_{\Sigma}}\left[\frac{1}{\omega_{mn}-\omega_2} \left(\frac{r^{b}_{mp}r_{pn}^{c}f_{pm}}{\omega_{mp}-\omega_\beta} -\frac{r^{c}_{pm}r_{pn}^{b}f_{np}}{\omega_{pn}-\omega_\beta} \right) \right]_{;d}& \nn \\ 
-i\sum_{nmp\v{k}}& \frac{r_{nm}^a}{\omega_{mn}-\omega_{\Sigma}} \left[ \left(\frac{r^{b}_{mp}f_{pm}}{\omega_{mp}-\omega_\beta} \right)_{;c}\frac{r^d_{pn}}{\omega_{mp}-\omega_2} - \frac{r^b_{mp}}{\omega_{pn}-\omega_2}\left(\frac{r^{b}_{pn}f_{np}}{\omega_{pn}-\omega_\beta} \right)_{;c}  \right]&\nn\\
- \sum_{nmpl\v{k}}& \frac{r^a_{nm}}{\omega_{mn}-\omega_{\Sigma}}\left[\frac{r^d_{mp}}{\omega_{pn}-\omega_2}\left( \frac{r^b_{pl}r^c_{lp}f_{lp}}{\omega_{pl}-\omega_{\beta}} - \frac{r^c_{pl}r^b_{ln}f_{nl}}{\omega_{lp}-\omega_{\beta}}\right) - 
\left( \frac{r^b_{ml}r^c_{lp}f_{lm}}{\omega_{ml}-\omega_{\beta}} - \frac{r^c_{ml}r^b_{lp}f_{pl}}{\omega_{lp}-\omega_{\beta}}\right)\frac{r^d_{pn}}{\omega_{mp}-\omega_2} \right] 
\end{align}
\end{widetext}
\begin{widetext}
\begin{align}
\frac{\chi_{3i}}{C_3}&\equiv \sum_{r=1}^{6}\chi_{3i r} \nn\\
&=\frac{1}{\omega_2 \omega^2_{\Sigma}}\sum_{nm\v{k}} \omega_{nm;a}f_{mn} \left(\frac{r^b_{nm}r^c_{mn} }{\omega_{nm}-\omega_{\beta}} \right)_{;d} \nn\\
&-\frac{1}{\omega^2_{\Sigma}}\sum_{nm\v{k}} \frac{\omega_{nm;a}r^d_{mn}}{\omega_{nm}-\omega_2}\left(\frac{r^b_{nm}f_{mn}}{\omega_{nm}-\omega_{\beta}} \right)_{;c} \hspace{-7pt}-\frac{i}{\omega^2_{\Sigma}}\sum_{nml\v{k}} \frac{\omega_{nm;a}r^d_{mn}}{\omega_{nm}-\omega_2} 
\left(\frac{r^b_{nl}r^c_{lm} f_{ln}}{\omega_{nl}-\omega_{\beta}} - \frac{r^c_{nl}r^b_{lm} f_{ml}}{\omega_{lm}-\omega_{\beta}} \right) \nn\\
&-\frac{i}{\omega_2 \omega_{\Sigma}} \sum_{nm\v{k}} \Omega^{ad}_{nm}\frac{r^b_{nm} r^c_{mn} f_{mn}}{\omega_{nm}-\omega_{\beta}} \nn \\
&+\frac{1}{\omega_{\Sigma}} \sum_{nm\v{k}} \frac{r^d_{mn;a}}{\omega_{nm}-\omega_2}\left(\frac{r^b_{nm}f_{mn}}{\omega_{nm}-\omega_{\beta}} \right)_{;c}
+\frac{i}{\omega_{\Sigma}} \sum_{nml\v{k}} \frac{r^d_{mn;a}}{\omega_{nm}-\omega_2}\left(\frac{r^b_{nl} r^c_{lm} f_{ln}}{\omega_{nl}-\omega_{\beta}} - \frac{r^c_{nl}r^b_{lm}f_{ml}}{\omega_{lm}-\omega_{\beta}}  \right).
\label{eq:chi3i}
\end{align}
\end{widetext}
We defined $C_3 \equiv e^4/\hbar^3V$, $\Omega^{ad}_{nm}\equiv \Omega^{ad}_{n} -\Omega^{ad}_{m}$, $\omega_{\Sigma} \equiv \omega_{\beta}+\omega_{\sigma}+\omega_{\Delta}$ and $\omega_{2} \equiv \omega_{\beta}+\omega_{\sigma}$. These expressions still need to be symmetrized with respect to pair-wise exchange of electric field indices $(b,\beta)$, $(c,\sigma)$, $(d,\Delta)$. We note that, it is easier to calculate $\chi_{3i}$  from the intraband current $\v{J}^{a(3)}_i$ rather than from $\v{P}^{(3)}_i$. 

Eq.~\ref{eq:chi3i} has a distinguishable structure, see Fig~\ref{fig:chi3i_conceptual}. The first three terms in $\chi_{3i}$ are derived from the combination $v^{a}_{n}\rho^{(3)}_{nn}$ (Eq.~\ref{eq:J_intra}). By analogy with $\chi_{2i}$ (Eq.~\ref{eq:chi2i}), we would expect these terms to be injection current-type of contributions with one caveat; the first term has no analog in $\chi_{2i}$  since it is proportional to \textit{three} powers of frequency, $\omega^{-2}_{\Sigma}\omega^{-1}_2$ and is the most divergent at zero frequency. The second and third terms, proportional to $\omega_{\Sigma}^{-2}$, seem standard injection coefficients similar to the first term in $\chi_{2i}$. 

The fourth term is proportional to $(\omega_{\Sigma}\omega_2)^{-1}$ and arises from the anomalous velocity $(\v{E}\times\pmb{\Omega})^a\rho^{(2)}_{nn}$. It is an injection current-type of coefficient. The fifth and sixth terms, proportional to $\omega^{-1}_{\Sigma}$, originate from $\v{E}\cdot \v{r}_{nm;a}\rho^{(2)}_{mn}$ and hence are expected to be shift current-type of contributions. 

The goal in the next three sections (D, E and F) is to calculate the coefficients $\iota_3,\eta_3,\sigma_3 $ in the expansion

\begin{align}
(-i\omega_{\Sigma})^3 \chi_{3i} = \iota_3 + (-i\omega_{\Sigma}) \eta_3 + (-i\omega_{\Sigma})^{2} \sigma_3 +\cdots 
\end{align}
To avoid cumbersome notation, we write the susceptibilities with the additional factors as 

\begin{align}
\frac{(-i\omega_{\Sigma})^3 \chi_{3i}^{abcd}}{C_3} \to  \chi_{3i}.
\end{align}
The strategy is to parametrize (the real part of) the external frequencies as

\begin{align}
\omega_{\beta} &= \omega + n_{\beta}\omega_{\Sigma}, \nn\\
\omega_{\sigma} &= -\omega + n_{\sigma}\omega_{\Sigma}, \nn\\
\omega_{\Delta} &= 0,
\label{eq:param_freq}
\end{align}
subject to $n_{\beta}+ n_{\sigma}=1$. Fig.~\ref{fig:chi3i_conceptual} summarizes the result.

\section{Derivation of $\iota_3$}
\label{app:iota3}

$\iota_3$ derives from $\chi_{3i1}$ and  $\chi_{3i2}$.

\subsection{First term of $\iota_3$}
Integrate by parts $\chi_{3i1}$ and symmetrize it with respect to pair-wise exchange of electric field indices $(b,\beta)$, $(c,\sigma)$, $(d,\Delta)$ to obtain
\begin{align}
\chi_{3i,1}&\equiv\sum_{l=1}^{3} \chi_{3i,1,l}\nn\\
&= -\frac{i\omega_{\Sigma}}{6}\sum_{nm\v{k}} \frac{\omega_{nm;ad} f_{mn} r_{nm}^b r_{mn}^c}{(\omega_{nm}-\omega_{\beta})(\omega_{nm}+\omega_{\sigma})}\nn\\
&-\frac{i\omega_{\Sigma}}{6}\sum_{nm\v{k}} \frac{\omega_{nm;ac} f_{mn} r_{nm}^b r_{mn}^d}{(\omega_{nm}-\omega_{\beta})(\omega_{nm}+\omega_{\Delta})}\nn\\
&-\frac{i\omega_{\Sigma}}{6}\sum_{nm\v{k}} \frac{\omega_{nm;ab} f_{mn} r_{nm}^d r_{mn}^c}{(\omega_{nm}-\omega_{\Delta})(\omega_{nm}+\omega_{\sigma})}.
\label{eq:chi3i1}
\end{align} 
The second and third terms will cancel against other terms as we show later, but the first term will contribute to $\iota_3$.  By partial fractions and writing  explicitly the imaginary parts of the frequencies, the first term gives
\begin{align}
\chi_{3i,1,1}&=\frac{-i\omega_{\Sigma}}{6(\omega_{\beta}+\omega_{\sigma})}\sum_{nm\v{k}} \frac{\omega_{nm;ad} f_{mn} r_{nm}^b r_{mn}^c}{(\omega_{nm}-\omega_{\beta}-i\epsilon)}\nn\\
&-\frac{i\omega_{\Sigma}}{6(\omega_{\beta}+\omega_{\sigma})}\sum_{nm\v{k}} \frac{\omega_{nm;ad} f_{mn} r_{nm}^c r_{mn}^b}{(\omega_{nm}-\omega_{\sigma}-i\epsilon)}.
\end{align}
Using Eq.~\ref{eq:param_freq}, $1/(x-i\epsilon)= 1/x + i\pi\delta(x)$, and expanding in powers of $\omega_{\Sigma}$ to first order we obtain 
\begin{align}
\chi_{3i,1,1}&=\frac{2\pi}{6}\sum_{nm\v{k}} \omega_{nm;ad} f_{mn} r^b_{nm}r^c_{mn}\delta(\omega_{nm}-\omega)\nn\\
&-\frac{i\omega_{\Sigma}}{6}\sum_{nm\v{k}} \omega_{nm;ad} f_{mn} r^b_{nm}r^c_{mn}\frac{\partial}{\partial \omega}\left(\frac{1}{\omega_{nm}-\omega}  \right)
\label{eq:l11}
\end{align}
The first term is independent of $\omega_{\Sigma}$ and vanishes for frequencies smaller than the energy band gap. This is the first term of $\iota_3$ in Eq.~\ref{eq:jerk}. The second nonresonant term will cancel against other terms.

\subsection{Second term of $\iota_3$}
This contribution is obtained from $\chi_{3i2}$. To see this, let us symmetrize the second term in Eq.~\ref{eq:chi3i}. After two integration by parts we obtain 
\begin{widetext}
\begin{align}
\chi_{3i2} \equiv\sum_{l=1}^{8} \chi_{3i2,l}&=\frac{i\omega_{\Sigma}}{6}\sum_{nm\v{k}}\frac{\omega_{nm;ac} r_{mn}^d r_{nm}^b f_{mn}}{(\omega_{nm}-\omega_{\beta}-\omega_{\sigma}) (\omega_{nm}-\omega_{\beta})} 
+\frac{i\omega_{\Sigma}}{6}\sum_{nm\v{k}} \frac{\omega_{nm;a} r_{nm}^b f_{mn}}{\omega_{nm}-\omega_{\beta}} \left(\frac{r_{mn}^d}{\omega_{nm}-\omega_{\beta}-\omega_{\sigma}} \right)_{;c} \nn \\
&+ \frac{i\omega_{\Sigma}}{6}\sum_{nm\v{k}}\frac{\omega_{nm;ab} r_{mn}^d r_{nm}^c f_{mn}}{(\omega_{nm}-\omega_{\beta}-\omega_{\sigma}) (\omega_{nm}-\omega_{\sigma})}
+\frac{i\omega_{\Sigma}}{6}\sum_{nm\v{k}} \frac{\omega_{nm;a} r_{nm}^c f_{mn}}{\omega_{nm}-\omega_{\sigma}} \left(\frac{r_{mn}^d}{\omega_{nm}-\omega_{\beta}-\omega_{\sigma}} \right)_{;b} \nn \\
&-\frac{i\omega_{\Sigma}}{6}\sum_{nm\v{k}} \frac{\omega_{nm;a} r_{mn}^c f_{mn}}{\omega_{nm}-\omega_{\beta}-\omega_{\Delta}} \left(\frac{r_{nm}^b}{\omega_{nm}-\omega_{\beta}} \right)_{;d} 
-\frac{i\omega_{\Sigma}}{6}\sum_{nm\v{k}} \frac{\omega_{nm;a} r_{mn}^b f_{mn}}{\omega_{nm}-\omega_{\Delta}-\omega_{\sigma}} \left(\frac{r_{nm}^d}{\omega_{nm}-\omega_{\Delta}} \right)_{;c} \nn\\
&-\frac{i\omega_{\Sigma}}{6}\sum_{nm\v{k}} \frac{\omega_{nm;a} r_{mn}^b f_{mn}}{\omega_{nm}-\omega_{\sigma}-\omega_{\Delta}} \left(\frac{r_{nm}^c}{\omega_{nm}-\omega_{\sigma}} \right)_{;d}
-\frac{i\omega_{\Sigma}}{6}\sum_{nm\v{k}} \frac{\omega_{nm;a} r_{mn}^c f_{mn}}{\omega_{nm}-\omega_{\Delta}-\omega_{\beta}} \left(\frac{r_{nm}^d}{\omega_{nm}-\omega_{\Delta}} \right)_{;b}.
\label{eq:chi3i2}
\end{align}
\end{widetext}
There are eight terms. To $\mathcal{O}(\omega_{\Sigma})$, the  $l=1,3$ terms cancel with identical second and third terms in Eq.~\ref{eq:chi3i1}. The terms $l=2,6$ and $l=4,8$ combine to give the third term of $\eta_3$ in Eq.~\ref{eq:eta3} (see next section). The $l=5,7$ terms contribute to $\iota_3$. 

Note that we can set $\omega_{\beta}+\omega_{\sigma}=0$ (or $\omega_{\Delta}=0$) (where this combination appears) since the pairs of poles in these expressions are distinct. This is not true in $l=5,7$ and we consider them separately. After differentiation the $l=5$ term we obtain 
\begin{align}
\chi_{3i2,5}&= -\frac{i\omega_{\Sigma}}{6}\sum_{nm\v{k}} \frac{\omega_{nm;a} r_{mn}^c r_{nm;d}^b f_{mn}}{(\omega_{nm}-\omega_2)(\omega_{nm}-\omega_{\beta})} \nn\\
&~~~+\frac{i\omega_{\Sigma}}{6}\sum_{nm\v{k}} \frac{\omega_{nm;a} r_{mn}^c r_{nm}^b f_{mn} \omega_{nm;d}}{(\omega_{nm}-\omega_2)(\omega_{nm}-\omega_{\beta})^2} 
\end{align}
here $\omega_2=\omega_{\beta}+\omega_{\Delta}$ and we used
\begin{align}
\left(\frac{r_{nm}^d}{\omega_{nm}-\omega_{\Delta}} \right)_{;c}& = \frac{r_{nm;c}^d}{\omega_{nm}-\omega_{\Delta}}
- \frac{r_{nm}^d\omega_{nm;c}}{(\omega_{nm}-\omega_{\Delta})^2}.
\label{eq:cov_der_quot}
\end{align}
Now obtain simple poles via partial fractions. The term with a square of frequencies in denominator can be handled by
\begin{align}
\frac{\omega_{nm;d}}{(\omega_{nm}-\omega_{\Delta})^2} = -\frac{\partial}{\partial k^d}(\omega_{nm}-\omega_{\beta})^{-1},
\end{align}
and a partial integration. Next, write the imaginary part of frequencies, use $1/(x-i\epsilon)= 1/x + i\pi\delta(x)$, and set $\omega_\beta = \omega + n_{\beta}\omega_{\Sigma}, \omega_\sigma = -\omega + n_{\sigma}\omega_{\Sigma}$, and $1=n_\beta+n_{\sigma}$. Note that with these definitions $\omega_2=\omega + (1+n_\beta)\omega_{\Sigma}$. Now expand to \textit{second} order in $\omega_{\Sigma}$ and set (without expanding) $\omega_{\Delta}=\omega_{\Sigma}$. After some algebra we obtain
\begin{align}
\chi_{3i2,5}=& \frac{i\omega_{\Sigma}}{6}\sum_{nm\v{k}} r_{mn}^c r_{nm;d}^b f_{mn} \frac{\partial}{\partial k^a}\left(\frac{1}{\omega_{nm}-\omega}\right) \nn \\
+\frac{i\omega_{\Sigma}}{12}&\sum_{nm\v{k}} \frac{\partial}{\partial k^d}( \omega_{nm;a} r_{mn}^c r_{nm}^b) f_{mn} \frac{\partial}{\partial \omega} \left( \frac{1}{\omega_{nm}-\omega}\right) \nn\\
+\frac{\pi}{6}&\sum_{nm\v{k}} \frac{\partial}{\partial k^d}( \omega_{nm;a} r_{mn}^c r_{nm}^b) f_{mn}\delta(\omega_{nm}-\omega)
\label{eq:chi_3i2_5}
\end{align}
In this calculation we have used the identity
\begin{align}
\frac{\partial}{\partial \omega}\left(\frac{1}{\omega_{nm}-\omega} \right) = -\frac{\partial}{\partial \omega_{nm}}\left(\frac{1}{\omega_{nm}-\omega} \right).
\end{align}
Note that the third term in (\ref{eq:chi_3i2_5}) contributes to $\iota_3$. The other two nonresonant terms will eventually cancel.
A similar calculation for the $l=7$ terms gives

\begin{align}
\chi_{3i2,7}=& \frac{i\omega_{\Sigma}}{6}\sum_{nm\v{k}} r_{mn}^b r_{nm;d}^c f_{mn} \frac{\partial}{\partial k^a}\left(\frac{1}{\omega_{nm}+\omega}\right) \nn \\
-\frac{i\omega_{\Sigma}}{12}&\sum_{nm\v{k}} \frac{\partial}{\partial k^d}( \omega_{nm;a} r_{mn}^b r_{nm}^c) f_{mn} \frac{\partial}{\partial \omega} \left( \frac{1}{\omega_{nm}+\omega}\right) \nn\\
+\frac{\pi}{6}&\sum_{nm\v{k}} \frac{\partial}{\partial k^d}( \omega_{nm;a} r_{mn}^b r_{nm}^c) f_{mn}\delta(\omega_{nm}+\omega)
\end{align}
Combining the $l=5$ and $l=7$ terms above and using ~\ref{eq:cov_der_prod} we obtain 
\begin{align}
\chi_{3i2,5}+&\chi_{3i2,7}=\nn\\
 \frac{i\omega_{\Sigma}}{6}&\sum_{nm\v{k}}  \omega_{nm;ad} r_{mn}^c r_{nm}^b f_{mn} \frac{\partial}{\partial \omega} \left( \frac{1}{\omega_{nm}-\omega}\right) \nn\\
+\frac{2\pi}{6}&\sum_{nm\v{k}} \frac{\partial}{\partial k^d}( \omega_{nm;a} r_{mn}^c r_{nm}^b) f_{mn}\delta(\omega_{nm}-\omega)
\end{align}
The first term is nonresonant and will cancel against the second term in Eq.~\ref{eq:l11}. The second term combined with the first term in Eq.~\ref{eq:l11} gives $\iota_3$ in Eq.~\ref{eq:jerk}. 

\section{Derivation of $\eta_3$}
\label{app:details_eta3}
We now derive each of the contributions to $\eta_3$ in Eq.~\ref{eq:eta3}.

\subsection{First term of $\eta_3$}
The first term in $\eta_3$ comes from $\chi_{3i4}$. Symmetrizing $\chi_{3i4}$ in Eq.~\ref{eq:chi3i2} and after partial fractions we obtain
\begin{align}
\chi_{3i,4}&\equiv \sum_{l=1}^{3} \chi_{3i,4,l}\\
&=\frac{\omega_{\Sigma}^2}{6(\omega_{\beta}+\omega_{\sigma})}\sum_{nm\v{k}} \Omega_{nm}^{ad} f_{mn} r_{nm}^b r_{mn}^c\bigg[ \frac{1}{\omega_{nm}-\omega_{\beta}}\nn\\
&~~~~~~~~~~~~~~~~~~~~~~~~~~~~~~~ - \frac{1}{\omega_{nm}+\omega_{\sigma}}\bigg] \nn\\
&+\frac{\omega_{\Sigma}^2}{6}\sum_{nm\v{k}} \frac{\Omega_{nm}^{ac} f_{mn} r_{nm}^b r_{mn}^d}{(\omega_{nm}-\omega_{\beta})(\omega_{nm}+\omega_{\Delta})}\nn\\
&+\frac{\omega_{\Sigma}^2}{6}\sum_{nm\v{k}} \frac{\Omega_{nm}^{ab} f_{mn} r_{nm}^d r_{mn}^c}{(\omega_{nm}-\omega_{\Delta})(\omega_{nm}+\omega_{\sigma})}
\label{eq:chi3i4}
\end{align} 
Only the first term contributes to $\eta_3$. Writing the imaginary parts of the frequencies, setting $\omega_\beta = \omega + n_{\beta}\omega_{\Sigma}$, $\omega_\sigma = -\omega + n_{\sigma}\omega_{\Sigma}$, and Taylor expanding, we obtain to leading order in $\omega_{\Sigma}$
\begin{align}
\chi_{3i,4,1}&= \frac{2i\pi \omega_{\Sigma}}{6}\sum_{nm\v{k}} \Omega_{nm}^{ad} f_{mn} r_{nm}^b r_{mn}^c \delta(\omega_{nm}-\omega)\nn\\
&+\frac{\omega_{\Sigma}^2}{6}\sum_{nm\v{k}} \Omega_{nm}^{ad} f_{mn} r_{nm}^b r_{mn}^c \frac{\partial}{\partial \omega}\bigg(\frac{1}{\omega_{nm}-\omega} \bigg)
\end{align}
Adding 1/2 of the first term to 1/2 of itself and letting $\v{k}\to -\v{k}$ in the second term we obtain the first contribution of $\eta_3$ in Eq.~\ref{eq:eta3}. The second term cancels against other nonresonant contributions.

\subsection{Second term of $\eta_3$}
This term arises from $\chi_{3i5}$. Symmetrizing we obtain 

\begin{widetext}
\begin{align}
\chi_{3i5} \equiv\sum_{l=1}^{6} \chi_{3i5,l}&=\frac{i\omega_{\Sigma}^2}{6}\sum_{nm\v{k}}\frac{ r_{mn;a}^d f_{mn}}{\omega_{nm}-\omega_{\beta}-\omega_{\sigma}}\left(\frac{r_{nm}^b}{\omega_{nm}-\omega_{\beta}}\right)_{;c}
%
%
+ \frac{i\omega_{\Sigma}^2}{6}\sum_{nm\v{k}}\frac{r_{mn;a}^d f_{mn}}{\omega_{nm}-\omega_{\beta}-\omega_{\sigma}}\left(\frac{r_{nm}^c }{\omega_{nm}-\omega_{\sigma}}\right)_{;b}\nn\\
%
%
&+ \frac{i\omega_{\Sigma}^2}{6}\sum_{nm\v{k}}\frac{r_{mn;a}^c f_{mn}}{\omega_{nm}-\omega_{\beta}-\omega_{\Delta}}\left(\frac{r_{nm}^b }{\omega_{nm}-\omega_{\beta}}\right)_{;d}
+ \frac{i\omega_{\Sigma}^2}{6}\sum_{nm\v{k}}\frac{r_{mn;a}^b f_{mn}}{\omega_{nm}-\omega_{\Delta}-\omega_{\sigma}}\left(\frac{r_{nm}^d }{\omega_{nm}-\omega_{\Delta}}\right)_{;c}\nn\\
&+ \frac{i\omega_{\Sigma}^2}{6}\sum_{nm\v{k}}\frac{r_{mn;a}^b f_{mn}}{\omega_{nm}-\omega_{\sigma}-\omega_{\Delta}}\left(\frac{r_{nm}^c }{\omega_{nm}-\omega_{\sigma}}\right)_{;d}
+ \frac{i\omega_{\Sigma}^2}{6}\sum_{nm\v{k}}\frac{r_{mn;a}^c f_{mn}}{\omega_{nm}-\omega_{\Delta}-\omega_{\delta}}\left(\frac{r_{nm}^d }{\omega_{nm}-\omega_{\Delta}}\right)_{;b}.
\label{eq:chi3i5n}
\end{align}
\end{widetext}
Let us consider $\chi_{3i5,3}$ first

\begin{align}
\chi_{3i5,3}= \frac{i\omega_{\Sigma}^2}{6}\sum_{nm\v{k}}\frac{r_{mn;a}^c f_{mn}}{\omega_{nm}-\omega_2}\left(\frac{r_{nm}^b }{\omega_{nm}-\omega_{\beta}}\right)_{;d},
\end{align}
where $\omega_2=\omega_{\beta}+\omega_{\Delta}$. Performing a partial fraction expansion, a substitution $1/(x-i\epsilon)=1/x + i\pi\delta(x)$, followed by a Taylor expansion (to second order) in  $\omega_{\Sigma}$ of the real part about ($\omega_{\beta},\omega_{\sigma})=(\omega,-\omega)$ using $\omega_\beta = \omega + n_{\beta}\omega_{\Sigma}, \omega_\sigma = -\omega + n_{\sigma}\omega_{\Sigma}$ such that $\omega_2=\omega + (1+n_{\beta})\omega_{\Sigma}$, we obtain

\begin{align}
\chi_{3i5,3}&= \frac{i\omega_{\Sigma}^2}{6}\sum_{nm\v{k}} r_{mn;a}^c r^{b}_{nm;d} f_{mn} \frac{\partial}{\partial \omega}\left( \frac{1 }{\omega_{nm}-\omega}\right)\nn\\
&- \frac{i\omega_{\Sigma}^2}{12}\sum_{nm\v{k}} (r_{mn;a}^c r^{b}_{nm})_{;d} f_{mn} \frac{\partial}{\partial \omega}\left( \frac{1 }{\omega_{nm}-\omega}\right)\nn\\
&+ \frac{i\omega_{\Sigma} (i\pi)}{6}\sum_{nm\v{k}} (r_{mn;a}^c r^{b}_{nm})_{;d} f_{mn} \delta(\omega_{nm}-\omega) 
\label{eq:chi3i5_3}
\end{align}
the first two terms are nonresonant contributions which cancel against other terms. A similar analysis of $\chi_{3i5,5}$ gives

\begin{align}
\chi_{3i5,5}&= -\frac{i\omega_{\Sigma}^2}{6}\sum_{nm\v{k}} r_{mn;a}^b r^{c}_{nm;d} f_{mn} \frac{\partial}{\partial \omega}\left( \frac{1 }{\omega_{nm}+\omega}\right)\nn\\
&+ \frac{i\omega_{\Sigma}^2}{12}\sum_{nm\v{k}} (r_{mn;a}^b r^{c}_{nm})_{;d} f_{mn} \frac{\partial}{\partial \omega}\left( \frac{1}{\omega_{nm}+\omega}\right)\nn\\
&+ \frac{i\omega_{\Sigma} (i\pi)}{6}\sum_{nm\v{k}} (r_{mn;a}^b r^{c}_{nm})_{;d} f_{mn} \delta(\omega_{nm}+\omega) 
\label{eq:chi3i5_5}
\end{align}
the first two terms are nonresonant contributions which cancel against other terms. After changing indices $n,m$ and $\v{k}\to -\v{k}$ we see that the third term in Eq.~\ref{eq:chi3i5_3} plus the third term in Eq.~\ref{eq:chi3i5_5} gives the second term of $\eta_3$ in Eq.~\ref{eq:eta3}. 

\subsection{Third term of $\eta_3$}
The third contribution to Eq.~\ref{eq:eta3} arises from $\chi_{3i2,2}+\chi_{3i2,6} + \chi_{3i2,4}+\chi_{3i2,8}$ in Eq.~\ref{eq:chi3i2}. Note that we can set $\omega_{\beta}+\omega_{\sigma}=0$ from the outset since the poles in these expressions are distinct. Setting $1/(x-i\epsilon)=1/x + i\pi\delta(x)$ and Taylor expanding about $(\omega_{\beta},\omega_{\sigma})=(\omega,-\omega)$ we see that to leading order the nonresonant parts vanish and we obtain  
\begin{align}
&\chi_{3i2,2}+\chi_{3i2,6} =\nn\\
 &-\frac{\omega_{\Sigma} \pi}{3}\sum_{bm\v{k}} \omega_{nm;a} \left( \frac{r^{d}_{mn}}{\omega_{nm}}\right)_{;c} r^{b}_{nm} f_{mn} \delta(\omega_{nm}-\omega).
\label{eq:3rd_eta_1}
\end{align}
Similar manipulations lead to vanishing nonresonant terms and to
 
\begin{align}
&\chi_{3i2,4}+\chi_{3i2,8} =\nn\\
&-\frac{\omega_{\Sigma} \pi}{3}\sum_{bm\v{k}} \omega_{nm;a} \left( \frac{r^{d}_{mn}}{\omega_{nm}}\right)_{;b} r^{c}_{nm} f_{mn} \delta(\omega_{nm}+\omega).
\end{align}
Relabeling of indices $n,m$, setting $\v{k}\to -\v{k}$, and adding to Eq.~\ref{eq:3rd_eta_1} we recover the third term of $\eta_3$.

\subsection{Fourth term of $\eta_3$}
The fourth term arises from $\chi_{3i3}$. Let us label the 12 terms obtained after symmetrization of $\chi_{3i3}$ as

\begin{widetext}
\begin{align}
\chi_{3i3}&\equiv \sum_{l}^{12} \chi_{3i3,l}\nn\\
&=\frac{\omega_{\Sigma}}{6}\sum_{nmo\v{k}} \frac{\omega_{nm;a} r^{d}_{mn}}{\omega_{nm}-\omega_{\beta}-\omega_{\sigma}}\left[\frac{r^{b}_{no} r^{c}_{om} f_{on}}{\omega_{no}-\omega_{\beta}} - \frac{r^{c}_{no} r^{b}_{om} f_{mo}}{\omega_{om}-\omega_{\beta}} \right]
+ \frac{\omega_{\Sigma}}{6}\sum_{nmo\v{k}} \frac{\omega_{nm;a} r^{d}_{mn}}{\omega_{nm}-\omega_{\sigma}-\omega_{\beta}}\left[\frac{r^{c}_{no} r^{b}_{om} f_{on}}{\omega_{no}-\omega_{\sigma}} - \frac{r^{b}_{no} r^{c}_{om} f_{mo}}{\omega_{om}-\omega_{\sigma}} \right]\nn\\
&+\frac{\omega_{\Sigma}}{6}\sum_{nmo\v{k}} \frac{\omega_{nm;a} r^{c}_{mn}}{\omega_{nm}-\omega_{\beta}-\omega_{\Delta}}\left[\frac{r^{b}_{no} r^{d}_{om} f_{on}}{\omega_{no}-\omega_{\beta}} - \frac{r^{d}_{no} r^{b}_{om} f_{mo}}{\omega_{om}-\omega_{\beta}} \right]
+ \frac{\omega_{\Sigma}}{6}\sum_{nmo\v{k}} \frac{\omega_{nm;a} r^{b}_{mn}}{\omega_{nm}-\omega_{\Delta}-\omega_{\sigma}}\left[\frac{r^{d}_{no} r^{c}_{om} f_{on}}{\omega_{no}-\omega_{\Delta}} - \frac{r^{c}_{no} r^{d}_{om} f_{mo}}{\omega_{om}-\omega_{\Delta}} \right]\nn\\
&+\frac{\omega_{\Sigma}}{6}\sum_{nmo\v{k}} \frac{\omega_{nm;a} r^{b}_{mn}}{\omega_{nm}-\omega_{\sigma}-\omega_{\Delta}}\left[\frac{r^{c}_{no} r^{d}_{om} f_{on}}{\omega_{no}-\omega_{\sigma}} - \frac{r^{d}_{no} r^{c}_{om} f_{mo}}{\omega_{om}-\omega_{\sigma}} \right]
+ \frac{\omega_{\Sigma}}{6}\sum_{nmo\v{k}} \frac{\omega_{nm;a} r^{c}_{mn}}{\omega_{nm}-\omega_{\Delta}-\omega_{\beta}}\left[\frac{r^{d}_{no} r^{b}_{om} f_{on}}{\omega_{no}-\omega_{\Delta}} - \frac{r^{b}_{no} r^{d}_{om} f_{mo}}{\omega_{om}-\omega_{\Delta}} \right].
\end{align}
\end{widetext}
We analyze the structure of $\chi_{3i3}$ by dividing its terms into two groups. The first group composed of the $l=1,2,3,4$ terms can be added together to give a simple result (see Eq.~\ref{eq:eta2_4rd_12}). The second group is composed of the  $l=5\text{-}12$ terms. The $l=5,6,9,10$ terms have pairs of poles separable by partial fractions and can be combined with the $l=12,11,8,7$ terms (respectively). Since we are interested in results to linear in $\omega_{\Sigma}$, it is useful to note we can set $\omega_{\beta}+\omega_{\sigma}=0$ or $\omega_{\Delta}=0$ in all terms from the outset. This is because the poles in each term are always distinct and separable by simple partial fractions. This should be contrasted with the $l=5,7$ terms of Eq.~\ref{eq:chi3i2}, or the l=3,5 terms in Eq.~\ref{eq:chi3i5n}, where the poles collide and they have to be treated separately.

The sum of the $l=1,2$ terms can be written as 

\begin{align}
&\chi_{3i3,1}+\chi_{3i3,2} =\nn\\
&=\frac{\omega_{\Sigma}}{6} \sum_{nmo\v{k}} \frac{\omega_{nm;a} r^{d}_{mn} r^{b}_{no} r^{c}_{om} f_{on}}{\omega_{nm}} F_{+}(\omega_{no},\omega_{\beta}),
\end{align}
where $F_{+}$ is defined as
\begin{align}
F_{+}(\omega_{no},\omega_{\beta}) &\equiv \frac{1}{\omega_{no}-\omega_{\beta}-i\epsilon} + \frac{1}{\omega_{no}+\omega_{\beta}+i\epsilon}\nn\\
&=H_{+}(\omega_{no},\omega_{\beta}) + i\pi D_{-}(\omega_{no},\omega_{\beta}),
\end{align}
and

\begin{align}
H_{\pm}(\omega_{no},\omega_{\beta})&\equiv \frac{1}{\omega_{no}-\omega_{\beta}} \pm \frac{1}{\omega_{no}+\omega_{\beta}}\nn\\
D_{\pm}(\omega_{no},\omega_{\beta})&\equiv \delta(\omega_{no}-\omega_{\beta}) \pm\delta(\omega_{no}+\omega_{\beta}).   
\end{align}
Similar manipulations for the sum of the $l=3,4$ terms leads to 

\begin{align}
&\chi_{3i3,3}+\chi_{3i3,4} =\nn\\
&=\frac{\omega_{\Sigma}}{6} \sum_{nmo\v{k}} \frac{\omega_{nm;a} r^{d}_{mn} r^{c}_{no} r^{b}_{om} f_{on}}{\omega_{nm}} F_{+}(\omega_{no},\omega_{\sigma}).
\end{align}
Adding the $l=1\text{-}4$ contributions we find

\begin{align}
&\sum_{l}^4 \chi_{3i3,l}=\nn\\
&=\frac{\omega_{\Sigma}}{6} \sum_{nmo\v{k}} \omega_{nm;a} \frac{ r^{d}_{mn}}{\omega_{nm}}( r^{b}_{no} r^{c}_{om} + r^{c}_{no} r^{b}_{om} ) f_{on} H_{+}(\omega_{no},\omega)\nn\\
&+\frac{i\pi \omega_{\Sigma}}{6} \sum_{nmo\v{k}} \omega_{nm;a} \frac{ r^{d}_{mn}}{\omega_{nm}}( r^{b}_{no} r^{c}_{om} - r^{c}_{no} r^{b}_{om} ) f_{on} D_{-}(\omega_{no},\omega)
\label{eq:eta2_4rd_12}
\end{align}
The first term will cancel against other nonresonant contributions. 

Next we consider the group of $l=5,6,9,10$. It is easy to show these terms can be written as 

\begin{widetext}
\begin{align}
\chi_{3i3,5}&\equiv \sum_{l}^{4} \chi_{3i3,5,l}=\frac{\omega_{\Sigma}}{6}\sum_{nmo\v{k}} \frac{\omega_{nm;a} r^{c}_{mn} r^{b}_{no} r^{d}_{om} f_{on}}{\omega_{mo}}\bigg[\frac{1}{\omega_{nm}-\omega} + i\pi \delta(\omega_{nm}-\omega) - \frac{1}{\omega_{no}-\omega} -i\pi \delta(\omega_{no}-\omega) \bigg], 
\label{eq:3i3_5}
\\
\chi_{3i3,6} &\equiv \sum_{l}^{4} \chi_{3i3,6,l}=\frac{\omega_{\Sigma}}{6}\hspace{-5pt}\sum_{nmo\v{k}} \frac{\omega_{nm;a} r^{c}_{mn} r^{d}_{no} r^{b}_{om} f_{mo}}{\omega_{no}}\bigg[\frac{1}{\omega_{nm}-\omega} + i\pi \delta(\omega_{nm}-\omega) - \frac{1}{\omega_{om}-\omega} - i\pi \delta(\omega_{om}-\omega) \bigg],
\label{eq:3i3_6}
\\
\chi_{3i3,9}&\equiv \sum_{l}^{4} \chi_{3i3,9,l}=\frac{\omega_{\Sigma}}{6}\sum_{nmo\v{k}} \frac{\omega_{nm;a} r^{b}_{mn} r^{c}_{no} r^{d}_{om} f_{on}}{\omega_{mo}}\bigg[\frac{1}{\omega_{nm}+\omega} + i\pi \delta(\omega_{nm}+\omega) - \frac{1}{\omega_{no}+\omega} - i\pi \delta(\omega_{no}+\omega) \bigg],
\label{eq:3i3_9}
\\
\chi_{3i3,10}&\equiv\hspace{-3pt} \sum_{l}^{4} \chi_{3i3,10,l}=\frac{\omega_{\Sigma}}{6} \hspace{-5pt} \sum_{nmo\v{k}} \frac{\omega_{nm;a} r^{b}_{mn} r^{d}_{no} r^{c}_{om} f_{mo}}{\omega_{no}}\bigg[\frac{1}{\omega_{nm}+\omega} + i\pi \delta(\omega_{nm}+\omega) - \frac{1}{\omega_{om}+\omega} - i\pi \delta(\omega_{om}+\omega) \bigg].
\label{eq:3i3_10}
\end{align}
\end{widetext} 

We now combine them with the resonant ($r$) and nonresonant ($nr$) parts of the $l=12,11,8,7$ terms (respectively). The result is 

\begin{widetext}
\begin{align}
\chi_{3i3,5,1} + (\chi_{3i3,12})_{nr}&= -\frac{\omega_{\Sigma}}{6}\sum_{nmo\v{k}}  \frac{\omega_{nm;a} r^{c}_{mn} r^{b}_{no} r^{d}_{om} f_{mn} }{\omega_{om}(\omega_{nm}-\omega)}, 
\label{eq:3i3_5_1_and_3i3_12}
\\
\chi_{3i3,5,2} + (\chi_{3i3,12})_{r} &=-\frac{i\pi \omega_{\Sigma}}{6}\sum_{nmo\v{k}}  \frac{\omega_{nm;a} r^{c}_{mn} r^{b}_{no} r^{d}_{om} f_{mn} }{\omega_{om}}\delta(\omega_{nm}-\omega), 
\label{eq:3i3_5_2_and_3i3_12} \\
\chi_{3i3,6,1} + (\chi_{3i3,11})_{nr}&= \frac{\omega_{\Sigma}}{6}\sum_{nmo\v{k}}  \frac{\omega_{nm;a} r^{c}_{mn} r^{d}_{no} r^{b}_{om} f_{mn} }{\omega_{no}(\omega_{nm}-\omega)},
\label{eq:3i3_6_1_and_3i3_11}
 \\
\chi_{3i3,6,2} + (\chi_{3i3,11})_{r} &=-\frac{i\pi \omega_{\Sigma}}{6}\sum_{nmo\v{k}}  \frac{\omega_{nm;a} r^{c}_{mn} r^{d}_{no} r^{b}_{om} f_{mn} }{\omega_{on}}\delta(\omega_{nm}-\omega),
\label{eq:3i3_6_2_and_3i3_11}\\
\chi_{3i3,9,1} + (\chi_{3i3,8})_{nr}&= -\frac{\omega_{\Sigma}}{6}\sum_{nmo\v{k}}  \frac{\omega_{nm;a} r^{b}_{mn} r^{c}_{no} r^{d}_{om} f_{mn} }{\omega_{om}(\omega_{nm}+\omega)},
\label{eq:3i3_9_1_and_3i3_8}
 \\
\chi_{3i3,9,2} + (\chi_{3i3,8})_{r} &=-\frac{i\pi \omega_{\Sigma}}{6}\sum_{nmo\v{k}}  \frac{\omega_{nm;a} r^{b}_{mn} r^{c}_{no} r^{d}_{om} f_{mn} }{\omega_{om}}\delta(\omega_{nm}+\omega),
\label{eq:3i3_9_2_and_3i3_8} \\
\chi_{3i3,10,1} + (\chi_{3i3,7})_{nr}&= \frac{\omega_{\Sigma}}{6}\sum_{nmo\v{k}}  \frac{\omega_{nm;a} r^{b}_{mn} r^{d}_{no} r^{c}_{om} f_{mn} }{\omega_{no}(\omega_{nm}+\omega)},
\label{eq:3i3_10_1_and_3i3_7}
 \\
\chi_{3i3,10,2} + (\chi_{3i3,7})_{r} &=\frac{i\pi \omega_{\Sigma}}{6}\sum_{nmo\v{k}}  \frac{\omega_{nm;a} r^{b}_{mn} r^{d}_{no} r^{c}_{om} f_{mn} }{\omega_{no}}\delta(\omega_{nm}+\omega).
\label{eq:3i3_10_2_and_3i3_7}
\end{align}
\end{widetext}
Now we want to show that to $\mathcal{O}(\omega_{\Sigma})$ the resonant part of the sum of the $l=1\text{-}4$ and $l=5\text{-}12$ groups gives the fourth term of $\eta_3$ and the nonresonant part vanishes. First the resonant contributions.

\subsubsection{Resonant contributions}
Let $n \leftrightarrow m$ and $\v{k}\to -\v{k}$ in $\chi_{3i3,6,4}$ and add to $\chi_{3i3,5,4}$ to obtain

\begin{align}
&\chi_{3i3,6,4} + \chi_{3i3,5,4} = \nn\\
&-\frac{i\pi \omega_{\Sigma}}{6} \sum_{nmo\v{k}} \omega_{nm;a} \frac{ r^{c}_{mn} r^{d}_{om} r^{b}_{no} f_{on}}{\omega_{mo}} D_{-}(\omega_{no},\omega).
\label{eq:eta2_4res1}
\end{align}
Similar manipulations on $\chi_{3i3,10,4}$ and  $\chi_{3i3,9,4}$ give 

\begin{align}
&\chi_{3i3,10,4} + \chi_{3i3,9,4} = \nn\\
&\frac{i\pi \omega_{\Sigma}}{6} \sum_{nmo\v{k}} \omega_{nm;a} \frac{ r^{b}_{mn} r^{d}_{om} r^{c}_{no} f_{on}}{\omega_{mo}} D_{-}(\omega_{no},\omega).
\label{eq:eta2_4res2}
\end{align}
Adding Eq.~\ref{eq:eta2_4res1} and \ref{eq:eta2_4res2} gives

\begin{align}
&\chi_{3i3,6,4} + \chi_{3i3,5,4} + \chi_{3i3,10,4} + \chi_{3i3,9,4} = \nn\\
&\frac{i\pi \omega_{\Sigma}}{6} \sum_{nmo\v{k}} \omega_{nm;a} \frac{r^{d}_{om}}{\omega_{mo}} ( r^{b}_{mn} r^{c}_{no} - r^{c}_{mn} r^{b}_{no}) f_{on} D_{-}(\omega_{no},\omega).
\label{eq:eta2_4res3}
\end{align}
Performing analogous manipulations, add Eq.~\ref{eq:3i3_5_2_and_3i3_12} to  Eq.~\ref{eq:3i3_6_2_and_3i3_11} and Eq.~\ref{eq:3i3_9_2_and_3i3_8} to Eq.~\ref{eq:3i3_10_2_and_3i3_7} to obtain

\begin{align}
&\chi_{3i3,5,2} + (\chi_{3i3,12})_{r}+ \chi_{3i3,6,2} + (\chi_{3i3,11})_{r}  = \nn\\
&-\frac{i\pi \omega_{\Sigma}}{6} \sum_{nmo\v{k}} \omega_{nm;a} \frac{ r^{c}_{mn} r^{b}_{no} r^{d}_{om} f_{mn}}{\omega_{om}} D_{-}(\omega_{nm},\omega),
\label{eq:eta2_4res4}
\end{align}
and
\begin{align}
&\chi_{3i3,10,2} + (\chi_{3i3,7})_{r}+\chi_{3i3,9,2} + (\chi_{3i3,8})_{r} = \nn\\
&\frac{i\pi \omega_{\Sigma}}{6} \sum_{nmo\v{k}} \omega_{nm;a} \frac{ r^{b}_{mn} r^{d}_{om} r^{c}_{no} f_{mn}}{\omega_{om}} D_{-}(\omega_{nm},\omega).
\label{eq:eta2_4res5}
\end{align}
respectively.  After $n \leftrightarrow l$, and $\v{k}\to -\v{k}$ in Eq.~\ref{eq:eta2_4res3} add to Eq.~\ref{eq:eta2_4rd_12} to obtain 

\begin{align}
&(\sum_{l}^4 \chi_{3i3,l})_{r} + \chi_{3i3,6,4} + \chi_{3i3,5,4}+\chi_{3i3,10,4} + \chi_{3i3,9,4}=\nn\\
&\frac{i\pi \omega_{\Sigma}}{6} \sum_{nmo\v{k}} \omega_{nl;a} \frac{ r^{d}_{mn}}{\omega_{nm}}( r^{b}_{no} r^{c}_{om} - r^{c}_{no} r^{b}_{om} ) f_{on} D_{-}(\omega_{no},\omega).
\label{eq:eta2_4rd_12_1}
\end{align}
Now add Eq.~\ref{eq:eta2_4res4} and Eq.~\ref{eq:eta2_4res5} to obtain 

\begin{align}
& \chi_{3i3,6,2} + (\chi_{3i3,11})_r + \chi_{3i3,5,2} + (\chi_{3i3,12})_{r}\nn\\
& + \chi_{3i3,10,2} + (\chi_{3i3,7})_{r}+\chi_{3i3,9,2} + (\chi_{3i3,8})_{r}= \nn\\
&\frac{i\pi \omega_{\Sigma}}{6} \sum_{nmo\v{k}} \omega_{no;a} \frac{ r^{d}_{mn}}{\omega_{nm}}( r^{b}_{no} r^{c}_{om} - r^{c}_{no} r^{b}_{om} ) f_{on} D_{-}(\omega_{no},\omega).
\label{eq:eta2_4rd_12_2}
\end{align}
Finally, the sum of all resonant terms in $\chi_{3i3}$ to linear order in $\omega_{\Sigma}$ amounts to adding Eq.~\ref{eq:eta2_4rd_12_1} to Eq.~\ref{eq:eta2_4rd_12_2}. The result is

\begin{align}
& \ref{eq:eta2_4rd_12_1} + \ref{eq:eta2_4rd_12_2} = \nn\\
&\frac{2i\pi \omega_{\Sigma}}{6} \sum_{nmo\v{k}} \omega_{no;a} \frac{ r^{d}_{mn}}{\omega_{nm}}( r^{b}_{no} r^{c}_{om} - r^{c}_{no} r^{b}_{on} ) f_{on} D_{-}(\omega_{no},\omega)
\label{eq:eta2_4rd}
\end{align}
which is the fourth term in $\eta_3$.

\subsubsection{Nonresonant contributions}
The sum of the (nonresonant) third terms in Eqs.~\ref{eq:3i3_5}, \ref{eq:3i3_6}, \ref{eq:3i3_9} and \ref{eq:3i3_10} gives

\begin{align}
&\chi_{3i3,6,3} + \chi_{3i3,5,3}+\chi_{3i3,9,3} + \chi_{3i3,10,3} = \nn\\
&-\frac{\omega_{\Sigma}}{6} \sum_{nmo\v{k}} \omega_{nm;a} \frac{ r^{d}_{om}( r^{c}_{mn} r^{b}_{no}+ r^{b}_{mn} r^{c}_{no})  f_{on}}{\omega_{mo}} H_{+}(\omega_{no},\omega).
\label{eq:eta2_4nr1}
\end{align}
Next, the sum of Eqs.~\ref{eq:3i3_5_1_and_3i3_12} and \ref{eq:3i3_6_1_and_3i3_11} and of Eq.~\ref{eq:3i3_9_1_and_3i3_8} and \ref{eq:3i3_10_1_and_3i3_7} gives
\begin{align}
&\chi_{3i3,5,1} + (\chi_{3i3,12})_{nr}+\chi_{3i3,6,1} + (\chi_{3i3,11})_{nr} = \nn\\
&-\frac{\omega_{\Sigma}}{6} \sum_{nmo\v{k}} \omega_{nm;a} \frac{ r^{c}_{mn} r^{b}_{no} r^{d}_{om} f_{mn}}{\omega_{om}} H_{+}(\omega_{nm},\omega),
\label{eq:eta2_4nr2}
\\
& \chi_{3i3,9,1} + (\chi_{3i3,8})_{nr} + \chi_{3i3,10,1} + (\chi_{3i3,7})_{nr}  = \nn\\
&-\frac{\omega_{\Sigma}}{6} \sum_{nmo\v{k}} \omega_{nm;a} \frac{ r^{b}_{mn} r^{c}_{no} r^{d}_{om} f_{mn}}{\omega_{om}} H_{+}(\omega_{nm},\omega).
\label{eq:eta2_4nr3}
\end{align}
After $l\leftrightarrow n$ and $\v{k}\to -\v{k}$ in Eq.~\ref{eq:eta2_4nr1} combined with the nonresonant part of Eq.~\ref{eq:eta2_4rd_12} we obtain 

\begin{align}
&(\sum_{l}^4 \chi_{3i3,l})_{nr}+ \chi_{3i3,6,3} + \chi_{3i3,5,3}+\chi_{3i3,9,3} + \chi_{3i3,10,3}=\nn\\
&\frac{\omega_{\Sigma}}{6} \sum_{nmo\v{k}} \omega_{no;a} \frac{ r^{d}_{mn}( r^{c}_{om} r^{b}_{no}+ r^{b}_{om} r^{c}_{no})  f_{on}}{\omega_{nm}} H_{+}(\omega_{no},\omega).
\label{eq:eta2_4nr4}
\end{align}
Adding Eq.~\ref{eq:eta2_4nr2} and Eq.~\ref{eq:eta2_4nr3} we obtain
\begin{align}
&\ref{eq:eta2_4nr2} + \ref{eq:eta2_4nr3} =\nn\\
&-\frac{\omega_{\Sigma}}{6} \sum_{nmo\v{k}} \omega_{nm;a} \frac{ r^{d}_{om}( r^{c}_{no} r^{b}_{mn} + r^{b}_{no} r^{c}_{mn})  f_{mn}}{\omega_{om}} H_{+}(\omega_{no},\omega),
\label{eq:eta2_4nr5}
\end{align}
which after $l \leftrightarrow n$ and $n \leftrightarrow m$, is seen to cancel Eq.~\ref{eq:eta2_4nr4}. This concludes the proof that to linear order on $\omega_{\Sigma}$ the nonresonant terms vanish.

\section{Derivation of $\sigma_3$}
\label{app:sigma3}

\subsection{First and second terms in $\sigma_3$}
Consider $\chi_{3i5,1}$ and $\chi_{3i5,2}$ in Eq.\ref{eq:chi3i5n}. In these terms we can set $\omega_{\beta}+\omega_{\sigma}=0$. Using $1/(x-i\epsilon)=1/x + i\pi\delta(x)$ and  

\begin{align}
\frac{\partial}{\partial k^c}\left(\frac{r^{d}_{mn;a}r^{b}_{nm}}{\omega_{nm}} \right) = \left( \frac{r^{d}_{mn;a}}{\omega_{nm}}\right)_{;c} r_{nm}^{b} + \left( \frac{r^{d}_{mn;a}}{\omega_{nm}}\right) r_{nm;c}^{b} 
\end{align}
the resonant parts are 

\begin{align}
&(\chi_{3i5,1} + \chi_{3i5,2})_{r} =\nn\\
& \frac{\pi \omega_{\Sigma}^2}{6} \sum_{nm\v{k}} f_{mn}\big[ \big(\frac{r^d_{mn;a}}{\omega_{nm}} \big)_{;c} r_{nm}^b 
+\big(\frac{r^d_{mn;a}}{\omega_{nm}} \big)_{;b} r_{nm}^c \big]\delta(\omega_{nm}-\omega)
\end{align}
Similar manipulations on $\chi_{3i5,4}$ and $\chi_{3i5,6}$ in Eq.~\ref{eq:chi3i5n} yield the rest of the terms in the square brackets in $\sigma_3$. The nonresonant parts can be shown to vanish. 

\subsection{Third and fourth terms in $\sigma_3$}
These contributions to $\sigma_3$ arise from $\chi_{3i6}$ in Eq. \ref{eq:chi3i}. It can be shown that the nonresonant parts vanish and the resonant part gives the third and fourth term in $\sigma_3$. Since the algebraic steps are very similar to those used in finding the third term in $\eta_3$ we omit the derivation.

\section{two-band model of single-layer GeS}
\label{sec:tb_GeS}
We consider a two-band, 2D model of single-layer GeS given by the Hamiltonian 

\begin{align}
H= f_0\sigma_0 + f_a \sigma_{a},
\end{align}
where $\sigma_a, a=x,y,z$ are the standard Pauli matrices and $\sigma_0$ is the $2\times 2$ identity matrix. In this section, summation over repeated indices is implied. The functions $f_a$ are given by the  hopping integrals of the model. The Hamiltonian has eigenvectors given by
\begin{align}
u_c &= A
\begin{pmatrix}
f_x-if_y  \\
\epsilon - f_z
\end{pmatrix} \\
u_v &= A
\begin{pmatrix}
f_z-\epsilon  \\
f_x+if_y
\end{pmatrix},
\end{align}
where $A^{-2}=2\epsilon(\epsilon-f_z)$ is the normalization and eigenvalues by $E_{c,v}=f_0 \pm \epsilon$ where $\epsilon=\sqrt{ f_a f_a}$ and $c,v$ denote the conduction and valence band respectively. An arbitrary phase factor in the eigenvectors has been omitted, since the final expressions are independent of this phase. The Bloch wave functions are constructed as

\begin{align}
\psi_{n\v{k}} = \sum_{\v{R}} &e^{i\v{k}\cdot \v{R}} [ u_{n}^{(1)} \phi(\v{r}-\v{R}) \nn \\
&+ e^{i\v{k}\cdot \v{r}_0} u_{n}^{(2)} \phi(\v{r}-\v{r}_0-\v{R})],
\end{align}
where $u_{n}^{(i)}$ denotes the eigenvector corresponding to eigenvalue $n=v,c$ (valence, conduction) and $i=1,2$ denotes the first and second components. $\v{r}_0=(a_0,0)$ is the position of site $B$ with respect to site $A$ which is taken to be the origin. $\phi(\v{r})$ are $p_z$-orbitals and $\v{R}$ runs over all lattice positions. Notice that the phase of the wave function at site $B$ is different than that at site $A$. 

The hopping parameters of the Hamiltonian are~\cite{Cook2017} 

\begin{align}
f_0 &= 2 t_{1}' [\cos\v{k}\cdot\v{a}_1 +\cos\v{k}\cdot\v{a}_2 ]\nn \\
&~~~~~~~~+ 2 t_2' \cos\v{k}\cdot(\v{a}_1-\v{a}_2),\\
f_x - i f_y &= e^{i\v{k}\cdot \v{r}_0} (t_1  + t_2 \Phi_{\v{k}}+  t_3 \Phi_{\v{k}}^{*}), \\
f_z &= \Delta,
\end{align}
where $\Phi_{\v{k}}\equiv e^{-i\v{k}\cdot \v{a}_1} + e^{-i\v{k}\cdot \v{a}_2}$, $\Delta$ is the onsite potential and $t_1,t_2,t_3,t_1',t_2'$ are hopping matrix elements as indicated in Fig.~\ref{fig:tb_model}(c).  $\v{a}_1 = (a_x,-a_y),\v{a}_2 = (a_x,a_y)$ are the primitive lattice vectors. 

For single-layer GeS the parameters are: $(a_x,a_y,d)=(4.53/2,3.63/2,2.56)$ \AA, where $d$ is the thickness of the slab, $a_0=0.62$ \AA, and $(t_1,t_2,t_3,t_1',t_2',\Delta)=(-2.33,0.61,0.13,0.07,-0.09,0.41)$ eV. It was shown that these parameters reproduce the band structure and geometry of the wavefunction in the vicinity of the Gamma point~\cite{Cook2017}. To compare with bulk values the results are multiplied by $2/d$. The factor of 2 takes into account the smaller unit cell of the tight-binding model.


\begin{thebibliography}{67}%
\makeatletter
\providecommand \@ifxundefined [1]{%
 \@ifx{#1\undefined}
}%
\providecommand \@ifnum [1]{%
 \ifnum #1\expandafter \@firstoftwo
 \else \expandafter \@secondoftwo
 \fi
}%
\providecommand \@ifx [1]{%
 \ifx #1\expandafter \@firstoftwo
 \else \expandafter \@secondoftwo
 \fi
}%
\providecommand \natexlab [1]{#1}%
\providecommand \enquote  [1]{``#1''}%
\providecommand \bibnamefont  [1]{#1}%
\providecommand \bibfnamefont [1]{#1}%
\providecommand \citenamefont [1]{#1}%
\providecommand \href@noop [0]{\@secondoftwo}%
\providecommand \href [0]{\begingroup \@sanitize@url \@href}%
\providecommand \@href[1]{\@@startlink{#1}\@@href}%
\providecommand \@@href[1]{\endgroup#1\@@endlink}%
\providecommand \@sanitize@url [0]{\catcode `\\12\catcode `\$12\catcode
  `\&12\catcode `\#12\catcode `\^12\catcode `\_12\catcode `\%12\relax}%
\providecommand \@@startlink[1]{}%
\providecommand \@@endlink[0]{}%
\providecommand \url  [0]{\begingroup\@sanitize@url \@url }%
\providecommand \@url [1]{\endgroup\@href {#1}{\urlprefix }}%
\providecommand \urlprefix  [0]{URL }%
\providecommand \Eprint [0]{\href }%
\providecommand \doibase [0]{http://dx.doi.org/}%
\providecommand \selectlanguage [0]{\@gobble}%
\providecommand \bibinfo  [0]{\@secondoftwo}%
\providecommand \bibfield  [0]{\@secondoftwo}%
\providecommand \translation [1]{[#1]}%
\providecommand \BibitemOpen [0]{}%
\providecommand \bibitemStop [0]{}%
\providecommand \bibitemNoStop [0]{.\EOS\space}%
\providecommand \EOS [0]{\spacefactor3000\relax}%
\providecommand \BibitemShut  [1]{\csname bibitem#1\endcsname}%
\let\auto@bib@innerbib\@empty
\bibitem [{\citenamefont {Karplus}\ and\ \citenamefont
  {Luttinger}(1954)}]{Karplus1954}%
  \BibitemOpen
  \bibfield  {author} {\bibinfo {author} {\bibfnamefont {R.}~\bibnamefont
  {Karplus}}\ and\ \bibinfo {author} {\bibfnamefont {J.~M.}\ \bibnamefont
  {Luttinger}},\ }\href@noop {} {\bibfield  {journal} {\bibinfo  {journal}
  {Phys. Rev.}\ }\textbf {\bibinfo {volume} {95}},\ \bibinfo {pages} {1154}
  (\bibinfo {year} {1954})}\BibitemShut {NoStop}%
\bibitem [{\citenamefont {Nagaosa}\ \emph {et~al.}(2010)\citenamefont
  {Nagaosa}, \citenamefont {Sinova}, \citenamefont {Onoda}, \citenamefont
  {MacDonald},\ and\ \citenamefont {Ong}}]{Nagaosa2010}%
  \BibitemOpen
  \bibfield  {author} {\bibinfo {author} {\bibfnamefont {N.}~\bibnamefont
  {Nagaosa}}, \bibinfo {author} {\bibfnamefont {J.}~\bibnamefont {Sinova}},
  \bibinfo {author} {\bibfnamefont {S.}~\bibnamefont {Onoda}}, \bibinfo
  {author} {\bibfnamefont {A.~H.}\ \bibnamefont {MacDonald}}, \ and\ \bibinfo
  {author} {\bibfnamefont {N.~P.}\ \bibnamefont {Ong}},\ }\href@noop {}
  {\bibfield  {journal} {\bibinfo  {journal} {Rev. Mod. Phys.}\ }\textbf
  {\bibinfo {volume} {82}},\ \bibinfo {pages} {1539} (\bibinfo {year}
  {2010})}\BibitemShut {NoStop}%
\bibitem [{\citenamefont {Armitage}\ \emph {et~al.}(2018)\citenamefont
  {Armitage}, \citenamefont {Mele},\ and\ \citenamefont
  {Vishwanath}}]{Armitage2018}%
  \BibitemOpen
  \bibfield  {author} {\bibinfo {author} {\bibfnamefont {N.~P.}\ \bibnamefont
  {Armitage}}, \bibinfo {author} {\bibfnamefont {E.~J.}\ \bibnamefont {Mele}},
  \ and\ \bibinfo {author} {\bibfnamefont {A.}~\bibnamefont {Vishwanath}},\
  }\href@noop {} {\bibfield  {journal} {\bibinfo  {journal} {Rev. Mod. Phys.}\
  }\textbf {\bibinfo {volume} {90}},\ \bibinfo {pages} {015001} (\bibinfo
  {year} {2018})}\BibitemShut {NoStop}%
\bibitem [{\citenamefont {Vishwanath}(2015)}]{Vishwanath2015}%
  \BibitemOpen
  \bibfield  {author} {\bibinfo {author} {\bibfnamefont {A.}~\bibnamefont
  {Vishwanath}},\ }\href@noop {} {\bibfield  {journal} {\bibinfo  {journal}
  {Physics}\ }\textbf {\bibinfo {volume} {8}},\ \bibinfo {pages} {84} (\bibinfo
  {year} {2015})}\BibitemShut {NoStop}%
\bibitem [{\citenamefont {Auston}\ \emph {et~al.}(1972)\citenamefont {Auston},
  \citenamefont {Glass},\ and\ \citenamefont {Ballman}}]{Auston1972}%
  \BibitemOpen
  \bibfield  {author} {\bibinfo {author} {\bibfnamefont {D.~H.}\ \bibnamefont
  {Auston}}, \bibinfo {author} {\bibfnamefont {A.~M.}\ \bibnamefont {Glass}}, \
  and\ \bibinfo {author} {\bibfnamefont {A.~A.}\ \bibnamefont {Ballman}},\
  }\href@noop {} {\bibfield  {journal} {\bibinfo  {journal} {Phys. Rev. Lett.}\
  }\textbf {\bibinfo {volume} {28}},\ \bibinfo {pages} {897} (\bibinfo {year}
  {1972})}\BibitemShut {NoStop}%
\bibitem [{\citenamefont {Glass}\ \emph {et~al.}(1974)\citenamefont {Glass},
  \citenamefont {von~der Linde},\ and\ \citenamefont {Negran}}]{Glass1974}%
  \BibitemOpen
  \bibfield  {author} {\bibinfo {author} {\bibfnamefont {A.~M.}\ \bibnamefont
  {Glass}}, \bibinfo {author} {\bibfnamefont {D.}~\bibnamefont {von~der
  Linde}}, \ and\ \bibinfo {author} {\bibfnamefont {T.~J.}\ \bibnamefont
  {Negran}},\ }\href@noop {} {\bibfield  {journal} {\bibinfo  {journal}
  {Applied Physics Letters}\ }\textbf {\bibinfo {volume} {25}},\ \bibinfo
  {pages} {233} (\bibinfo {year} {1974})}\BibitemShut {NoStop}%
\bibitem [{\citenamefont {Koch}\ \emph {et~al.}(1976)\citenamefont {Koch},
  \citenamefont {Munser}, \citenamefont {Ruppel},\ and\ \citenamefont
  {Würfel}}]{Koch1976}%
  \BibitemOpen
  \bibfield  {author} {\bibinfo {author} {\bibfnamefont {W.~T.~H.}\
  \bibnamefont {Koch}}, \bibinfo {author} {\bibfnamefont {R.}~\bibnamefont
  {Munser}}, \bibinfo {author} {\bibfnamefont {W.}~\bibnamefont {Ruppel}}, \
  and\ \bibinfo {author} {\bibfnamefont {P.}~\bibnamefont {Würfel}},\
  }\href@noop {} {\bibfield  {journal} {\bibinfo  {journal} {Ferroelectrics}\
  }\textbf {\bibinfo {volume} {13}},\ \bibinfo {pages} {305} (\bibinfo {year}
  {1976})}\BibitemShut {NoStop}%
\bibitem [{\citenamefont {Belinicher}\ and\ \citenamefont
  {Sturman}(1980)}]{Belinicher1980}%
  \BibitemOpen
  \bibfield  {author} {\bibinfo {author} {\bibfnamefont {V.~I.}\ \bibnamefont
  {Belinicher}}\ and\ \bibinfo {author} {\bibfnamefont {B.~I.}\ \bibnamefont
  {Sturman}},\ }\href@noop {} {\bibfield  {journal} {\bibinfo  {journal}
  {Physics-Uspekhi}\ }\textbf {\bibinfo {volume} {23}},\ \bibinfo {pages} {199}
  (\bibinfo {year} {1980})}\BibitemShut {NoStop}%
\bibitem [{\citenamefont {Sturman}\ and\ \citenamefont
  {Fridkin}(1992)}]{Sturman1992}%
  \BibitemOpen
  \bibfield  {author} {\bibinfo {author} {\bibfnamefont {B.~I.}\ \bibnamefont
  {Sturman}}\ and\ \bibinfo {author} {\bibfnamefont {V.~M.}\ \bibnamefont
  {Fridkin}},\ }\href@noop {} {\emph {\bibinfo {title} {The Photovoltaic and
  Photorefractive Effects in Non-CentrosymmetricMaterials}}}\ (\bibinfo
  {publisher} {Gordon and Breach Science Publishers, Philadelphia},\ \bibinfo
  {year} {1992})\BibitemShut {NoStop}%
\bibitem [{\citenamefont {Belinicher}\ and\ \citenamefont
  {Sturman}(1988)}]{Belinicher1988}%
  \BibitemOpen
  \bibfield  {author} {\bibinfo {author} {\bibfnamefont {V.~I.}\ \bibnamefont
  {Belinicher}}\ and\ \bibinfo {author} {\bibfnamefont {B.~I.}\ \bibnamefont
  {Sturman}},\ }\href@noop {} {\bibfield  {journal} {\bibinfo  {journal}
  {Ferroelectrics}\ }\textbf {\bibinfo {volume} {83}},\ \bibinfo {pages} {29}
  (\bibinfo {year} {1988})}\BibitemShut {NoStop}%
\bibitem [{\citenamefont {von Baltz}\ and\ \citenamefont
  {Kraut}(1981)}]{Baltz1981}%
  \BibitemOpen
  \bibfield  {author} {\bibinfo {author} {\bibfnamefont {R.}~\bibnamefont {von
  Baltz}}\ and\ \bibinfo {author} {\bibfnamefont {W.}~\bibnamefont {Kraut}},\
  }\href@noop {} {\bibfield  {journal} {\bibinfo  {journal} {Phys. Rev. B}\
  }\textbf {\bibinfo {volume} {23}},\ \bibinfo {pages} {5590} (\bibinfo {year}
  {1981})}\BibitemShut {NoStop}%
\bibitem [{\citenamefont {Laman}\ \emph {et~al.}(1999)\citenamefont {Laman},
  \citenamefont {Shkrebtii}, \citenamefont {Sipe},\ and\ \citenamefont {van
  Driel}}]{Laman1999}%
  \BibitemOpen
  \bibfield  {author} {\bibinfo {author} {\bibfnamefont {N.}~\bibnamefont
  {Laman}}, \bibinfo {author} {\bibfnamefont {A.~I.}\ \bibnamefont
  {Shkrebtii}}, \bibinfo {author} {\bibfnamefont {J.~E.}\ \bibnamefont {Sipe}},
  \ and\ \bibinfo {author} {\bibfnamefont {H.~M.}\ \bibnamefont {van Driel}},\
  }\href@noop {} {\bibfield  {journal} {\bibinfo  {journal} {Applied Physics
  Letters}\ }\textbf {\bibinfo {volume} {75}},\ \bibinfo {pages} {2581}
  (\bibinfo {year} {1999})}\BibitemShut {NoStop}%
\bibitem [{\citenamefont {Laman}\ \emph {et~al.}(2005)\citenamefont {Laman},
  \citenamefont {Bieler},\ and\ \citenamefont {van Driel}}]{Laman2005}%
  \BibitemOpen
  \bibfield  {author} {\bibinfo {author} {\bibfnamefont {N.}~\bibnamefont
  {Laman}}, \bibinfo {author} {\bibfnamefont {M.}~\bibnamefont {Bieler}}, \
  and\ \bibinfo {author} {\bibfnamefont {H.~M.}\ \bibnamefont {van Driel}},\
  }\href@noop {} {\bibfield  {journal} {\bibinfo  {journal} {Journal of Applied
  Physics}\ }\textbf {\bibinfo {volume} {98}},\ \bibinfo {pages} {103507}
  (\bibinfo {year} {2005})}\BibitemShut {NoStop}%
\bibitem [{\citenamefont {Sipe}\ and\ \citenamefont
  {Shkrebtii}(2000)}]{Sipe2000}%
  \BibitemOpen
  \bibfield  {author} {\bibinfo {author} {\bibfnamefont {J.~E.}\ \bibnamefont
  {Sipe}}\ and\ \bibinfo {author} {\bibfnamefont {A.~I.}\ \bibnamefont
  {Shkrebtii}},\ }\href@noop {} {\bibfield  {journal} {\bibinfo  {journal}
  {Phys. Rev. B}\ }\textbf {\bibinfo {volume} {61}},\ \bibinfo {pages} {5337}
  (\bibinfo {year} {2000})}\BibitemShut {NoStop}%
\bibitem [{\citenamefont {van Driel}\ and\ \citenamefont
  {Sipe}(2001)}]{Driel2001}%
  \BibitemOpen
  \bibfield  {author} {\bibinfo {author} {\bibfnamefont {H.~M.}\ \bibnamefont
  {van Driel}}\ and\ \bibinfo {author} {\bibfnamefont {J.~E.}\ \bibnamefont
  {Sipe}},\ }\enquote {\bibinfo {title} {Coherence control of photocurrents in
  semiconductors},}\ \ (\bibinfo  {publisher} {Springer, New York, NY},\
  \bibinfo {year} {2001})\ Chap.~\bibinfo {chapter} {5}, pp.\ \bibinfo {pages}
  {261--306}\BibitemShut {NoStop}%
\bibitem [{\citenamefont {C\^{o}t\'{e}}\ \emph {et~al.}(2002)\citenamefont
  {C\^{o}t\'{e}}, \citenamefont {Laman},\ and\ \citenamefont {van
  Driel}}]{Cote2002}%
  \BibitemOpen
  \bibfield  {author} {\bibinfo {author} {\bibfnamefont {D.}~\bibnamefont
  {C\^{o}t\'{e}}}, \bibinfo {author} {\bibfnamefont {N.}~\bibnamefont {Laman}},
  \ and\ \bibinfo {author} {\bibfnamefont {H.~M.}\ \bibnamefont {van Driel}},\
  }\href@noop {} {\bibfield  {journal} {\bibinfo  {journal} {Applied Physics
  Letters}\ }\textbf {\bibinfo {volume} {80}},\ \bibinfo {pages} {905}
  (\bibinfo {year} {2002})}\BibitemShut {NoStop}%
\bibitem [{\citenamefont {Ghalgaoui}\ \emph {et~al.}(2018)\citenamefont
  {Ghalgaoui}, \citenamefont {Reimann}, \citenamefont {Woerner}, \citenamefont
  {Elsaesser}, \citenamefont {Flytzanis},\ and\ \citenamefont
  {Biermann}}]{Ghalgaoui2018}%
  \BibitemOpen
  \bibfield  {author} {\bibinfo {author} {\bibfnamefont {A.}~\bibnamefont
  {Ghalgaoui}}, \bibinfo {author} {\bibfnamefont {K.}~\bibnamefont {Reimann}},
  \bibinfo {author} {\bibfnamefont {M.}~\bibnamefont {Woerner}}, \bibinfo
  {author} {\bibfnamefont {T.}~\bibnamefont {Elsaesser}}, \bibinfo {author}
  {\bibfnamefont {C.}~\bibnamefont {Flytzanis}}, \ and\ \bibinfo {author}
  {\bibfnamefont {K.}~\bibnamefont {Biermann}},\ }\href@noop {} {\bibfield
  {journal} {\bibinfo  {journal} {Phys. Rev. Lett.}\ }\textbf {\bibinfo
  {volume} {121}},\ \bibinfo {pages} {266602} (\bibinfo {year}
  {2018})}\BibitemShut {NoStop}%
\bibitem [{\citenamefont {Bieler}\ \emph {et~al.}(2005)\citenamefont {Bieler},
  \citenamefont {Laman}, \citenamefont {van Driel},\ and\ \citenamefont
  {Smirl}}]{Bieler2005}%
  \BibitemOpen
  \bibfield  {author} {\bibinfo {author} {\bibfnamefont {M.}~\bibnamefont
  {Bieler}}, \bibinfo {author} {\bibfnamefont {N.}~\bibnamefont {Laman}},
  \bibinfo {author} {\bibfnamefont {H.~M.}\ \bibnamefont {van Driel}}, \ and\
  \bibinfo {author} {\bibfnamefont {A.~L.}\ \bibnamefont {Smirl}},\ }\href@noop
  {} {\bibfield  {journal} {\bibinfo  {journal} {Applied Physics Letters}\
  }\textbf {\bibinfo {volume} {86}},\ \bibinfo {pages} {061102} (\bibinfo
  {year} {2005})}\BibitemShut {NoStop}%
\bibitem [{\citenamefont {Bieler}\ \emph {et~al.}(2007)\citenamefont {Bieler},
  \citenamefont {Pierz}, \citenamefont {Siegner},\ and\ \citenamefont
  {Dawson}}]{Bieler2007}%
  \BibitemOpen
  \bibfield  {author} {\bibinfo {author} {\bibfnamefont {M.}~\bibnamefont
  {Bieler}}, \bibinfo {author} {\bibfnamefont {K.}~\bibnamefont {Pierz}},
  \bibinfo {author} {\bibfnamefont {U.}~\bibnamefont {Siegner}}, \ and\
  \bibinfo {author} {\bibfnamefont {P.}~\bibnamefont {Dawson}},\ }\href@noop {}
  {\bibfield  {journal} {\bibinfo  {journal} {Phys. Rev. B}\ }\textbf {\bibinfo
  {volume} {76}},\ \bibinfo {pages} {161304} (\bibinfo {year}
  {2007})}\BibitemShut {NoStop}%
\bibitem [{\citenamefont {Rioux}\ \emph {et~al.}(2011)\citenamefont {Rioux},
  \citenamefont {Burkard},\ and\ \citenamefont {Sipe}}]{Rioux2011}%
  \BibitemOpen
  \bibfield  {author} {\bibinfo {author} {\bibfnamefont {J.}~\bibnamefont
  {Rioux}}, \bibinfo {author} {\bibfnamefont {G.}~\bibnamefont {Burkard}}, \
  and\ \bibinfo {author} {\bibfnamefont {J.~E.}\ \bibnamefont {Sipe}},\
  }\href@noop {} {\bibfield  {journal} {\bibinfo  {journal} {Phys. Rev. B}\
  }\textbf {\bibinfo {volume} {83}},\ \bibinfo {pages} {195406} (\bibinfo
  {year} {2011})}\BibitemShut {NoStop}%
\bibitem [{\citenamefont {Rioux}\ and\ \citenamefont {Sipe}(2012)}]{Rioux2012}%
  \BibitemOpen
  \bibfield  {author} {\bibinfo {author} {\bibfnamefont {J.}~\bibnamefont
  {Rioux}}\ and\ \bibinfo {author} {\bibfnamefont {J.}~\bibnamefont {Sipe}},\
  }\href@noop {} {\bibfield  {journal} {\bibinfo  {journal} {Physica E:
  Low-dimensional Systems and Nanostructures}\ }\textbf {\bibinfo {volume}
  {45}},\ \bibinfo {pages} {1} (\bibinfo {year} {2012})}\BibitemShut {NoStop}%
\bibitem [{\citenamefont {Somma}\ \emph {et~al.}(2014)\citenamefont {Somma},
  \citenamefont {Reimann}, \citenamefont {Flytzanis}, \citenamefont
  {Elsaesser},\ and\ \citenamefont {Woerner}}]{Somma2014}%
  \BibitemOpen
  \bibfield  {author} {\bibinfo {author} {\bibfnamefont {C.}~\bibnamefont
  {Somma}}, \bibinfo {author} {\bibfnamefont {K.}~\bibnamefont {Reimann}},
  \bibinfo {author} {\bibfnamefont {C.}~\bibnamefont {Flytzanis}}, \bibinfo
  {author} {\bibfnamefont {T.}~\bibnamefont {Elsaesser}}, \ and\ \bibinfo
  {author} {\bibfnamefont {M.}~\bibnamefont {Woerner}},\ }\href@noop {}
  {\bibfield  {journal} {\bibinfo  {journal} {Phys. Rev. Lett.}\ }\textbf
  {\bibinfo {volume} {112}},\ \bibinfo {pages} {146602} (\bibinfo {year}
  {2014})}\BibitemShut {NoStop}%
\bibitem [{\citenamefont {Nakamura}\ \emph {et~al.}(2016)\citenamefont
  {Nakamura}, \citenamefont {Kagawa}, \citenamefont {Tanigaki}, \citenamefont
  {Park}, \citenamefont {Matsuda}, \citenamefont {Shindo}, \citenamefont
  {Tokura},\ and\ \citenamefont {Kawasaki}}]{Nakamura2016}%
  \BibitemOpen
  \bibfield  {author} {\bibinfo {author} {\bibfnamefont {M.}~\bibnamefont
  {Nakamura}}, \bibinfo {author} {\bibfnamefont {F.}~\bibnamefont {Kagawa}},
  \bibinfo {author} {\bibfnamefont {T.}~\bibnamefont {Tanigaki}}, \bibinfo
  {author} {\bibfnamefont {H.~S.}\ \bibnamefont {Park}}, \bibinfo {author}
  {\bibfnamefont {T.}~\bibnamefont {Matsuda}}, \bibinfo {author} {\bibfnamefont
  {D.}~\bibnamefont {Shindo}}, \bibinfo {author} {\bibfnamefont
  {Y.}~\bibnamefont {Tokura}}, \ and\ \bibinfo {author} {\bibfnamefont
  {M.}~\bibnamefont {Kawasaki}},\ }\href@noop {} {\bibfield  {journal}
  {\bibinfo  {journal} {Phys. Rev. Lett.}\ }\textbf {\bibinfo {volume} {116}},\
  \bibinfo {pages} {156801} (\bibinfo {year} {2016})}\BibitemShut {NoStop}%
\bibitem [{\citenamefont {Holtz}\ \emph {et~al.}(2016)\citenamefont {Holtz},
  \citenamefont {Hauf}, \citenamefont {Hern\'andez~Salvador}, \citenamefont
  {Costard}, \citenamefont {Woerner},\ and\ \citenamefont
  {Elsaesser}}]{Holtz2016}%
  \BibitemOpen
  \bibfield  {author} {\bibinfo {author} {\bibfnamefont {M.}~\bibnamefont
  {Holtz}}, \bibinfo {author} {\bibfnamefont {C.}~\bibnamefont {Hauf}},
  \bibinfo {author} {\bibfnamefont {A.-A.}\ \bibnamefont
  {Hern\'andez~Salvador}}, \bibinfo {author} {\bibfnamefont {R.}~\bibnamefont
  {Costard}}, \bibinfo {author} {\bibfnamefont {M.}~\bibnamefont {Woerner}}, \
  and\ \bibinfo {author} {\bibfnamefont {T.}~\bibnamefont {Elsaesser}},\
  }\href@noop {} {\bibfield  {journal} {\bibinfo  {journal} {Phys. Rev. B}\
  }\textbf {\bibinfo {volume} {94}},\ \bibinfo {pages} {104302} (\bibinfo
  {year} {2016})}\BibitemShut {NoStop}%
\bibitem [{\citenamefont {Rappe}\ \emph {et~al.}(2017)\citenamefont {Rappe},
  \citenamefont {Grinberg},\ and\ \citenamefont {Spanier}}]{Rappe2017}%
  \BibitemOpen
  \bibfield  {author} {\bibinfo {author} {\bibfnamefont {A.~M.}\ \bibnamefont
  {Rappe}}, \bibinfo {author} {\bibfnamefont {I.}~\bibnamefont {Grinberg}}, \
  and\ \bibinfo {author} {\bibfnamefont {J.~E.}\ \bibnamefont {Spanier}},\
  }\href@noop {} {\bibfield  {journal} {\bibinfo  {journal} {Proceedings of the
  National Academy of Sciences}\ }\textbf {\bibinfo {volume} {114}},\ \bibinfo
  {pages} {7191} (\bibinfo {year} {2017})}\BibitemShut {NoStop}%
\bibitem [{\citenamefont {Spanier}\ \emph {et~al.}(2016)\citenamefont
  {Spanier}, \citenamefont {Fridkin}, \citenamefont {Rappe}, \citenamefont
  {Akbashev}, \citenamefont {Polemi}, \citenamefont {Qi}, \citenamefont {Gu},
  \citenamefont {Young}, \citenamefont {Hawley}, \citenamefont {Imbrenda},
  \citenamefont {Xiao}, \citenamefont {Bennett-Jackson},\ and\ \citenamefont
  {Johnson}}]{Spanier2016}%
  \BibitemOpen
  \bibfield  {author} {\bibinfo {author} {\bibfnamefont {J.~E.}\ \bibnamefont
  {Spanier}}, \bibinfo {author} {\bibfnamefont {V.~M.}\ \bibnamefont
  {Fridkin}}, \bibinfo {author} {\bibfnamefont {A.~M.}\ \bibnamefont {Rappe}},
  \bibinfo {author} {\bibfnamefont {A.~R.}\ \bibnamefont {Akbashev}}, \bibinfo
  {author} {\bibfnamefont {A.}~\bibnamefont {Polemi}}, \bibinfo {author}
  {\bibfnamefont {Y.}~\bibnamefont {Qi}}, \bibinfo {author} {\bibfnamefont
  {Z.}~\bibnamefont {Gu}}, \bibinfo {author} {\bibfnamefont {S.~M.}\
  \bibnamefont {Young}}, \bibinfo {author} {\bibfnamefont {C.~J.}\ \bibnamefont
  {Hawley}}, \bibinfo {author} {\bibfnamefont {D.}~\bibnamefont {Imbrenda}},
  \bibinfo {author} {\bibfnamefont {G.}~\bibnamefont {Xiao}}, \bibinfo {author}
  {\bibfnamefont {A.~L.}\ \bibnamefont {Bennett-Jackson}}, \ and\ \bibinfo
  {author} {\bibfnamefont {C.~L.}\ \bibnamefont {Johnson}},\ }\href@noop {}
  {\bibfield  {journal} {\bibinfo  {journal} {Nature Photonics}\ }\textbf
  {\bibinfo {volume} {10}},\ \bibinfo {pages} {611} (\bibinfo {year}
  {2016})}\BibitemShut {NoStop}%
\bibitem [{\citenamefont {Tan}\ \emph {et~al.}(2016)\citenamefont {Tan},
  \citenamefont {Zheng}, \citenamefont {Young}, \citenamefont {Wang},
  \citenamefont {Liu},\ and\ \citenamefont {Rappe}}]{Tan2016}%
  \BibitemOpen
  \bibfield  {author} {\bibinfo {author} {\bibfnamefont {L.~Z.}\ \bibnamefont
  {Tan}}, \bibinfo {author} {\bibfnamefont {F.}~\bibnamefont {Zheng}}, \bibinfo
  {author} {\bibfnamefont {S.~M.}\ \bibnamefont {Young}}, \bibinfo {author}
  {\bibfnamefont {F.}~\bibnamefont {Wang}}, \bibinfo {author} {\bibfnamefont
  {S.}~\bibnamefont {Liu}}, \ and\ \bibinfo {author} {\bibfnamefont {A.~M.}\
  \bibnamefont {Rappe}},\ }\href@noop {} {\bibfield  {journal} {\bibinfo
  {journal} {npj Comput. Mater.}\ }\textbf {\bibinfo {volume} {2}},\ \bibinfo
  {pages} {16026} (\bibinfo {year} {2016})}\BibitemShut {NoStop}%
\bibitem [{\citenamefont {Rangel}\ \emph {et~al.}(2017)\citenamefont {Rangel},
  \citenamefont {Fregoso}, \citenamefont {Mendoza}, \citenamefont {Morimoto},
  \citenamefont {Moore},\ and\ \citenamefont {Neaton}}]{Rangel2017}%
  \BibitemOpen
  \bibfield  {author} {\bibinfo {author} {\bibfnamefont {T.}~\bibnamefont
  {Rangel}}, \bibinfo {author} {\bibfnamefont {B.~M.}\ \bibnamefont {Fregoso}},
  \bibinfo {author} {\bibfnamefont {B.~S.}\ \bibnamefont {Mendoza}}, \bibinfo
  {author} {\bibfnamefont {T.}~\bibnamefont {Morimoto}}, \bibinfo {author}
  {\bibfnamefont {J.~E.}\ \bibnamefont {Moore}}, \ and\ \bibinfo {author}
  {\bibfnamefont {J.~B.}\ \bibnamefont {Neaton}},\ }\href@noop {} {\bibfield
  {journal} {\bibinfo  {journal} {Phys. Rev. Lett.}\ }\textbf {\bibinfo
  {volume} {119}},\ \bibinfo {pages} {067402} (\bibinfo {year}
  {2017})}\BibitemShut {NoStop}%
\bibitem [{\citenamefont {Iba\~nez Azpiroz}\ \emph {et~al.}(2018)\citenamefont
  {Iba\~nez Azpiroz}, \citenamefont {Tsirkin},\ and\ \citenamefont
  {Souza}}]{Ibanez-Azpiroz2018}%
  \BibitemOpen
  \bibfield  {author} {\bibinfo {author} {\bibfnamefont {J.}~\bibnamefont
  {Iba\~nez Azpiroz}}, \bibinfo {author} {\bibfnamefont {S.~S.}\ \bibnamefont
  {Tsirkin}}, \ and\ \bibinfo {author} {\bibfnamefont {I.}~\bibnamefont
  {Souza}},\ }\href@noop {} {\bibfield  {journal} {\bibinfo  {journal} {Phys.
  Rev. B}\ }\textbf {\bibinfo {volume} {97}},\ \bibinfo {pages} {245143}
  (\bibinfo {year} {2018})}\BibitemShut {NoStop}%
\bibitem [{\citenamefont {Panday}\ \emph {et~al.}()\citenamefont {Panday},
  \citenamefont {Barraza-Lopez}, \citenamefont {Rangel},\ and\ \citenamefont
  {Fregoso}}]{Panday}%
  \BibitemOpen
  \bibfield  {author} {\bibinfo {author} {\bibfnamefont {S.~R.}\ \bibnamefont
  {Panday}}, \bibinfo {author} {\bibfnamefont {S.}~\bibnamefont
  {Barraza-Lopez}}, \bibinfo {author} {\bibfnamefont {T.}~\bibnamefont
  {Rangel}}, \ and\ \bibinfo {author} {\bibfnamefont {B.~M.}\ \bibnamefont
  {Fregoso}},\ }\href@noop {} {\enquote {\bibinfo {title} {Injection current in
  ferroelectric group-iv monochalcogenide monolayers},}\ }\bibinfo {note}
  {ArXiv:1811.06474 [cond-mat.mes-hall]}\BibitemShut {NoStop}%
\bibitem [{\citenamefont {Wang}\ and\ \citenamefont {Qian}()}]{Wang}%
  \BibitemOpen
  \bibfield  {author} {\bibinfo {author} {\bibfnamefont {H.}~\bibnamefont
  {Wang}}\ and\ \bibinfo {author} {\bibfnamefont {X.}~\bibnamefont {Qian}},\
  }\href@noop {} {\enquote {\bibinfo {title} {Quantum nonlinear ferroic optical
  hall effect},}\ }\bibinfo {note} {ArXiv:1811.03133
  [cond-mat.mes-hall]}\BibitemShut {NoStop}%
\bibitem [{\citenamefont {Fregoso}\ \emph {et~al.}(2017)\citenamefont
  {Fregoso}, \citenamefont {Morimoto},\ and\ \citenamefont
  {Moore}}]{Fregoso2017}%
  \BibitemOpen
  \bibfield  {author} {\bibinfo {author} {\bibfnamefont {B.~M.}\ \bibnamefont
  {Fregoso}}, \bibinfo {author} {\bibfnamefont {T.}~\bibnamefont {Morimoto}}, \
  and\ \bibinfo {author} {\bibfnamefont {J.~E.}\ \bibnamefont {Moore}},\
  }\href@noop {} {\bibfield  {journal} {\bibinfo  {journal} {Phys. Rev. B}\
  }\textbf {\bibinfo {volume} {96}},\ \bibinfo {pages} {075421} (\bibinfo
  {year} {2017})}\BibitemShut {NoStop}%
\bibitem [{\citenamefont {Kushnir}\ \emph {et~al.}(2017)\citenamefont
  {Kushnir}, \citenamefont {Wang}, \citenamefont {Fitzgerald}, \citenamefont
  {Koski},\ and\ \citenamefont {Titova}}]{Kushnir2017}%
  \BibitemOpen
  \bibfield  {author} {\bibinfo {author} {\bibfnamefont {K.}~\bibnamefont
  {Kushnir}}, \bibinfo {author} {\bibfnamefont {M.}~\bibnamefont {Wang}},
  \bibinfo {author} {\bibfnamefont {P.~D.}\ \bibnamefont {Fitzgerald}},
  \bibinfo {author} {\bibfnamefont {K.~J.}\ \bibnamefont {Koski}}, \ and\
  \bibinfo {author} {\bibfnamefont {L.~V.}\ \bibnamefont {Titova}},\
  }\href@noop {} {\bibfield  {journal} {\bibinfo  {journal} {ACS Energy
  Letters}\ }\textbf {\bibinfo {volume} {2}},\ \bibinfo {pages} {1429}
  (\bibinfo {year} {2017})}\BibitemShut {NoStop}%
\bibitem [{\citenamefont {Nakamura}\ \emph {et~al.}(2017)\citenamefont
  {Nakamura}, \citenamefont {Horiuchi}, \citenamefont {Kagawa}, \citenamefont
  {Ogawa}, \citenamefont {Kurumaji}, \citenamefont {Tokura},\ and\
  \citenamefont {Kawasaki}}]{Nakamura2017}%
  \BibitemOpen
  \bibfield  {author} {\bibinfo {author} {\bibfnamefont {M.}~\bibnamefont
  {Nakamura}}, \bibinfo {author} {\bibfnamefont {S.}~\bibnamefont {Horiuchi}},
  \bibinfo {author} {\bibfnamefont {F.}~\bibnamefont {Kagawa}}, \bibinfo
  {author} {\bibfnamefont {N.}~\bibnamefont {Ogawa}}, \bibinfo {author}
  {\bibfnamefont {T.}~\bibnamefont {Kurumaji}}, \bibinfo {author}
  {\bibfnamefont {Y.}~\bibnamefont {Tokura}}, \ and\ \bibinfo {author}
  {\bibfnamefont {M.}~\bibnamefont {Kawasaki}},\ }\href@noop {} {\bibfield
  {journal} {\bibinfo  {journal} {Nature Communication}\ }\textbf {\bibinfo
  {volume} {8}},\ \bibinfo {pages} {281} (\bibinfo {year} {2017})}\BibitemShut
  {NoStop}%
\bibitem [{\citenamefont {Ogawa}\ \emph {et~al.}(2017)\citenamefont {Ogawa},
  \citenamefont {Sotome}, \citenamefont {Kaneko}, \citenamefont {Ogino},\ and\
  \citenamefont {Tokura}}]{Ogawa2017}%
  \BibitemOpen
  \bibfield  {author} {\bibinfo {author} {\bibfnamefont {N.}~\bibnamefont
  {Ogawa}}, \bibinfo {author} {\bibfnamefont {M.}~\bibnamefont {Sotome}},
  \bibinfo {author} {\bibfnamefont {Y.}~\bibnamefont {Kaneko}}, \bibinfo
  {author} {\bibfnamefont {M.}~\bibnamefont {Ogino}}, \ and\ \bibinfo {author}
  {\bibfnamefont {Y.}~\bibnamefont {Tokura}},\ }\href@noop {} {\bibfield
  {journal} {\bibinfo  {journal} {Phys. Rev. B}\ }\textbf {\bibinfo {volume}
  {96}},\ \bibinfo {pages} {241203} (\bibinfo {year} {2017})}\BibitemShut
  {NoStop}%
\bibitem [{\citenamefont {Kushnir}\ \emph {et~al.}(2019)\citenamefont
  {Kushnir}, \citenamefont {Qin}, \citenamefont {Shen}, \citenamefont {Li},
  \citenamefont {Fregoso}, \citenamefont {Tongay},\ and\ \citenamefont
  {Titova}}]{Kushnir2019}%
  \BibitemOpen
  \bibfield  {author} {\bibinfo {author} {\bibfnamefont {K.}~\bibnamefont
  {Kushnir}}, \bibinfo {author} {\bibfnamefont {Y.}~\bibnamefont {Qin}},
  \bibinfo {author} {\bibfnamefont {Y.}~\bibnamefont {Shen}}, \bibinfo {author}
  {\bibfnamefont {G.}~\bibnamefont {Li}}, \bibinfo {author} {\bibfnamefont
  {B.~M.}\ \bibnamefont {Fregoso}}, \bibinfo {author} {\bibfnamefont
  {S.}~\bibnamefont {Tongay}}, \ and\ \bibinfo {author} {\bibfnamefont {L.~V.}\
  \bibnamefont {Titova}},\ }\href@noop {} {\bibfield  {journal} {\bibinfo
  {journal} {ACS Applied Materials \& Interfaces}\ }\textbf {\bibinfo {volume}
  {11}},\ \bibinfo {pages} {5492} (\bibinfo {year} {2019})}\BibitemShut
  {NoStop}%
\bibitem [{\citenamefont {Burger}\ \emph {et~al.}(2019)\citenamefont {Burger},
  \citenamefont {Agarwal}, \citenamefont {Aprelev}, \citenamefont {Schruba},
  \citenamefont {Gutierrez-Perez}, \citenamefont {Fridkin},\ and\ \citenamefont
  {Spanier}}]{Burger2019}%
  \BibitemOpen
  \bibfield  {author} {\bibinfo {author} {\bibfnamefont {A.~M.}\ \bibnamefont
  {Burger}}, \bibinfo {author} {\bibfnamefont {R.}~\bibnamefont {Agarwal}},
  \bibinfo {author} {\bibfnamefont {A.}~\bibnamefont {Aprelev}}, \bibinfo
  {author} {\bibfnamefont {E.}~\bibnamefont {Schruba}}, \bibinfo {author}
  {\bibfnamefont {A.}~\bibnamefont {Gutierrez-Perez}}, \bibinfo {author}
  {\bibfnamefont {V.~M.}\ \bibnamefont {Fridkin}}, \ and\ \bibinfo {author}
  {\bibfnamefont {J.~E.}\ \bibnamefont {Spanier}},\ }\href@noop {} {\bibfield
  {journal} {\bibinfo  {journal} {Science Adv.}\ }\textbf {\bibinfo {volume}
  {5}},\ \bibinfo {pages} {eaau5588} (\bibinfo {year} {2019})}\BibitemShut
  {NoStop}%
\bibitem [{\citenamefont {Sotome}\ \emph {et~al.}(2019)\citenamefont {Sotome},
  \citenamefont {Nakamura}, \citenamefont {Fujioka}, \citenamefont {Ogino},
  \citenamefont {Kaneko}, \citenamefont {Morimoto}, \citenamefont {Zhang},
  \citenamefont {Kawasaki}, \citenamefont {Nagaosa}, \citenamefont {Tokura},\
  and\ \citenamefont {Ogawa}}]{Sotome2019}%
  \BibitemOpen
  \bibfield  {author} {\bibinfo {author} {\bibfnamefont {M.}~\bibnamefont
  {Sotome}}, \bibinfo {author} {\bibfnamefont {M.}~\bibnamefont {Nakamura}},
  \bibinfo {author} {\bibfnamefont {J.}~\bibnamefont {Fujioka}}, \bibinfo
  {author} {\bibfnamefont {M.}~\bibnamefont {Ogino}}, \bibinfo {author}
  {\bibfnamefont {Y.}~\bibnamefont {Kaneko}}, \bibinfo {author} {\bibfnamefont
  {T.}~\bibnamefont {Morimoto}}, \bibinfo {author} {\bibfnamefont
  {Y.}~\bibnamefont {Zhang}}, \bibinfo {author} {\bibfnamefont
  {M.}~\bibnamefont {Kawasaki}}, \bibinfo {author} {\bibfnamefont
  {N.}~\bibnamefont {Nagaosa}}, \bibinfo {author} {\bibfnamefont
  {Y.}~\bibnamefont {Tokura}}, \ and\ \bibinfo {author} {\bibfnamefont
  {N.}~\bibnamefont {Ogawa}},\ }\href@noop {} {\bibfield  {journal} {\bibinfo
  {journal} {Applied Physics Letters}\ }\textbf {\bibinfo {volume} {114}},\
  \bibinfo {pages} {151101} (\bibinfo {year} {2019})}\BibitemShut {NoStop}%
\bibitem [{\citenamefont {de~Juan}\ \emph {et~al.}(2017)\citenamefont
  {de~Juan}, \citenamefont {Grushin}, \citenamefont {Morimoto},\ and\
  \citenamefont {Moore}}]{Juan2017}%
  \BibitemOpen
  \bibfield  {author} {\bibinfo {author} {\bibfnamefont {F.}~\bibnamefont
  {de~Juan}}, \bibinfo {author} {\bibfnamefont {A.~G.}\ \bibnamefont
  {Grushin}}, \bibinfo {author} {\bibfnamefont {T.}~\bibnamefont {Morimoto}}, \
  and\ \bibinfo {author} {\bibfnamefont {J.~E.}\ \bibnamefont {Moore}},\
  }\href@noop {} {\bibfield  {journal} {\bibinfo  {journal} {Nature
  Communications}\ }\textbf {\bibinfo {volume} {8}},\ \bibinfo {pages} {15995}
  (\bibinfo {year} {2017})}\BibitemShut {NoStop}%
\bibitem [{\citenamefont {Rees}\ \emph {et~al.}()\citenamefont {Rees},
  \citenamefont {Manna}, \citenamefont {Lu}, \citenamefont {Morimoto},
  \citenamefont {Borrmann}, \citenamefont {Felser}, \citenamefont {Moore},
  \citenamefont {Torchinsky},\ and\ \citenamefont {Orenstein}}]{Rees}%
  \BibitemOpen
  \bibfield  {author} {\bibinfo {author} {\bibfnamefont {D.}~\bibnamefont
  {Rees}}, \bibinfo {author} {\bibfnamefont {K.}~\bibnamefont {Manna}},
  \bibinfo {author} {\bibfnamefont {B.}~\bibnamefont {Lu}}, \bibinfo {author}
  {\bibfnamefont {T.}~\bibnamefont {Morimoto}}, \bibinfo {author}
  {\bibfnamefont {H.}~\bibnamefont {Borrmann}}, \bibinfo {author}
  {\bibfnamefont {C.}~\bibnamefont {Felser}}, \bibinfo {author} {\bibfnamefont
  {J.}~\bibnamefont {Moore}}, \bibinfo {author} {\bibfnamefont {D.~H.}\
  \bibnamefont {Torchinsky}}, \ and\ \bibinfo {author} {\bibfnamefont
  {J.}~\bibnamefont {Orenstein}},\ }\href@noop {} {\enquote {\bibinfo {title}
  {Quantized photocurrents in the chiral multifold fermion system rhsi},}\
  }\bibinfo {note} {ArXiv:1902.03230 [cond-mat.mes-hall]}\BibitemShut {NoStop}%
\bibitem [{\citenamefont {Boyd}(2003)}]{Boyd2008}%
  \BibitemOpen
  \bibfield  {author} {\bibinfo {author} {\bibfnamefont {R.~W.}\ \bibnamefont
  {Boyd}},\ }\href@noop {} {\emph {\bibinfo {title} {Nonlinear Optics}}}\
  (\bibinfo  {publisher} {Academic Press; 2nd edition, San Diego, USA},\
  \bibinfo {year} {2003})\BibitemShut {NoStop}%
\bibitem [{\citenamefont {Aversa}\ and\ \citenamefont
  {Sipe}(1995)}]{Aversa1995}%
  \BibitemOpen
  \bibfield  {author} {\bibinfo {author} {\bibfnamefont {C.}~\bibnamefont
  {Aversa}}\ and\ \bibinfo {author} {\bibfnamefont {J.~E.}\ \bibnamefont
  {Sipe}},\ }\href@noop {} {\bibfield  {journal} {\bibinfo  {journal} {Phys.
  Rev. B}\ }\textbf {\bibinfo {volume} {52}},\ \bibinfo {pages} {14636}
  (\bibinfo {year} {1995})}\BibitemShut {NoStop}%
\bibitem [{\citenamefont {Fregoso}\ \emph {et~al.}(2018)\citenamefont
  {Fregoso}, \citenamefont {Muniz},\ and\ \citenamefont {Sipe}}]{Fregoso2018}%
  \BibitemOpen
  \bibfield  {author} {\bibinfo {author} {\bibfnamefont {B.~M.}\ \bibnamefont
  {Fregoso}}, \bibinfo {author} {\bibfnamefont {R.~A.}\ \bibnamefont {Muniz}},
  \ and\ \bibinfo {author} {\bibfnamefont {J.~E.}\ \bibnamefont {Sipe}},\
  }\href@noop {} {\bibfield  {journal} {\bibinfo  {journal} {Phys. Rev. Lett.}\
  }\textbf {\bibinfo {volume} {121}},\ \bibinfo {pages} {176604} (\bibinfo
  {year} {2018})}\BibitemShut {NoStop}%
\bibitem [{\citenamefont {Aversa}\ and\ \citenamefont
  {Sipe}(1996)}]{Aversa1996}%
  \BibitemOpen
  \bibfield  {author} {\bibinfo {author} {\bibfnamefont {C.}~\bibnamefont
  {Aversa}}\ and\ \bibinfo {author} {\bibfnamefont {J.~E.}\ \bibnamefont
  {Sipe}},\ }\href@noop {} {\bibfield  {journal} {\bibinfo  {journal} {IEEE
  Journal of Quantum Electronics}\ }\textbf {\bibinfo {volume} {32}},\ \bibinfo
  {pages} {1570} (\bibinfo {year} {1996})}\BibitemShut {NoStop}%
\bibitem [{\citenamefont {King-Smith}\ and\ \citenamefont
  {Vanderbilt}(1993)}]{King-Smith1993}%
  \BibitemOpen
  \bibfield  {author} {\bibinfo {author} {\bibfnamefont {R.~D.}\ \bibnamefont
  {King-Smith}}\ and\ \bibinfo {author} {\bibfnamefont {D.}~\bibnamefont
  {Vanderbilt}},\ }\href@noop {} {\bibfield  {journal} {\bibinfo  {journal}
  {Phys. Rev. B}\ }\textbf {\bibinfo {volume} {47}},\ \bibinfo {pages} {1651}
  (\bibinfo {year} {1993})}\BibitemShut {NoStop}%
\bibitem [{\citenamefont {Resta}(1994)}]{Resta1994}%
  \BibitemOpen
  \bibfield  {author} {\bibinfo {author} {\bibfnamefont {R.}~\bibnamefont
  {Resta}},\ }\href@noop {} {\bibfield  {journal} {\bibinfo  {journal} {Rev.
  Mod. Phys.}\ }\textbf {\bibinfo {volume} {66}},\ \bibinfo {pages} {899}
  (\bibinfo {year} {1994})}\BibitemShut {NoStop}%
\bibitem [{com()}]{comment_1}%
  \BibitemOpen
  \href@noop {} {}\bibinfo {note} {In the standard notation of
  susceptibilities~\cite{Boyd2008} a permittivity of free space, $\epsilon_0$,
  is factored out of $\chi_n$. For clarity of notation we dont factor this
  term.}\BibitemShut {Stop}%
\bibitem [{\citenamefont {Aspnes}(1972)}]{Aspnes1972}%
  \BibitemOpen
  \bibfield  {author} {\bibinfo {author} {\bibfnamefont {D.~E.}\ \bibnamefont
  {Aspnes}},\ }\href@noop {} {\bibfield  {journal} {\bibinfo  {journal} {Phys.
  Rev. B}\ }\textbf {\bibinfo {volume} {6}},\ \bibinfo {pages} {4648} (\bibinfo
  {year} {1972})}\BibitemShut {NoStop}%
\bibitem [{\citenamefont {Culcer}\ \emph {et~al.}(2017)\citenamefont {Culcer},
  \citenamefont {Sekine},\ and\ \citenamefont {MacDonald}}]{Culcer2017}%
  \BibitemOpen
  \bibfield  {author} {\bibinfo {author} {\bibfnamefont {D.}~\bibnamefont
  {Culcer}}, \bibinfo {author} {\bibfnamefont {A.}~\bibnamefont {Sekine}}, \
  and\ \bibinfo {author} {\bibfnamefont {A.~H.}\ \bibnamefont {MacDonald}},\
  }\href@noop {} {\bibfield  {journal} {\bibinfo  {journal} {Phys. Rev. B}\
  }\textbf {\bibinfo {volume} {96}},\ \bibinfo {pages} {035106} (\bibinfo
  {year} {2017})}\BibitemShut {NoStop}%
\bibitem [{\citenamefont {Bass}\ \emph {et~al.}(1962)\citenamefont {Bass},
  \citenamefont {Franken}, \citenamefont {Ward},\ and\ \citenamefont
  {Weinreich}}]{Bass1962}%
  \BibitemOpen
  \bibfield  {author} {\bibinfo {author} {\bibfnamefont {M.}~\bibnamefont
  {Bass}}, \bibinfo {author} {\bibfnamefont {P.~A.}\ \bibnamefont {Franken}},
  \bibinfo {author} {\bibfnamefont {J.~F.}\ \bibnamefont {Ward}}, \ and\
  \bibinfo {author} {\bibfnamefont {G.}~\bibnamefont {Weinreich}},\ }\href@noop
  {} {\bibfield  {journal} {\bibinfo  {journal} {Phys. Rev. Lett.}\ }\textbf
  {\bibinfo {volume} {9}},\ \bibinfo {pages} {446} (\bibinfo {year}
  {1962})}\BibitemShut {NoStop}%
\bibitem [{\citenamefont {Nastos}\ and\ \citenamefont
  {Sipe}(2010)}]{Nastos2010}%
  \BibitemOpen
  \bibfield  {author} {\bibinfo {author} {\bibfnamefont {F.}~\bibnamefont
  {Nastos}}\ and\ \bibinfo {author} {\bibfnamefont {J.~E.}\ \bibnamefont
  {Sipe}},\ }\href@noop {} {\bibfield  {journal} {\bibinfo  {journal} {Phys.
  Rev. B}\ }\textbf {\bibinfo {volume} {82}},\ \bibinfo {pages} {235204}
  (\bibinfo {year} {2010})}\BibitemShut {NoStop}%
\bibitem [{\citenamefont {Jepsen}\ \emph {et~al.}(1996)\citenamefont {Jepsen},
  \citenamefont {Jacobsen},\ and\ \citenamefont {Keiding}}]{Jepsen1996}%
  \BibitemOpen
  \bibfield  {author} {\bibinfo {author} {\bibfnamefont {P.~U.}\ \bibnamefont
  {Jepsen}}, \bibinfo {author} {\bibfnamefont {R.~H.}\ \bibnamefont
  {Jacobsen}}, \ and\ \bibinfo {author} {\bibfnamefont {S.~R.}\ \bibnamefont
  {Keiding}},\ }\href@noop {} {\bibfield  {journal} {\bibinfo  {journal} {J.
  Opt. Soc. Am. B}\ }\textbf {\bibinfo {volume} {13}},\ \bibinfo {pages} {2424}
  (\bibinfo {year} {1996})}\BibitemShut {NoStop}%
\bibitem [{\citenamefont {Li}\ \emph {et~al.}(2018)\citenamefont {Li},
  \citenamefont {Kushnir}, \citenamefont {Wang}, \citenamefont {Dong},
  \citenamefont {Chertopalov}, \citenamefont {Rao}, \citenamefont {Mochalin},
  \citenamefont {Podila}, \citenamefont {Koski},\ and\ \citenamefont
  {Titova}}]{Li2018}%
  \BibitemOpen
  \bibfield  {author} {\bibinfo {author} {\bibfnamefont {G.}~\bibnamefont
  {Li}}, \bibinfo {author} {\bibfnamefont {K.}~\bibnamefont {Kushnir}},
  \bibinfo {author} {\bibfnamefont {M.}~\bibnamefont {Wang}}, \bibinfo {author}
  {\bibfnamefont {Y.}~\bibnamefont {Dong}}, \bibinfo {author} {\bibfnamefont
  {S.}~\bibnamefont {Chertopalov}}, \bibinfo {author} {\bibfnamefont {A.~M.}\
  \bibnamefont {Rao}}, \bibinfo {author} {\bibfnamefont {V.~N.}\ \bibnamefont
  {Mochalin}}, \bibinfo {author} {\bibfnamefont {R.}~\bibnamefont {Podila}},
  \bibinfo {author} {\bibfnamefont {K.}~\bibnamefont {Koski}}, \ and\ \bibinfo
  {author} {\bibfnamefont {L.~V.}\ \bibnamefont {Titova}},\ }in\ \href@noop {}
  {\emph {\bibinfo {booktitle} {2018 43rd International Conference on Infrared,
  Millimeter, and Terahertz Waves (IRMMW-THz)}}}\ (\bibinfo {year}
  {2018})\BibitemShut {NoStop}%
\bibitem [{\citenamefont {Blount}(1962)}]{Blount1962}%
  \BibitemOpen
  \bibfield  {author} {\bibinfo {author} {\bibfnamefont {E.~I.}\ \bibnamefont
  {Blount}},\ }\href@noop {} {\emph {\bibinfo {title} {Solid State Physics:
  Advances in Research and Applications}}},\ Vol.\ \bibinfo {volume} {vol 13}\
  (\bibinfo  {publisher} {Academic Press},\ \bibinfo {year} {1962})\BibitemShut
  {NoStop}%
\bibitem [{\citenamefont {Haldane}(2004)}]{Haldane2004}%
  \BibitemOpen
  \bibfield  {author} {\bibinfo {author} {\bibfnamefont {F.~D.~M.}\
  \bibnamefont {Haldane}},\ }\href@noop {} {\bibfield  {journal} {\bibinfo
  {journal} {Phys. Rev. Lett.}\ }\textbf {\bibinfo {volume} {93}},\ \bibinfo
  {pages} {206602} (\bibinfo {year} {2004})}\BibitemShut {NoStop}%
\bibitem [{\citenamefont {Sodemann}\ and\ \citenamefont
  {Fu}(2015)}]{Sodemann2015}%
  \BibitemOpen
  \bibfield  {author} {\bibinfo {author} {\bibfnamefont {I.}~\bibnamefont
  {Sodemann}}\ and\ \bibinfo {author} {\bibfnamefont {L.}~\bibnamefont {Fu}},\
  }\href@noop {} {\bibfield  {journal} {\bibinfo  {journal} {Phys. Rev. Lett.}\
  }\textbf {\bibinfo {volume} {115}},\ \bibinfo {pages} {216806} (\bibinfo
  {year} {2015})}\BibitemShut {NoStop}%
\bibitem [{\citenamefont {Moore}\ and\ \citenamefont
  {Orenstein}(2010)}]{Moore2010}%
  \BibitemOpen
  \bibfield  {author} {\bibinfo {author} {\bibfnamefont {J.~E.}\ \bibnamefont
  {Moore}}\ and\ \bibinfo {author} {\bibfnamefont {J.}~\bibnamefont
  {Orenstein}},\ }\href@noop {} {\bibfield  {journal} {\bibinfo  {journal}
  {Phys. Rev. Lett.}\ }\textbf {\bibinfo {volume} {105}},\ \bibinfo {pages}
  {026805} (\bibinfo {year} {2010})}\BibitemShut {NoStop}%
\bibitem [{\citenamefont {Xiao}\ \emph {et~al.}(2010)\citenamefont {Xiao},
  \citenamefont {Chang},\ and\ \citenamefont {Niu}}]{Xiao2010}%
  \BibitemOpen
  \bibfield  {author} {\bibinfo {author} {\bibfnamefont {D.}~\bibnamefont
  {Xiao}}, \bibinfo {author} {\bibfnamefont {M.-C.}\ \bibnamefont {Chang}}, \
  and\ \bibinfo {author} {\bibfnamefont {Q.}~\bibnamefont {Niu}},\ }\href@noop
  {} {\bibfield  {journal} {\bibinfo  {journal} {Rev. Mod. Phys.}\ }\textbf
  {\bibinfo {volume} {82}},\ \bibinfo {pages} {1959} (\bibinfo {year}
  {2010})}\BibitemShut {NoStop}%
\bibitem [{\citenamefont {Matsyshyn}\ and\ \citenamefont
  {Sodemann}()}]{Matsyshyn}%
  \BibitemOpen
  \bibfield  {author} {\bibinfo {author} {\bibfnamefont {O.}~\bibnamefont
  {Matsyshyn}}\ and\ \bibinfo {author} {\bibfnamefont {I.}~\bibnamefont
  {Sodemann}},\ }\href@noop {} {\enquote {\bibinfo {title} {The non-linear hall
  acceleration and the quantum rectification sum rule},}\ }\bibinfo {note}
  {ArXiv:1907.02532 [cond-mat.mes-hall]}\BibitemShut {NoStop}%
\bibitem [{\citenamefont {Gomes}\ and\ \citenamefont
  {Carvalho}(2015)}]{Gomes2015}%
  \BibitemOpen
  \bibfield  {author} {\bibinfo {author} {\bibfnamefont {L.~C.}\ \bibnamefont
  {Gomes}}\ and\ \bibinfo {author} {\bibfnamefont {A.}~\bibnamefont
  {Carvalho}},\ }\href@noop {} {\bibfield  {journal} {\bibinfo  {journal}
  {Phys. Rev. B}\ }\textbf {\bibinfo {volume} {92}},\ \bibinfo {pages} {085406}
  (\bibinfo {year} {2015})}\BibitemShut {NoStop}%
\bibitem [{\citenamefont {Cook}\ \emph {et~al.}(2017)\citenamefont {Cook},
  \citenamefont {M.~Fregoso}, \citenamefont {de~Juan}, \citenamefont {Coh},\
  and\ \citenamefont {Moore}}]{Cook2017}%
  \BibitemOpen
  \bibfield  {author} {\bibinfo {author} {\bibfnamefont {A.~M.}\ \bibnamefont
  {Cook}}, \bibinfo {author} {\bibfnamefont {B.}~\bibnamefont {M.~Fregoso}},
  \bibinfo {author} {\bibfnamefont {F.}~\bibnamefont {de~Juan}}, \bibinfo
  {author} {\bibfnamefont {S.}~\bibnamefont {Coh}}, \ and\ \bibinfo {author}
  {\bibfnamefont {J.~E.}\ \bibnamefont {Moore}},\ }\href@noop {} {\bibfield
  {journal} {\bibinfo  {journal} {Nature Communications}\ }\textbf {\bibinfo
  {volume} {8}},\ \bibinfo {pages} {14176} (\bibinfo {year}
  {2017})}\BibitemShut {NoStop}%
\bibitem [{\citenamefont {Panday}\ and\ \citenamefont
  {Fregoso}(2017)}]{Panday2017}%
  \BibitemOpen
  \bibfield  {author} {\bibinfo {author} {\bibfnamefont {S.~R.}\ \bibnamefont
  {Panday}}\ and\ \bibinfo {author} {\bibfnamefont {B.~M.}\ \bibnamefont
  {Fregoso}},\ }\href@noop {} {\bibfield  {journal} {\bibinfo  {journal}
  {Journal of Physics: Condensed Matter}\ }\textbf {\bibinfo {volume} {29}},\
  \bibinfo {pages} {43LT01} (\bibinfo {year} {2017})}\BibitemShut {NoStop}%
\bibitem [{\citenamefont {Wang}\ and\ \citenamefont {Qian}(2017)}]{Wang2017}%
  \BibitemOpen
  \bibfield  {author} {\bibinfo {author} {\bibfnamefont {H.}~\bibnamefont
  {Wang}}\ and\ \bibinfo {author} {\bibfnamefont {X.}~\bibnamefont {Qian}},\
  }\href@noop {} {\bibfield  {journal} {\bibinfo  {journal} {Nano Letters}\
  }\textbf {\bibinfo {volume} {17}},\ \bibinfo {pages} {5027} (\bibinfo {year}
  {2017})}\BibitemShut {NoStop}%
\bibitem [{\citenamefont {Sun}\ \emph {et~al.}(2010)\citenamefont {Sun},
  \citenamefont {Divin}, \citenamefont {Rioux}, \citenamefont {Sipe},
  \citenamefont {Berger}, \citenamefont {de~Heer}, \citenamefont {First},\ and\
  \citenamefont {Norris}}]{Sun2010}%
  \BibitemOpen
  \bibfield  {author} {\bibinfo {author} {\bibfnamefont {D.}~\bibnamefont
  {Sun}}, \bibinfo {author} {\bibfnamefont {C.}~\bibnamefont {Divin}}, \bibinfo
  {author} {\bibfnamefont {J.}~\bibnamefont {Rioux}}, \bibinfo {author}
  {\bibfnamefont {J.~E.}\ \bibnamefont {Sipe}}, \bibinfo {author}
  {\bibfnamefont {C.}~\bibnamefont {Berger}}, \bibinfo {author} {\bibfnamefont
  {W.~A.}\ \bibnamefont {de~Heer}}, \bibinfo {author} {\bibfnamefont {P.~N.}\
  \bibnamefont {First}}, \ and\ \bibinfo {author} {\bibfnamefont {T.~B.}\
  \bibnamefont {Norris}},\ }\href@noop {} {\bibfield  {journal} {\bibinfo
  {journal} {Nano Letters}\ }\textbf {\bibinfo {volume} {10}},\ \bibinfo
  {pages} {1293} (\bibinfo {year} {2010})}\BibitemShut {NoStop}%
\bibitem [{\citenamefont {Bas}\ \emph {et~al.}(2015)\citenamefont {Bas},
  \citenamefont {Vargas-Velez}, \citenamefont {Babakiray}, \citenamefont
  {Johnson}, \citenamefont {Borisov}, \citenamefont {Stanescu}, \citenamefont
  {Lederman},\ and\ \citenamefont {Bristow}}]{Bas2015}%
  \BibitemOpen
  \bibfield  {author} {\bibinfo {author} {\bibfnamefont {D.~A.}\ \bibnamefont
  {Bas}}, \bibinfo {author} {\bibfnamefont {K.}~\bibnamefont {Vargas-Velez}},
  \bibinfo {author} {\bibfnamefont {S.}~\bibnamefont {Babakiray}}, \bibinfo
  {author} {\bibfnamefont {T.~A.}\ \bibnamefont {Johnson}}, \bibinfo {author}
  {\bibfnamefont {P.}~\bibnamefont {Borisov}}, \bibinfo {author} {\bibfnamefont
  {T.~D.}\ \bibnamefont {Stanescu}}, \bibinfo {author} {\bibfnamefont
  {D.}~\bibnamefont {Lederman}}, \ and\ \bibinfo {author} {\bibfnamefont
  {A.~D.}\ \bibnamefont {Bristow}},\ }\href@noop {} {\bibfield  {journal}
  {\bibinfo  {journal} {Applied Physics Letters}\ }\textbf {\bibinfo {volume}
  {106}},\ \bibinfo {pages} {041109} (\bibinfo {year} {2015})}\BibitemShut
  {NoStop}%
\bibitem [{\citenamefont {Bas}\ \emph {et~al.}(2016)\citenamefont {Bas},
  \citenamefont {Muniz}, \citenamefont {Babakiray}, \citenamefont {Lederman},
  \citenamefont {Sipe},\ and\ \citenamefont {Bristow}}]{Bas2016}%
  \BibitemOpen
  \bibfield  {author} {\bibinfo {author} {\bibfnamefont {D.~A.}\ \bibnamefont
  {Bas}}, \bibinfo {author} {\bibfnamefont {R.~A.}\ \bibnamefont {Muniz}},
  \bibinfo {author} {\bibfnamefont {S.}~\bibnamefont {Babakiray}}, \bibinfo
  {author} {\bibfnamefont {D.}~\bibnamefont {Lederman}}, \bibinfo {author}
  {\bibfnamefont {J.~E.}\ \bibnamefont {Sipe}}, \ and\ \bibinfo {author}
  {\bibfnamefont {A.~D.}\ \bibnamefont {Bristow}},\ }\href@noop {} {\bibfield
  {journal} {\bibinfo  {journal} {Opt. Express}\ }\textbf {\bibinfo {volume}
  {24}},\ \bibinfo {pages} {23583} (\bibinfo {year} {2016})}\BibitemShut
  {NoStop}%
\bibitem [{\citenamefont {Atanasov}\ \emph {et~al.}(1996)\citenamefont
  {Atanasov}, \citenamefont {Hach\'e}, \citenamefont {Hughes}, \citenamefont
  {van Driel},\ and\ \citenamefont {Sipe}}]{Atanasov1996}%
  \BibitemOpen
  \bibfield  {author} {\bibinfo {author} {\bibfnamefont {R.}~\bibnamefont
  {Atanasov}}, \bibinfo {author} {\bibfnamefont {A.}~\bibnamefont {Hach\'e}},
  \bibinfo {author} {\bibfnamefont {J.~L.~P.}\ \bibnamefont {Hughes}}, \bibinfo
  {author} {\bibfnamefont {H.~M.}\ \bibnamefont {van Driel}}, \ and\ \bibinfo
  {author} {\bibfnamefont {J.~E.}\ \bibnamefont {Sipe}},\ }\href@noop {}
  {\bibfield  {journal} {\bibinfo  {journal} {Phys. Rev. Lett.}\ }\textbf
  {\bibinfo {volume} {76}},\ \bibinfo {pages} {1703} (\bibinfo {year}
  {1996})}\BibitemShut {NoStop}%
\end{thebibliography}

%

\end{document}